\documentclass[12pt]{article}
\usepackage{amsmath}
\usepackage{multirow}
\usepackage{amsfonts}
\usepackage{amssymb,graphics,psfrag}
\usepackage{array,epsfig,multirow,stmaryrd,graphicx}
\usepackage{comment}
\usepackage{feynmp}

\usepackage{hyperref}

\def\hybrid{\topmargin -20pt    \oddsidemargin 0pt
        \headheight 0pt \headsep 0pt
        \textwidth 6.25in      
        \textheight 9 in      
        \marginparwidth .875in
        \parskip 5pt plus 1pt
          \jot = 1.5ex
  }
\hybrid

\newcommand{\be}{\begin{equation}}
\newcommand{\ee}{\end{equation}} 
\newcommand{\bea}{\begin{eqnarray}}
\newcommand{\eea}{\end{eqnarray}}

\newcommand{\nn}{\nonumber}

\newcommand{\beq}{\begin{equation}}
\newcommand{\eeq}{\end{equation}}

\newcommand{\cC}{\mathcal{C}}
\newcommand{\cD}{\mathcal{D}}

\newcommand{\cK}{\mathcal{K}}
\newcommand{\cN}{\mathcal{N}}

\newcommand{\cA}{\mathcal{A}}

\newcommand{\cB}{\mathcal{B}}
\newcommand{\cF}{\mathcal{F}}

\newcommand{\bbP}{\mathbb{P}}

\newcommand{\cref}{{\bf [check ref]}}

\newcommand{\repr}{\mathbf{R}}
\newcommand{\repA}{\mathbf{r}}

\newcommand{\tr}{\mathrm{Tr}\:}

\def\blfootnote{\xdef\@thefnmark{}\@footnotetext}
\long\def\symbolfootnote[#1]#2{\begingroup%
\def\thefootnote{\fnsymbol{footnote}}\footnote[#1]{#2}\endgroup}

\begin{document}

\baselineskip=15pt

\begin{titlepage}
\begin{flushright}
\parbox[t]{1.08in}{UPR-1243-T\\MPP-2012-126}
\end{flushright}

\begin{center}

\vspace*{ 1.2cm}

{\Large \bf Anomaly Cancellation And \\[.2cm]
    Abelian Gauge Symmetries In F-theory}

\vskip 1.2cm

\renewcommand{\thefootnote}{}
\begin{center}
 {Mirjam Cveti\v{c}$^{1,2}$, Thomas W.~Grimm$^3$ and Denis Klevers$^1$}
\end{center}
\vskip .2cm
\renewcommand{\thefootnote}{\arabic{footnote}} 

$\,^1$ {Department of Physics and Astronomy,\\
University of Pennsylvania, Philadelphia, PA 19104-6396, USA} \\[.3cm]

$\,^2$ {Center for Applied Mathematics and Theoretical Physics,\\ University of Maribor, Maribor, Slovenia}\\[.3cm]

$\,^3$ {Max-Planck-Institut f\"ur Physik, \\
F\"ohringer Ring 6, 80805 Munich, Germany} \\[.3cm]

{cvetic\ \textsf{at}\ cvetic.hep.upenn.edu, grimm\ \textsf{at}\ mpp.mpg.de, klevers\ \textsf{at}\ sas.upenn.edu } 

 \vspace*{0.8cm}

\end{center}

\vskip 0.2cm
 
\begin{center} {\bf ABSTRACT } \end{center}

We study 4D F-theory compactifications on singular Calabi-Yau fourfolds 
with fluxes. The resulting $\mathcal{N}=1$ effective theories can admit 
non-Abelian and $U(1)$ gauge groups as well as charged chiral matter. 
In these setups we analyze anomaly cancellation and the generalized 
Green-Schwarz mechanism.
This requires the study of 3D $\mathcal{N}=2$ theories obtained by a 
circle compactification and their M-theory duals. Reducing M-theory on
resolved Calabi-Yau fourfolds corresponds to considering the effective 
theory on the 3D Coulomb branch in which certain massive states are 
integrated out. Both 4D gaugings and 3D one-loop corrections of these 
massive states induce Chern-Simons terms.   
All 4D anomalies are captured by the one-loop terms. The ones 
corresponding to the mixed gauge-gravitational anomalies depend on the 
Kaluza-Klein vector and  are induced by integrating out Kaluza-Klein modes 
of the $U(1)$ charged matter. In M-theory all Chern-Simons terms 
classically arise from $G_4$-flux. We find that F-theory fluxes implement 
automatically the 4D Green-Schwarz mechanism if non-trivial geometric 
relations for the resolved Calabi-Yau fourfold are satisfied. We confirm 
these relations in various explicit examples and elucidate the 
general construction of $U(1)$ symmetries in F-theory. We also compare 
anomaly cancellation in F-theory with its analog in Type IIB orientifold 
setups.

\hfill {October, 2012}
\end{titlepage}

\section{Introduction and Summary}

Local symmetries are the guiding principle for formulating 
field theories as well as the theory of gravity. 
While manifest in the classical theory these symmetries can, however,
be broken at the quantum level and lead to inconsistencies by the violation
of essential current conservation laws on the quantum level. Such 
inconsistencies manifest themselves already
at one-loop level and are known as anomalies \cite{Anomaly-reviews}. 
In particular, four-dimensional quantum field theories can admit
anomalies which signal the breaking of the gauge symmetry transformations
acting on chiral fermions. The cancellation of these anomalies is thus
crucial to determine consistent theories and imposes constraints on the
spectrum and couplings of the theory. Anomalous transformations
of the chiral fermions either cancel among each other or require
the implementation of a generalized Green-Schwarz mechanism \cite{Green:1984sg,Sagnotti:1992qw}. 
In the latter case the one-loop anomalies are
canceled by a tree-level diagram involving
a $U(1)$ gauged axion-like scalar.
In this work we study the manifestation of the  
anomaly cancellation mechanisms in four-dimensional F-theory 
compactifications. 

To extract four-dimensional observed physics from string theory one is aiming 
to find a compactification scenario that describes a very broad class of 
consistent string vacua and naturally incorporates the ingredients of the 
Standard Model of particle physics and its extensions. One 
promising scheme is to consider F-theory compactifications with space-time 
filling seven-branes supporting non-Abelian gauge groups. 
F-theory geometrizes the complexified coupling constant 
of Type IIB string theory as the complex structure modulus 
of an additional two-torus. Much of the in general non-perturbative physics of 
seven-branes is encoded 
in the singular geometry of a torus-fibered compact manifold. Requiring 
$\cN=1$ supersymmetry in the four-dimensional effective theory enforces this 
space to be an elliptically  fibered Calabi-Yau fourfold $X_4$. 
The complex three-dimensional base $B_3$ of $X_4$ is the physical 
compactification space of Type IIB string theory. 
The singularities of this fourfold along co-dimension one 
loci in the base signal the presence of non-Abelian gauge groups, while 
additional sections of the elliptic fibration signal extra $U(1)$
gauge group factors \cite{Vafa:1996xn,Morrison:1996pp}. 

A non-trivial four-dimensional chiral spectrum is only induced if the 
seven-branes also support gauge fluxes on their internal world volume. 
The interplay of geometry, fluxes and the low-energy 
effective action of the four-dimensional gauge theory is crucial 
in the study of anomaly cancellation. However, the fact that F-theory does 
not posses a low-energy effective action in twelve dimensions prevents one 
from deriving the four-dimensional low-energy effective theory directly. 
To nevertheless study 
F-theory effective physics one has to use the M-theory dual description in one
dimension lower. In fact, M-theory on the same Calabi-Yau fourfold $X_4$ is 
dual to F-theory on $X_4 \times S^1$. The new modulus, the radius of 
$S^1$, is in this identification related to 
the inverse volume of the torus fiber in M-theory. In this duality, 
the familiar F-theory limit of a shrinking volume of the torus-fiber maps to 
the decompactification limit of the $S^1$ in which one grows a large extra 
dimension and recovers four dimensional F-theory physics. All 
F-theory objects can then be followed through this three-dimensional 
duality and can be studied directly in the dual M-theory. In this work
we will mainly be interested in questions about the low energy physics
of F-theory, most prominently anomalies, and it will be one of our results to 
reformulate and answer these questions in M-theory.

At low energies the theory of interest is an effective 
three-dimensional (3D) $\cN=2$ gauge theory with a number of chiral multiplets 
coupled to supergravity. An important property of the 3D
$\cN=2$ gauge theories is that they admit, in contrast to a 
$\mathcal{N}=1$ gauge theory in four dimensions (4D), a Coulomb branch where 
the non-Abelian gauge group $G$ breaks to its maximal torus, 
$G \rightarrow U(1)^{r}$, with massless 3D gauge fields $A^\Lambda$, 
$\Lambda=1,\ldots,r=\text{rank}(G)$. In this work we will
mainly analyze the theory on this 3D Coulomb branch and understand corrections 
to the effective action both from the F-theory perspective and from the 
M-theory perspective. 

There is one essential difference between the two 
descriptions of the 3D theory, that is crucial for this discussion. 
By reducing F-theory from 4D to 3D we can describe the 3D $\mathcal{N}=2$ 
gauge theory also away from the Coulomb branch. The
Coulomb branch is understood purely field theoretically by giving a vacuum
expectation value (vev) to the scalars $\zeta^\Lambda$ along the Cartan 
generators of $G$ in the 3D vector 
multiples coming from the direction of the 4D vector along the $S^1$. Fields 
become massive due to 
this vev and at low energies have to be integrated out quantum 
mechanically correcting the effective theory on the Coulomb branch. In 
contrast, although M-theory as a fundamental theory should also describe the 
full 3D $\mathcal{N}=2$ gauge theory, it is only explicitly known how to 
describe the 3D Coulomb branch. The description is given by M-theory on the 
smooth $\hat{X}_4$ fourfold with all singularities in $X_4$ resolved inducing
shrinkable rational curves, i.e.~two-spheres $\bbP^1$, in the geometry. 
Geometrically, the resolution process corresponds to giving a vev
in the field theory perspective with $\zeta^\Lambda$ being related to the 
volume of these shrinkable $\bbP^1$'s.
Furthermore, at large volume of $\hat{X}_4$ it 
is consistent to consider the long wavelength approximation of 
eleven-dimensional supergravity and dimensionally reduce it on $\hat{X}_4$ to 3D. 
In this description certain microscopic M-theory
degrees of freedom corresponding to certain wrapped M2-branes are massive. 
However, these states have already been integrated out consistently in the 
eleven-dimensional supergravity, which
follows from the validity and consistency of the supergravity approximation at 
low energies. The corrections to the 3D low-energy theory on the Coulomb 
branch are thus visible as \textit{classical} effects in the dimensional 
reduction on $\hat{X}_4$. Morally speaking, the information about the massless  
microscopic states on $X_4$ has been traded for the new shrinkable $\bbP^1$'s 
on $\hat{X}_4$, that have not been present on $X_4$. 
These corrections are the same as those that arise 
\textit{quantum mechanically} in F-theory and it will be one key observation 
of this paper to map these corrections in the M-/F-theory duality in 3D.

The geometrically massive modes that have to be integrated out on the 3D 
Coulomb branch and are relevant for our discussion of anomalies are from an F- 
and M-theory point of view 
\begin{itemize}
	\item massive 3D W-bosons in F-theory from the breaking 
	$G\rightarrow U(1)^r$, or massive M2-branes on shrinking 
	$\bbP^1$'s in $\hat{X}_4\rightarrow X_4$ over co-dimension one in $B_3$ 
	in M-theory.
    \item charged 3D fermions massive on the 3D Coulomb branch in F-theory, or 
    massive M2-branes on $\bbP^1$'s fibered over co-dimension two in $B_3$, 
    i.e.~curves from intersections of seven-branes in F-theory later denoted 
    as matter curves, in M-theory.
    \item massive Kaluza-Klein states of 4D charged fermions in the reduction 
    of F-theory from 4D to 3D, or M2-branes wrapping a shrinking $\bbP^1$ over 
    a matter curve once and multiply the elliptic fiber of $\hat{X}_4$.
\end{itemize}

The 3D couplings that are corrected by integrating out these massive states 
and that will be of most use for our discussion of 4D 
anomalies in F-theory are the 3D Chern-Simons (CS) terms for the Abelian 
vector fields $A^\Lambda$ on the 3D Coulomb branch,
\beq 
	S^{(3)}_{CS}=-\frac{1}
	{2}\int \Theta_{\Lambda\Gamma}A^\Lambda \wedge F^\Gamma\,. 
\eeq
Again, the crucial point is that from the F-theory perspective these 
CS-terms are purely quantum mechanically and 
generated only at one-loop of massive fermions charged under the vectors 
$A^\Lambda$. In contrast  on the M-theory
side they are generated classically by $G_4$-flux on $\hat{X}_4$. Thus, the
CS-terms still carry the signature of the states that have been integrated 
out in F-theory but are efficiently calculated in M-theory. More precisely 
we will use in this work as a tool to derive the chiral 
index $\chi(\mathbf{R})$ of 4D charged matter in a representation 
$\mathbf{R}$ the observation of \cite{Grimm:2011fx} that 
certain $G_4$-fluxes induce particular classical M-theory CS-terms 
which are induced in F-theory by one-loop diagrams of the 3D massive 
charged fermions from the 4D chiral matter multiplets reduced on $S^1$. 
The identification of classical M- and one-loop F-theory CS-terms 
$\Theta_{\Lambda\Sigma}^M$ respectively $\Theta_{\Lambda\Sigma}^F$ we will 
employ takes the form
\beq
\frac12\Theta_{\Lambda\Sigma}^M=\frac14\int_{\hat{X}_4}G_4\wedge\omega_\Lambda\wedge\omega_\Sigma
\equiv-\frac12\sum_{\mathbf{R}}\chi(\mathbf{R})\sum_{q\in \mathbf{R}}q_\Lambda 
q_\Sigma \ \text{sign}(q\cdot \zeta)=-\Theta^{F}_{\Lambda\Sigma}\,,
\eeq 
where $\omega_{\Lambda}$, $\omega_{\Sigma}$ denote $(1,1)$-forms on 
$\hat{X}_4$ generated by resolving the singularities in $X_4$ and the
sum is taken over all 4D matter representations $\mathbf{R}$ with Dynkin 
labels $q$. The sign-function is applied to the scalar product of the 
charges and the Coulomb branch parameters, 
$q\cdot \zeta=q_\Lambda\, \zeta^\Lambda$.
In general, M-theory $G_4$-flux on the fully resolved Calabi-Yau 
$\hat{X}_4$ that is consistent with the F-theory limit is the description 
of seven-brane gauge flux in F-theory. Thus, as is known 
that gauge fluxes on seven-branes induce chirality of 4D matter in Type 
IIB $G_4$-flux induces chirality in F-theory. The relation of such fluxes 
to the 4D matter spectrum has been recently studied intensively 
\cite{Donagi:2008ca,Beasley:2008dc,Beasley:2008kw,Hayashi:2008ba,Braun:2011zm,Marsano:2011hv,Krause:2011xj,Grimm:2011fx,Krause:2012yh,Kuntzler:2012bu}.

Besides using the CS-terms in 3D as tools for determining 4D chirality we
will discover the intricate relations between different 3D CS-terms that are
implied by 4D anomaly cancellation. Let us summarize our findings where we 
distinguish F-theory compactifications without Abelian gauge symmetries and 
those with Abelian gauge factors. In the first case it is 
required that the non-Abelian gauge anomaly vanishes which puts one constraint 
on the spectrum for each gauge group factor $G_{(I)}$ in $G$ whereas in the second 
case anomaly cancellation is more involved due to mixed anomalies. 
The cancellation of these anomalies may require a Green-Schwarz (GS) mechanism 
in F-theory. Then fluxes on space-time filling seven-brane have to both induce 
a non-trivial chiral spectrum and corresponding gaugings of the axions for a 
working GS-mechanism. Anomaly cancellation is then 
tightly linked with tadpole cancellation that imposes constraints on the brane 
configuration and allowed fluxes. For setups at weak string coupling this link has 
been discussed and reviewed, for example, in \cite{Cvetic:2001nr,MarchesanoBuznego:2003hp,Blumenhagen:2005mu,Blumenhagen:2006ci,Plauschinn:2008yd,Cvetic:2011vz}.

In contrast to the weakly coupled Type IIB theory, where brane fluxes are constrained by tadpole conditions, 
for $G_4$ fluxes lifting from M-theory to F-theory it is expected that all anomalies are canceled 
and the Green-Schwarz mechanism is implemented automatically. This can be anticipated 
since in M-theory no additional consistency conditions on $G_4$, apart from the for 
anomaly cancellation irrelevant M2-brane tadpole matching, 
have to be imposed. In particular the seven-brane tadpole is 
canceled geometrically in F-theory by considering a compact Calabi-Yau 
fourfold $X_4$ and 5-brane tadpoles are canceled by considering closed fluxes $dG_4=0$ 
signalling the absence of a net M5-brane charge. Therefore, by 
consistency of the underlying M-theory compactification also the low-energy effective action 
should be consistent. 

From anomaly cancellation in combination with F-/M-theory duality in 3D we 
discover the following constraints for and links to 3D CS-terms:
\begin{itemize}
	\item Without $U(1)$-factors in 4D, the CS-terms for 
	the vectors $A^\Lambda$ encoding chiralities have to obey non-trivial 
	relations such that the cubic non-Abelian anomalies are canceled.
	\item With $U(1)$-factors in 4D, the CS-terms for the 4D chiral indices 	      		 
	      have to be related to the CS-terms determining the gaugings of the
	      4D axions so that all mixed anomalies are canceled.
	\item We show the cancellation of 4D mixed Abelian-gravitational anomalies in 
	      3D using F-/M-theory duality. We discover that the 4D mixed 
	      Abelian-gravitational anomaly is the coefficient $\Theta^{\text{loop}}_{0m}$ 
	      of the 3D Chern-Simons term $A^0 \wedge F^m$ for the 
	      Kaluza-Klein vector $A^0$ and the corresponding 3D $U(1)$ vector field $A^m$,
	\beq
	\Theta^{\text{loop}}_{0m}=-\frac1{12}\sum_{\underline{q}} 
	n(\underline{q})q_m\,,
	 \eeq
	 where the sum is taken over all $U(1)$-charges $
	 \underline{q}$ under the 4D $U(1)$-vectors $A^m$ and $n(\underline{q})$ is 
	 the number of fermions with a given charge $\underline{q}$.
	 We obtain this result since $\Theta_{0m}$ is one-loop induced in F-theory 
	with the full Kaluza-Klein tower of 4D charged fermions in 
	the loop. F-/M-theory duality implies that this expression is related to 
	the classical flux-integral of $G_4$ in M-theory, which is precisely the 
	tree-level contribution to the anomaly from the GS-mechanism,
	\beq
		\Theta^{\text{loop}}_{0m}\equiv-\frac{1}{2}\Theta_{0m}=\frac{1}{4}
	 K^\alpha\Theta_{\alpha m}\,.
	\eeq
	Here $K^\alpha$ is a vector determining a direction in the space of 4D  	  
	axions and $\Theta_{m\alpha}$ is the $G_4$-flux induced gauging of the
	$\alpha$th axion  under $A^m$.
\end{itemize}
In addition, we discuss in general terms the construction of $U(1)$ 
symmetries in F-theory on fourfolds by considering a non-trivial 
Mordell-Weil group of rational sections\footnote{We neglect the torsion 
subgroup here for simplicity.} of $\hat{X}_4$. This construction has been 
applied successfully on elliptically fibered threefolds in 
\cite{Morrison:1996pp,Aspinwall:1998xj,Park:2011ji,Morrison:2012ei} to 
study non-simply laced gauge groups and $U(1)$'s as well as Abelian 
anomalies in six-dimensional F-theory. The discussion
we provide in this note is a first step to being able to construct also 
four-dimensional F-theory compactifications with more than one $U(1)$ and 
a matter sector with more general $U(1)$-charges extending the analysis of 
\cite{Grimm:2010ez}.

Furthermore, for consistent $G_4$-fluxes in the F-theory limit, we 
discover geometric relations that need to be satisfied in any 
resolved elliptically fibered fourfold  $\hat{X}_4$,
\bea  \label{eq:geo_rel}
	\frac{1}{3}\sum_{S_{\mathbf{R}}}\sum_{c\,\subset S_\mathbf{R}}(S_\mathbf{R}\cdot [G_4])(c\cdot D_\Lambda)(c\cdot D_\Sigma)
	(c\cdot D_\Gamma)&=&\frac{1}{2}[G_4]\cdot D_{(\Gamma}\cdot
    \pi_*(D_\Lambda\cdot D_{\Sigma)})\ ,\\
    \frac1{3}\sum_{S_{\mathbf{R}}}\sum_{c\,\subset S_\mathbf{R}}(S_\mathbf{R}\cdot [G_4])(c\cdot D_\Lambda)&=&[G_4]\cdot 
    [c_1(B_3)]\cdot D_\Lambda\, .
\eea
Here $G_4$ is the $G_4$-flux in M-theory and $S_{\mathbf{R}}$ is a 
four-cycle denoted the matter surfaces obtained by fibering the shrinking 
$\bbP^1$'s in  $\hat{X}_4$ that are the weights of a representation 
$\mathbf{R}$ of $G$ over matter curves. The divisors 
$D_\Lambda$, $D_\Sigma$ and $D_\Gamma$ are the exceptional divisors 
obtained by fibering the shrinkable curves $\bbP^1$ over 
co-dimension one loci in the base $B_3$, i.e.~the seven-brane divisors in 
$B_3$. The curves $c$ are any shrinkable holomorphic curve in the fiber 
$S_{\mathbf{R}}$. The map $\pi_*$ is induced by the 
projection to the base $\pi\,:\,\hat{X}_4\rightarrow B_3$ of the elliptic 
fibration of $\hat{X}_4$ and $c_1(B_3)$ denotes its first Chern-class.
These geometric relations are indeed valid for concrete 
resolutions as we show for several examples with non-Abelian gauge groups. 

We note that the geometric relations \eqref{eq:geo_rel} and the 
corresponding statements about anomaly cancellation should even apply in a 
broader context than considered in this paper. In particular one expects 
that \eqref{eq:geo_rel} is valid for more general $G_4$-flux that 
is not necessarily given by a sum of products of $(1,1)$-forms, 
i.e.~$G_4$-flux that is not vertical. A phenomenologically interesting 
example of these fluxes is the F-theory analog of hypercharge flux $F_Y$ 
considered in \cite{Marsano:2010sq,Dudas:2010zb,Palti:2012dd} 
in the context of anomalies. In order to understand anomaly cancellation 
also in these cases, it is crucial to note that the right hand side of 
\eqref{eq:geo_rel} is in general only non-trivial for vertical $G_4$-flux 
and vanishes otherwise. Constraints on the spectrum then arise from the 
vanishing of the left-hand side of this equality. A better understanding 
of the global geometry of $\hat X_4$ and the matter surfaces 
$S_{\mathbf R}$ might reveal that these constrains are also automatic 
fulfilled for concrete resolved Calabi-Yau fourfolds with no further 
restriction on the $G_4$-flux. It would be interesting to investigate this 
further. 

Finally,  we compare our analysis of anomaly cancellation in F-theory to anomaly cancellation
in Type IIB Calabi-Yau orientifold compactifications with O7-planes and intersecting D7-branes 
\cite{MarchesanoBuznego:2003hp,Blumenhagen:2005mu,Blumenhagen:2006ci,Plauschinn:2008yd}. We show that in general the Type IIB anomaly cancellation
reduces to the F-theory anomaly cancellation when projecting to the sector of geometrically massless $U(1)$'s
in Type IIB. This demonstrates on the levels of anomalies that the geometrically massless $U(1)$'s correspond directly 
to the $U(1)$'s engineered in F-theory, whereas the geometrically massive $U(1)$'s in Type IIB are captured by the 
residual discriminant of the elliptic fibration in F-theory, cf.~\cite{Grimm:2011tb} for a more general discussion 
of geometrically massive $U(1)$'s.

The paper is organized as follows. In section \ref{sec:anomlies+cancellation} we recall some basic facts about anomaly cancellation in 
4D and introduce the Green-Schwarz mechanism. In order to determine the Green-Schwarz counter terms in 
F-theory we discuss the geometric structure of resolved Calabi-Yau fourfolds used in the 
duality of M-theory to F-theory in section \ref{sec:AnomaliesInFGUTs}. It will be crucial to allow for  
$U(1)$ gauge group factors and their geometric F-theory manifestation in our analysis.
In section \ref{sec:ChiralityInF} we turn to the analysis of one-loop Chern-Simons terms in the 
3D effective theory. We recall how they encode the 4D chiral index for the matter spectrum 
and show that they can also capture 4D mixed Abelian/gravitational anomalies. 
Anomaly cancellation in F-theory is discussed in section \ref{sec:anomalycancellationinF} were we also derive the 
geometric conditions \eqref{eq:geo_rel}. We contrast the F-theory analysis with the description 
of anomaly cancellation in weakly coupled Type IIB setups with D7-branes and O7-planes 
in section \ref{sec:orientifold_anomalies}. Explicit examples of resolved Calabi-Yau fourfolds are introduced 
in section \ref{sec:examples}. Our work has two appendices supplementing additional information 
about 4D anomalies and $SU(5)$ representations.

\section{Four-Dimensional Anomaly Cancellation}  
\label{sec:anomlies+cancellation}

In this section we present the basic techniques for characterizing 
anomalies of gauge symmetries and their cancellation via the 
generalized Green-Schwarz mechanism in a four dimensions. More details 
can be found in appendix \ref{app:ReviewOfAnomalies} and the reviews \cite{Anomaly-reviews}. 

We will consider a 4D $\mathcal{N}=1$ supersymmetric effective theory 
with a vector multiplet transforming in the adjoint of a general 
non-Abelian gauge group, 
\beq \label{eq:Gsplit}
	G=G_{(1)}\times\cdots \times G_{(n_G)}\times U(1)_1\times\cdots\times U(1)_{n_{U(1)}}\,.
\eeq
Here the group factors $G_{(I)}$ denote $n_G$ arbitrary simple Lie-groups 
and we allow for a number $n_{U(1)}$ of $U(1)_m$-factors.
In the following we will use indices
\beq
   I, J= 1,\ldots, n_G\ , \qquad \quad m,n =1,\ldots, n_{U(1)}\ ,
\eeq
to label non-Abelian and Abelian group factors.
We consider matter in chiral multiplets that transform in 
representations denoted 
\beq
	\mathbf{R}=(\mathbf{r}^1,\ldots,
	\mathbf{r}^{n_G})_{\underline{q}}\,
\eeq
of $G$. Here $\mathbf{r}^I$ denote the representations of the 
non-Abelian factors of $G_{(I)}$ and $\underline{q}=(q_1,
\ldots ,q_{n_{U(1)}})$ denotes the corresponding $U(1)$-charges 
arranged as a column vector. Furthermore, since we are considering
a theory with $\mathcal{N}=1$ supersymmetry we have one 
gravity multiplet containing a single gravitino.

In general an anomaly of a symmetry denotes the effect that a
symmetry of the classical theory is not promoted to a symmetry of
the quantum theory. An anomaly of a local symmetry, i.e.~a gauge 
symmetry, spoils the consistency of the quantum theory due to the
quantum mechanical violation of current conservation laws. Thus in a 
consistent quantum theory anomalies of gauge symmetries have to be absent. 
The breakdown of a gauge symmetry is encoded
in a gauge invariant and regulator independent way in the anomaly 
polynomial. In  four dimensions this polynomial is a cubic 
polynomial in the gauge field strength of the gauge group $G$ 
and the 4D Riemann tensor. Thus, it is a formal six-form denoted by 
$I_6$. The polynomial $I_6$
is the sum of the anomaly polynomials for all fields contributing 
to the anomaly. These are massless Weyl-fermions, gravitinos and 
self-dual tensors and their corresponding anomaly polynomials have been 
worked out in every space-time dimension in the seminal work of 
\cite{AlvarezGaume:1983ig}. The theory is anomaly free if the full anomaly 
polynomial vanishes identically, i.e.~if all 
coefficients of the various monomials in the field strength and the 
Riemann tensor are zero. As we will discuss next, the full anomaly 
polynomial consistent of $I_6$ from the quantum anomalies in the matter 
sector and another contribution from a tree-level effect, the Green-
Schwarz mechanism.

In four dimensions, the possible anomalies are gauge and mixed anomalies 
only, since pure gravitational anomalies are absent 
by symmetry. For the fields in the standard $\mathcal{N}=1$ supergravity 
theory described above the anomaly polynomial reads
\beq 
\label{eq:I6general}
	I_6=\sum_{\mathbf{R}}n(\mathbf{R})I_{1/2}(\mathbf{R})\,,
\eeq
where $I_{1/2}(\mathbf{R})$ denotes the anomaly polynomial 
\eqref{I12-4d} of a left-chiral Weyl fermion that 
occurs with multiplicity $n(\mathbf{R})$ in the 4D spectrum.

In general, the anomaly polynomial \eqref{eq:I6general} of the matter
sector does not need to vanish identically if it is a sum of factorizable 
contributions. 
In this case, the residual anomalies can be canceled by a certain 
tree-level mechanism known as the Green-Schwarz mechanism 
\cite{Green:1984sg}. We consider the higher derivative effective 
action, referred to as the Green-Schwarz counter terms, 
\beq \label{eq:GSterm}
     S^{(4)}_{\rm GS} = -\frac{1}{8}\int \frac{2}{\lambda_I} 
     b_I^\alpha \rho_{\alpha}  \
     \text{tr}_{\textbf{f}}\,( F^I\wedge F^I)+2 b_{mn}^{\alpha} 
     \rho_{\alpha}\   F^m\wedge F^n  -  \frac{1}{2} 
     a^\alpha \rho_\alpha\ \text{tr}( R\wedge R )\ .
\eeq
Here $F^I$ and $F^m$ denote the field strengths in the adjoint of 
$G_{(I)}$ respectively of the $m$-th $U(1)$ and $\text{tr}_{\mathbf{f}}$ 
denote the traces in the corresponding fundamental representations. Furthermore, we have used 
$\lambda_I=2c_{G_{(I)}}/V(\mathbf{adj})=\frac{2}{\langle \alpha_0,\alpha_0\rangle}$ with $c_{G_{(I)}}$ the 
dual Coxeter number of $G_{(I)}$ and $V(\mathbf{adj})$ defined in 
\eqref{trF^3}, whereas $\langle \alpha_0,\alpha_0\rangle$ denotes the length squared of the root of maximal length $\alpha_0$. 
The real scalar fields $\rho_\alpha$ are axions that are gauged by the 
Abelian vectors $A^m$ of $G$ as
\beq \label{eq:rhoGauging}
   \cD \rho_\alpha = d \rho_\alpha + 
   \Theta_{\alpha m}\, A^m\,. 
\eeq
The combinations of axions $\rho_\alpha$ in the various terms in 
\eqref{eq:GSterm} are determined by parameters $b^\alpha_I$, 
$b^\alpha_{mn}$ and $a^\alpha$, which can be determined 
by the underlying microscopic theory together with the matrix 
$\Theta_{\alpha m}$ in \eqref{eq:rhoGauging}. We will
explicitly determine these parameters for F-theory compactifications 
in section \ref{sec:GStermsInF}.
The gauging \eqref{eq:rhoGauging} induces an anomalous variation 
of the action \eqref{eq:GSterm} that lifts to a contribution to the
anomaly polynomial. In factorizable situations this can cancel a 
non-vanishing $I_6$ in \eqref{eq:I6general} from the matter sector of the
theory.

Before we continue by writing down the anomaly conditions we pause for a 
brief discussion of additional terms in the effective action that can have 
an anomalous variation. As discussed in \cite{Anastasopoulos:2006cz} there 
can be generalized Chern-Simons terms in four dimensions of the form 
$E_{mnk}A^m\wedge A^n\wedge F^k$ for the $U(1)$'s respectively non-Abelian 
terms $E_{mI}A^m\wedge \omega_3^I$ with $\omega_3^I$ denoting the 
Chern-Simons form of $\text{tr}(F^I\wedge F^I)$. As discussed also in
\cite{Anastasopoulos:2006cz} these terms depend on the chosen 
regularization scheme and it is possible to work in a scheme where all 
these terms are absent. Equivalently these terms parametrize the 
ambiguities, namely the exact forms, in the descend equations 
that is used to relate the unambiguous, i.e.~scheme-independent anomaly 
polynomial $I_6$, with the anomalous and scheme dependent counter terms in 
the quantum effective action. Thus the generalized
Chern-Simons terms are irrelevant for anomaly cancellation and will be 
neglected in the rest of our discussion.

Adding up the two contributions to the total anomaly polynomial from the 
matter sector, $I_6$ in \eqref{eq:I6general}, and from the Green-Schwarz 
mechanism, $I_6^{\text GS}$ in \eqref{I6-GS}, and requiring 
the sum to vanish, the conditions for cancellation read, as 
reviewed in more detail in appendix 
\ref{app:ReviewOfAnomalies},
\bea 
	\text{tr}_\mathbf{f} (F^I)^3&:\qquad 
        & \sum_\mathbf{\mathbf{r}^I} n(\mathbf{r}^I)V(\mathbf{r}^I)=0\,, \label{eq:Anomalies:purenonAb}\\
	F^m F^n F^k&:\qquad  
        & \frac{1}{6}\sum_{\underline{q}}n(\underline{q})q_{(m} q_n q_{k)}=\frac1{4}b_{(mn}^\alpha\, \Theta^{\phantom{\cA}}_{k) \alpha} \label{eq:Anomalies:pureAb}  \\
        F^m\text{tr}_{\mathbf{f}}(F^I)^2&:\qquad 
        & \frac12\sum_{r^I}\sum_{\underline{q}} n(\repA^I_{\underline{q}})U(\repA^I)q_m
=\frac{1}{4\lambda_I} b_I^\alpha\, \Theta_{\alpha m} \label{eq:Anomalies:AbnonAb} \\
        F^m\text{tr}R^2&:\qquad 
       &\frac{1}{48}\sum_{\underline{q}}n(\underline{q})q_m=-\frac1{16}a^\alpha\Theta_{m \alpha}\,, \label{eq:Anomalies:Abgravitational}
\eea
where in the second line we have symmetrized in the indices $m,n,k$.
Here, we have to sum over each representation $\mathbf{r}^I$ of 
the simple group $G_{(I)}$ and over all possible $U(1)$-charges 
$\underline{q}$ with which the representation $\mathbf{r}^I$ 
occurs in the 4D spectrum. In addition, we introduced 
$n(\mathbf{r}^I_{\underline{q}})$ denoting the number of chiral 
multiplets in the representation $\mathbf{r}^I_{\underline{q}}$ 
and the number of chiral multiplets $n(\underline{q})$ with 
charges $\underline{q}$. The latter can be written as
\beq
	n(\underline{q})=\sum_{\mathbf{r}_{\underline{q}'}}n(\mathbf{r}^I_{\underline{q}'})
	\text{dim}(\mathbf{r}^I)\delta_{\underline{q}\,\underline{q}'}\,,
\eeq
where $\delta_{\underline{q}\,\underline{q}'}=1$ if $\underline{q}=\underline{q}'$ and 
zero otherwise.
Furthermore, we made use of the group theory relations
\beq \label{eq:traceRelations}
     \text{tr}_{\repA^I} F^3 = V(\repA^I)  \text{tr}_{\mathbf{f}} 
     F^3\,,\qquad
	\text{tr}_{\repA^I}
	F^2=U(\repA^I)\text{tr}_{\mathbf{f}}F^2\,.
\eeq
The conditions in the order in which they appear in 
\eqref{eq:Anomalies:purenonAb}--\eqref{eq:Anomalies:Abgravitational} 
are the purely non-Abelian anomaly, that has to cancel by itself, 
the purely Abelian anomaly and the two mixed anomalies, the 
Abelian-non-Abelian and Abelian-gravitational anomalies.

\section{Fluxes and Green-Schwarz-Terms in F-Theory}
\label{sec:AnomaliesInFGUTs}

In this section we prepare the ground for our analysis of anomalies in 
four-dimensional F-theory compactifications. We start in section 
\ref{sec:F+MresolvedFF} with a brief review of the geometry of smooth 
fourfolds $\hat{X}_4$, obtained by resolving the singularities
of the elliptic fibration of $X_4$, and describe the construction 
of $G_4$-flux on $\hat{X}_4$. Recalling the duality of 
3D M-theory compactifications on $\hat X_4$ with F-theory on $X_4\times S^1$
will be crucial. This 3D perspective on 4D F-theory physics is the backbone of most 
considerations in this paper. In section \ref{sec:flux_review} we discuss 
the conditions on $G_4$-fluxes in M-theory to be viable fluxes in F-theory.
Finally, we show that
4D Green-Schwarz terms in F-theory are determined by the M-theory 
compactification geometry $\hat{X}_4$ in section \ref{sec:GStermsInF}. 

\subsection{F-theory as M-theory \& the geometry of resolved fourfolds}
\label{sec:F+MresolvedFF}

An F-theory compactification to four dimensions is specified by
an, in general singular, elliptically fibered Calabi-Yau fourfold
$\pi:\,X_4\rightarrow B_3$ over a K\"ahler threefold base $B_3$. 
We will consider cases in which the fibration has at least one
section $\sigma$ and a number of rational sections $\hat{\sigma}_m$. 

Since there exists no twelve-dimensional 
low-energy effective action of F-theory the 4D physics of F-theory 
compactifications has to be extracted via its M-theory 
dual. More precisely, one considers the 4D theory on 
an additional $S^1$ and pushes the resulting 3D theory onto 
its Coulomb branch. In the dual picture this 3D 
theory is described by compactifying M-theory on the smooth 
Calabi-Yau fourfold $\hat{X}_4$, that is obtained by resolving all 
singularities in $X_4$. Since we will make extensive use of these 
two dual perspectives, we summarize them 
schematically as
\beq \label{eq:3dDuality}
	\text{F-theory on }X_4 \quad \rightarrow \quad \begin{array}{c}
	                                                \text{F-theory on }X_4\times S^1\\
                                                        \text{in 3D Coulomb branch}
	                                               \end{array} \ \equiv\ \text{M-theory on }  \hat{X}_4\,.
\eeq

In general the identification of F-theory on $X_4 \times S^1$ with 
M-theory on $X_4$ will also hold at the origin of the Coulomb branch
where the non-Abelian gauge group is restored. 
This provides a microscopic definition of F-theory in the UV by M-theory.
However, due to our poor understanding of the microscopics of M-theory 
the equality in \eqref{eq:3dDuality} is most explicitly evaluated on the Coulomb 
branch. On the M-theory side this corresponds to 
using the resolved $\hat{X}_4$ in the compactification of 
11D supergravity to three dimensions.  
The resulting low energy effective theory is valid below the Kaluza-Klein scale and the 
energy scales defined by wrapped M2-branes on (shrinkable) 
cycles in $\hat{X}_4$. Using 11D supergravity on 
$\hat X_4$ the M2-branes and Kaluza-Klein modes have effectively been integrated out.
On the F-theory side, W-bosons and matter fields in 4D F-theory precisely arise from such  
M2-brane states. For the matching \eqref{eq:3dDuality} yet to work, we thus have to go to 
the 3D Coulomb branch where these states become massive and have been integrated 
out in the IR. In addition we have to choose the circle radius $S^1$ to be 
in a regime so that Kaluza-Klein modes and winding modes of F-theory are above the cut-off scale. 
Only then the light degrees of freedom in the 3D effective theories of  
F-theory on $X_4\times S^1$ and M-theory on $\hat{X}_4$ in \eqref{eq:3dDuality} match  
in the IR and the identification of the effective actions can be performed.

There are further motivations to consider compactifications on the 
resolved fourfold~$\hat X_4$ that should be stressed here. Firstly, 
from a mathematical point of view
the singularities of the fourfold $X_4$ can be 
classified on the smooth $\hat{X}_4$ by analyzing the local 
resolution geometry. Resolution of 
the singularities leads to new exceptional divisors $E_i$ in $\hat{X}_4$, whose 
intersections specify the type of the original singularities. 
Secondly, the resolved fourfold allows to include $G_4$-fluxes, i.e.~topologically 
non-trivial background values of the M-theory 
three-form field strength. Such fluxes are elements  
of the cohomology $H^4(\hat{X}_4,\mathbb{Z}/2)$, where half-integrality 
can be consistent with the quantization condition discussed in \eqref{G4half} below. 
This cohomology group also contains new classes due to the resolution of $\hat X_4$.
Precisely the fluxes arising in the expansion with respect to these forms correspond to seven-brane
gauge fluxes and are the key to understand chiral anomalies and their cancellation
as we will discuss in more detail in the following. 

Having a closer look at the geometry of a resolved fourfold $\hat X_4$ one encounters four 
different types of divisors. We denote a basis of divisors and their Poincar\'e
dual two-forms by $D_A$ and $\omega_A$ with $A=1,\ldots,h^{1,1}(\hat{X}_4)$.
In the following we discuss the intersection numbers of these divisors and two-forms in 
detail. They are denoted by
\beq \label{eq:fourfoldintersect}
   \cK_{ABCD} = D_A\cdot D_B \cdot D_C \cdot D_D = \int_{\hat X_4} \omega_A \wedge \omega_B \wedge \omega_C \wedge \omega_D\ .
\eeq
The four types of divisors and their Poincar\'e dual two-forms in $\hat X_4$ are
\beq \label{eq:basisH^11}
	D_A=(B,D_\alpha,D_i,D_m)\,,\qquad \omega_A=( \omega_{\hat 0},\omega_\alpha,\omega_i,\omega_m)\,,
\eeq  
that we characterize as follows:
\begin{itemize}
\item The zero section $B$: The single divisor $B$ is the zero section of the elliptic fibration of $\hat{X}_4$. It is the class of the 
base $B_3$ with its Poincar\'e dual $\omega_{\hat 0}$\footnote{The hat on the index $0$ is introduced since, as we will discuss in detail later,
it is more natural for the  description of the 3D effective action to redefine $B$ and $\omega_{\hat{0}}$.}.
The section obeys the intersection property
\beq \label{eq:B^2}
	B\cdot B=-c_1(B_3)\cdot B\,,
\eeq
as can be seen from application of the adjunction formula.

\item The vertical divisors $D_\alpha$: There are $h^{(1,1)}(B_3)$ divisors $D_\alpha=\pi^{-1}(D_\alpha^{\rm b})$, with 
dual two-forms $\omega_\alpha$, that are lifted from divisors $D_\alpha^{\rm b}$ of the base $B_3$ to the fourfold 
$\hat{X}_4$ and are thus inherited from the singular fourfold $X_4$. 

\item The Cartan divisors $D_i$: There are 
$\text{rank}(G)-n_{U(1)}$ divisors $D_i$ with their dual two-forms $\omega_i$ that are related to the exceptional divisors $E_i$
resolving the singularities in the elliptic fibration of $X_4$. The $D_i$ are denoted as Cartan divisors and their intersections encode the 
types of singularities  in $X_4$ that correspond to the non-Abelian 
gauge symmetry in F-theory.  

\item Rational sections $\hat{\sigma}_m$ and $U(1)$ Cartan divisors $D_m$: In 
general there are $n_{U(1)}$ extra 
divisor classes $D_m$, denoted the Cartan divisors of the $m$th $U(1)$ 
symmetry in F-theory, with Poincar\'e duals $\omega_m$. Geometrically 
these are related, as discussed below, to a non-trivial Mordell-Weil group of 
rational sections $\hat{\sigma}_m$ of the elliptic fibration of $\hat{X}_4$ 
\cite{Morrison:1996pp,Aspinwall:1998xj}. For our purposes we are mostly interested in the intersection 
properties of these sections. Most notably, the $\hat{\sigma}_m$ obey a 
relation like \eqref{eq:B^2},
\beq \label{eq:defPropSections}
	\hat{\sigma}_m^2\cdot D_\alpha\cdot D_\beta=-[c_1(B_3)]\cdot\hat{\sigma}_m\cdot D_\alpha\cdot D_\beta
\eeq 
for all $m$ and all vertical divisors $D_\alpha$, $D_\beta$, 
where $[\cdot]$ indicates Poincar\'e duality.
\end{itemize}

Let us next explain the intersection properties of these four different
types of divisors. As we will see in this discussion, these intersections 
reflect on the one hand the geometry of the elliptic fibration 
of $\hat{X}_4$ and on the other hand  the physical structure of the 3D 
effective theory. These intersection properties can be checked
by performing an explicit resolution, employing compact toric methods 
in 
\cite{Candelas:1996su,Candelas:1997eh,Blumenhagen:2009yv,Grimm:2009yu,Cvetic:2010rq,Chen:2010ts,Knapp:2011wk,Braun:2011ux} 
and the local methods and their extensions in 
\cite{Esole:2011sm,Marsano:2011hv}.

We begin with the divisor $B$. Instead of stating its intersections we first 
perform a basis change.
It is was noted in \cite{Grimm:2011sk,Park:2011ji,Bonetti:2011mw} that it is 
necessary in the reduction of M-theory on $\hat{X}_4$ to 
shift the 3D M-theory fields for the correct identification by the duality 
\eqref{eq:3dDuality} with 4D F-theory fields obtained by the 
reduction on a circle to 3D. Furthermore, we will find in this work that 
this redefinition is the key to discover a simple  
interpretation of the mixed Abelian-gravitational anomalies in the three-
dimensional effective theory. 
The coordinate shift of the 3D fields can be translated into a redefinition of the basis \eqref{eq:basisH^11} as
\beq \label{eq:omega0tilde}
	\tilde{\omega}_0=\omega_{\hat 0}+\frac12 c_1(B_3)\,,\qquad \tilde{B}=B+\frac12 [c_1(B_3)]\,,	
\eeq
with $\omega_\alpha,\omega_\Lambda$ respectively $D_\alpha,D_\Lambda$ 
unchanged. The brackets $[\cdot]$
indicate the Poincar\'e dual cycle of the cohomology class $c_1(B_3)$. 
The interesting intersection properties of this new basis are
\beq \label{eq:newIntsTildeB}
	\tilde{B}^2=\frac14 [c_1(B_3)]^2\,,\qquad  \cK_{00 \alpha \beta}=0\,,
\eeq
where we stress that the intersection numbers with indices $0$ without a hat 
are invoking $\tilde \omega_0$.
The first equation follows from \eqref{eq:B^2} for the section $B$, 
and the second equation is a consequence of the first and 
\eqref{eq:tripleintersect}.

Next we turn to the intersection numbers of the vertical divisors $D_\alpha$. By the fibration structure of $\hat{X}_4$ the intersections of 
three and four $D_\alpha$ are given by 
\beq \label{eq:tripleintersect}
   \cK_{\hat{0} \alpha \beta \gamma} = \cK_{\alpha \beta \gamma} = D^{\rm b}_\alpha \cdot D^{\rm b}_\beta \cdot D^{\rm b}_\gamma\ , \qquad \qquad \cK_{\alpha \beta \gamma \delta} = 0\ ,   
\eeq
where we introduced the triple intersections $\cK_{\alpha \beta \gamma}$ of the divisors $D^{\rm b}_\alpha$  in the base $B_3$.

Let us next turn to the intersection numbers involving the Cartan divisors $D_i$. 
We consider stacks of non-Abelian seven-branes wrapped on divisors $S^{\rm b}_{(I)}$ in the base $B_3$.
Each such divisor class can be expanded in the basis $D_\alpha^{\rm b}$ as
\beq \label{eq:def-Calpha}
   S_{(I)}^{\rm b} = \delta^\alpha_{(I)} D_\alpha^{\rm b} \ ,
\eeq
where $\delta^\alpha_{(I)}$ are constant coefficients.
In order to incorporate the split \eqref{eq:Gsplit} of $G$ into simple Lie-groups 
we divide the $D_i$ as
\beq
   D_i = ( D_{i_I})\ , \qquad \quad i_I = 1,\ldots \text{rank}(G_{(I)})\ , 
\eeq
where the index $i_I$ labels the Cartan divisors for the group factor $G_{(I)}$.
We say that a singularity in $X_4$ at codimension one in the base $B_3$ is of type $G_{(I)}$, if the subset of 
divisors $D_{i_I}$ resolving that particular 
singularity intersect on $\hat{X}_4$ as
\beq
\label{eq:Cartan}
	 D_{i_I}\cdot D_{j_J}\cdot D_\alpha\cdot D_\beta=-\delta_{IJ}\mathcal{C}^{(I)}_{i_Ij_I} S_{(I)}\cdot B\cdot D_\alpha\cdot D_\beta
              \ ,  \qquad \quad  D_{i_I}\cdot D_\alpha\cdot D_\beta\cdot D_{\gamma}=0
        \,,
\eeq
where no sum is taken over $I,J$. 
Here we introduced, starting with the divisors $S_{(I)}^{\rm b}$ in 
\eqref{eq:def-Calpha}, vertical 
divisors $S_{(I)} = \pi^{-1}(S_{(I)}^{\rm b})$ defined on the fourfold 
$\hat X_4$. Thus, they are related to the $S_{(I)}^{\rm b}$ in the base
by $S_{(I)}\cdot B=S_{(I)}^{\rm b}$. The matrices $\mathcal{C}^{(I)}_{i_Ij_I}$ 
characterize the type of the singularity over $S_{(I)}^{\rm b}$ and can agree 
with the inner product of coroots of simple Lie-groups including the ones of 
ADE type. For simply laced Lie-groups, i.e.~precisely for ADE groups,
the coroot inner product agrees with the Cartan matrix 
$\mathcal{C}^{(I)}_{i_Ij_I}=C^{(I)}_{i_Ij_I}$ since all roots have length 
$2$ and $\lambda_I$. In general we have the relation
\beq
	\mathcal{C}^{(I)}_{i_Ij_I}
	=\frac{2}{\lambda_{I}\langle \alpha_J,\alpha_J\rangle}C_{i_Ij_I}^{(I)}\,,
\eeq
where $\alpha_{i_I}$ are the simple roots of the Lie-algebra with inner 
product $\langle\cdot,\cdot\rangle$.
This motivates the name Cartan divisors of $G_{(I)}$ for the 
divisors $D_{i_I}$ since these divisors can be identified with the
negative of the simple roots of an ADE gauge group\footnote{For a non-simply laced Lie-
group the $D_{i_I}$ are to be identified with the negative of the simple 
coroots $\alpha^{\vee}_{i_I}=\frac{2}{\langle\alpha_{i_I},
\alpha_{i_I}\rangle}\alpha_{i_I}$, 
i.e.~the simple roots of the coroot lattice.}, 
$-\alpha_i^{(I)}$. In other words, the $D_{i_I}$ span the negative of the 
root lattice of $G_{(I)}$ that consequently is embedded 
into the K\"ahler cone of $\hat{X}_4$, more precisely the complement of 
the K\"ahler cone of the singular fourfold $X_4$ in $\hat{X}_4$, that is 
denoted the relative K\"ahler cone. Note that given the $D_{i_I}$ one can 
reverse the logic and use their intersections to unambiguously define the 
$S_{(I)}^{\rm b}$ by equation \eqref{eq:Cartan} as discussed in section \ref{sec:GStermsInF}.

The vertical divisors $S_{(I)} $ on the fourfold $\hat X_4$ are related to the inverse images $\hat{S}_{(I)}$ 
of the $S^{\rm b}_{(I)}$ under the projection to the base on the \textit{singular} fourfold $X_4$ 
by the shift
\beq
\label{eq:Shat}
	\hat{S}_{(I)}=S_{(I)}-\sum_{i_I} a_{i_I} D_{i_I}\,,
\eeq
where $a_{i_I}$ are the dual Coxeter labels of $G_{(I)}$ and no sum is taken over $I$. We note that due 
to this shift the divisors $\hat{S}_{(I)}$ are 
not elements in the base $B_3$ in the resolved fourfold 
$\hat{X}_4$.  The divisors $D_{i_I}$ project onto the divisors $\pi(\hat{S}_{(I)})$ in $B_3$ 
in the blow-down map $\hat{X}_4\rightarrow X_4$.

Finally we discuss the intersections of the Cartan divisors $D_m$ of the 
$U(1)$-factors. First we introduce analogously to \eqref{eq:def-Calpha} 
divisors $S_{(m)}^{\rm b}$ that indicate the location of the seven-branes 
supporting the $U(1)$'s in the base $B_3$. We expand
\beq \label{eq:def-Calpham}
  S_{(m)}^{\rm b} = \delta^\alpha_{(m)} D_\alpha^{\rm b}
\eeq
and introduce the corresponding vertical divisors 
$S_{(m)}=\pi^{-1}(S^{\rm b}_{(m)})$. Next we construct divisors 
$\tilde D_m$ starting from a given basis of rational sections $\hat{\sigma}_m$
by the Shioda map \cite{ShiodaI,ShiodaII}   
\beq \label{eq:ShiodaMap}	\tilde{D}_m = \hat{\sigma}_m-\tilde{B}-(\hat{\sigma}_m\cdot\tilde{B}\cdot\mathcal{C}^\alpha)
	\eta^{-1\beta}_\alpha	D_\beta+
	\sum_{I}(\hat{\sigma}_m\cdot D_{i_I}\cdot \mathcal{C}^\alpha)
	(\delta^\beta_{(I)}\eta_{\beta}^{\,\ \alpha})^{-1}(C_{(I)}^{-1})^{i_Ij_I}D_{j_I}\ , 
\eeq 
where we denoted the inverse Cartan matrix of $G_{(I)}$
by $C_{(I)}^{-1}$. We are not summing over $\alpha$, but rather fix
one particular value $\alpha$ so that 
$\delta^\beta_{(I)}\eta_\beta^{\,\ \alpha}\neq0$ respectively a different 
$\alpha$ so that $\delta^\beta_{(n)}\eta_\beta^{\,\ \alpha}\neq0$ to evaluate 
\eqref{eq:ShiodaMap}.\footnote{We note that this step intrinsically introduces 
the exceptional curves $c$ which are the negative of a simple 
root $-\alpha_{i_I}$ of $G_{(I)}$ respectively the Cartans of the $m$th $U(1)$,
\beq
	 c_{-\alpha_{i_I}}\equiv D_{i_I}\cdot \mathcal{C}^\alpha
	(\delta^\beta_{(I)}\eta_{\beta}^{\,\ \alpha})^{-1}\,,\quad c_{m}\equiv D_{m}\cdot \mathcal{C}^\alpha
	(\delta^\beta_{(m)}\eta_{\beta}^{\,\ \alpha})^{-1}
\eeq
This is clear by calculating $c_{-\alpha_{i_I}}\!\!\cdot D_{j_J}$ and 
$c_{m}\cdot D_n$ yielding $-\delta_{IJ}C^{(I)}_{ij}$ respectively $-1$ for $m=n$ and an in general model-dependent number for $n\neq m$.}
We also have to introduce a basis of $h^{1,1}(B_3)$ 
vertical four cycles $\mathcal{C}^\alpha$ in $\hat{X}_4$ as
\beq 
    \cC^\alpha = \pi^{-1}(\cC^\alpha_{\rm b})\ ,
\eeq 
with Poincar\'e dual four-forms $\tilde \omega^\alpha$.
These are inherited from curves $\cC_{\rm b}^\alpha$ in $B_3$\footnote{\label{footnt1}Technically  these curves are formed by 
finding the linearly independent combinations of intersections of two $D^{\rm b}_\alpha$ in the base as $D^{\rm b}_\alpha\cdot 
D^{\rm b}_\beta=\tilde{\mathcal{K}}_{\alpha\beta\gamma}\mathcal{C}_{\rm b}^\gamma$, where we introduced the three-point function 
${\mathcal{K}}_{\alpha\beta\gamma}=\tilde {\mathcal{K}}_{\alpha\beta\delta}\eta_{\gamma}^{\ \, \delta}\,$ 
on $B_3$. The metric $\eta$ is defined in \eqref{eq:intmatrix}.}. In general, these curves have a full rank intersection matrix
\beq \label{eq:intmatrix}
	\eta_{\alpha}^{\ \, \beta} = D^{\rm b}_\alpha\cdot \cC_{\rm b}^\beta\ ,
\eeq
with the divisors $D^{\rm b}_\alpha$ in $B_3$. Finally, the Cartan divisors $D_m$ are obtained 
from $\tilde D_m$ by a simple basis transformation which diagonalizes the intersection 
numbers $\tilde D_m \cdot \tilde D_n \cdot D_\alpha \cdot D_\beta$ in the indices $m,n$
as in \eqref{eq:intsD_m}.

The Shioda map \eqref{eq:ShiodaMap} is a map from the Mordell-Weil group to 
$H_6(\hat{X}_4)$ and constructed such that the following intersections vanish,
\beq \label{eq:def_properties_Shioda}
	\tilde{D}_m\cdot D_\alpha \cdot D_\beta\cdot D_\gamma=0\,,\quad 
	\tilde{D}_m\cdot \tilde{B} \cdot \mathcal{C}^\alpha=0\,,\quad 
	\tilde{D}_m\cdot c=0\,,
\eeq
for all exceptional curves $c$ introduced in the resolution of the 
singularities of type $G_{(I)}$\footnote{More precisely, these are curves that 
are associated to the roots in the root lattice of any $G_{(I)}$.} in 
$\hat{X}_4$. The first relation follows since 
$\hat{\sigma}_m\cdot D_\alpha \cdot 
D_\beta\cdot D_\gamma=\tilde{B}\cdot D_\alpha \cdot D_\beta\cdot D_\gamma$ as 
both are sections and the second and third relation are obvious from 
\eqref{eq:ShiodaMap} and \eqref{eq:tripleintersect}, \eqref{eq:Cartan}. 
The Shioda map has been applied for the 
construction of $U(1)$-symmetries in six-dimensional F-theory 
compactifications \cite{Morrison:1996pp,Aspinwall:1998xj,Park:2011ji,Morrison:2012ei} on 
elliptically fibered 
Calabi-Yau threefolds with rational sections. The map
\eqref{eq:ShiodaMap} is the natural extension of the conventional Shioda map
to Calabi-Yau fourfolds. Both $\tilde{D}_m$ and $D_m$ define 
$U(1)$-symmetries in F-theory. However, the definition 
of $D_m$ ensures in addition that the $D_m$ do not mutually intersect,
whereas the intersections of the $\tilde{D}_m$ can be in general non-diagonal.
This is clear since the $\tilde{D}_m$ are fibrations of the curves $c_m$,
see footnote \ref{footnt1}, 
over divisors in the base $B_3$.

Then the divisors $D_m$, $S_{(m)}^{\rm b} $ describe four-dimensional $U(1)$ 
gauge symmetries with the following intersection properties, that are in 
complete analogy with
\eqref{eq:Cartan},
\beq \label{eq:intsD_m}
	D_m\cdot D_n\cdot D_\alpha\cdot D_\beta=-\delta_{mn}S_{(n)}\cdot B\cdot D_\alpha\cdot D_\beta\,,\quad \quad 
        D_m\cdot D_{i_I}\cdot D_\alpha\cdot D_\beta=0\,,\quad  
        D_m\cdot D_\alpha\cdot D_\beta\cdot D_{\gamma}=0 \,.    
\eeq
The second and third equalities are a direct consequence of the 
defining properties \eqref{eq:def_properties_Shioda} of the Shioda map, 
the definition of $D_m$,
and the fact that the $D_i$ are fibrations of shrinking curves $c$ over 
divisors in the base $B_3$. 
Given a set of $D_m$ the first equation in \eqref{eq:intsD_m} can be 
viewed as the defining equation for the $S^{\rm b}_{(m)}$ 
and the vertical divisors  $S_{(m)}$. We will show
in concrete examples how to construct a basis of $D_m$ and the divisors
$S_{(m)}$ obeying the properties \eqref{eq:intsD_m}. We note that the 
association of divisors $S_{(m)}$ to $U(1)$'s in F-theory leads to a new
perspective on the interpretation of $U(1)$'s in F-theory.

That the conditions \eqref{eq:intsD_m} have to hold can also be inferred 
physically from the analysis of the 3D gauge kinetic terms
and their F-theory lift by extending the discussion \cite{Grimm:2010ks} to 
include $U(1)$-gauge symmetries. In 
this context the first relation ensures that the gauge kinetic function of 
the $U(1)$'s is diagonal, which is always achievable in field theory, and 
the second that the gauge couplings is also diagonal between the $U(1)$'s 
and the non-Abelian group $G_{(I)}$.

We can summarize equations \eqref{eq:Cartan} and \eqref{eq:intsD_m} in a more compact way in terms of the
quartic intersections \eqref{eq:fourfoldintersect} as
\beq \label{eq:vanishingInts_0}
		\cK_{i_I\, j_J\, \alpha \beta} =- \delta_{IJ}\, \mathcal{C}^{(I)}_{ij} \delta^\gamma_{(I)}   \cK_{\gamma \alpha \beta} \,,\qquad\cK_{mn\alpha \beta} = -\delta_{mn} \delta^\gamma_{(m)} \cK_{\gamma \alpha \beta}\ , \qquad 
        \cK_{\Lambda \alpha \beta \gamma} = \cK_{i_I m \alpha \beta}= 0 \ .
\eeq
Here $\cK_{\alpha \beta \gamma}$ are the triple intersections \eqref{eq:tripleintersect} in the base $B_3$ and $\delta^\gamma_{(I)}$ respectively  
$\delta^\gamma_{(m)}$ restrict to the $I$'s seven-brane stack defined in \eqref{eq:def-Calpha} respectively the $m$'s $U(1)$ seven-brane 
defined in \eqref{eq:def-Calpham}. 
We also introduced for later convenience the notation 
\beq 
   D_\Lambda=(D_i,D_m)\ ,\qquad \quad  \omega_\Lambda=(\omega_i,\omega_m)\ , \qquad \Lambda \in\{i,m\}
\eeq
unifying all divisors associated to gauge symmetries in M- and F-theory.
To complete the discussion of intersection relations
let us also note that on $\hat{X}_4$ one has
\beq \label{eq:vanishingInts_1}
 \cK_{\hat 0 \Lambda A B}=  B \cdot D_{\Lambda} \cdot D_A \cdot D_B  = 0\,.
\eeq

\subsection{Four-form fluxes in M-theory to F-theory duality} \label{sec:flux_review}

Next we construct the four-form flux $G_4$ on 
$\hat{X}_4$. The flux $G_4$ is an element in the fourth cohomology 
group $H^{4}(\hat{X}_4,\mathbb{Z}/2)$ due to the quantization 
condition \cite{Witten:1996md}
\beq \label{G4half}
	G_4+\frac12 c_2(\hat{X}_4)\,\,\in\,\, H^{4}(\hat{X}_4,\mathbb{Z})\,.
\eeq 
These conditions have been investigated in the F-theory context in 
\cite{Collinucci:2010gz,Collinucci:2012as}.
Splitting into Hodge types, there are two different types of fluxes due to the even complex 
dimension of $\hat{X}_4$ \cite{Greene:1993vm,Mayr:1996sh,Klemm:1996ts}. The first type are 
$G_4$-fluxes in the vertical cohomology group 
$H_V^{(2,2)}(\hat{X}_4,\mathbb{Z}/2)$ that is generated by the product of two forms in 
$H^{2}(\hat{X}_4,\mathbb{Z})$. Thus fluxes in $H_V^{(2,2)}(\hat{X}_4,\mathbb{Z}/2)$ can be specified as
\beq \label{eq:verticalFluxGeneral}
	G_4=m^{AB}\omega_A\wedge \omega_B\,,
\eeq
for appropriate constant coefficients $m^{AB}$. Note that it is 
crucial here to know the 
cohomology $H^{2}(\hat{X}_4,\mathbb{Z})$ explicitly which is not 
straightforward for the singular 
geometry $X_4$. These vertical fluxes are crucial in generating 
chirality in F-theory as discussed in section \ref{sec:ChiralityInF}.

The second type are fluxes $G_4$ in the normal space to the 
vertical cohomology. They lie in 
the horizontal cohomology $H^{4}_H(\hat{X}_4,\mathbb{Z}/2)$ that is 
obtained from complex structure 
variations of the $(4,0)$-form on $\hat{X}_4$. Physically they give
rise to a non-trivial classical flux superpotential \cite{Gukov:1999ya}, that corresponds in weak coupling
to D7-brane and Type IIB flux superpotentials. See 
\cite{Lust:2005bd,Grimm:2009ef,Grimm:2009sy,Jockers:2009ti,Alim:2010za,Klevers:2011xs,Braun:2011zm,Intriligator:2012ue} 
for a list of some works on 
the physical interpretations of the flux superpotential in F-theory and studies of properties 
of horizontal $G_4$-fluxes.

It can be shown that the vertical fluxes 
\eqref{eq:verticalFluxGeneral} induce Chern-Simons terms for 
vectors $A^A$ in the 3D effective action that are obtained by reducing the M-theory three-form $C_3$. For more details on the complete reduction of M-theory as well as on the lift back to 4D F-theory see \cite{Grimm:2010ks}.
Reducing $C_3$ in the M-theory compactification on $\hat{X}_4$ with 
respect to the forms $\omega_A$ introduced in \eqref{eq:basisH^11} 
we expand into 3D vectors $A^A$ as
\beq
\label{eq:C3expansion}
	C_3=A^A\wedge \omega_A=A^0\wedge \tilde \omega_0+ A^\alpha \wedge \omega_\alpha +A^\Lambda\wedge \omega_\Lambda\, ,
\eeq
with $\tilde \omega_0$ defined in \eqref{eq:omega0tilde}.
From counting indices we thus obtain 
$1+h^{1,1}(B_3)+\text{rank}(G)
=h^{1,1}(\hat{X}_4)$ Abelian $U(1)$ vector fields. The vectors $A^\Lambda$ are the 
remaining massless vectors on the 3D Coulomb branch, that are, 
from the F-theory perspective in \eqref{eq:3dDuality}, 
the gauge fields in the maximal torus of the non-Abelian gauge group $G$. The additional vectors
$A^\alpha$ are the 3D dual to the imaginary part of the K\"ahler moduli of $B_3$. 
It will be important for us that $A^0$ is identified with the  
Kaluza-Klein vector from reducing the 4D metric of the F-theory effective 
action on $S^1$, also denoted as the 3D graviphoton. It arises as the 
component $A^0_\mu \propto g_{\mu 3}$, where the index $3$ indicates the 
$S^1$-direction. 
 
Performing the dimensional reduction of the eleven-dimensional action 
in a background with the $G_4$-flux \eqref{eq:verticalFluxGeneral} one 
obtains a 3D Chern-Simons action for the vectors $A^A$ of the form \cite{Haack:2001jz} 
\beq \label{eq:theta_AB}
	S^{(3)}_{CS}=-\int \frac{1}
	{2}\Theta_{AB}A^A \wedge F^B\,,\qquad 
	\Theta_{AB}=\frac12\int_{\hat{X}_4}G_4\wedge \omega_A\wedge 
	\omega_B\,,
\eeq
where we used the conventions of \cite{Grimm:2010ks}.
Here we employ in the definition of the Chern-Simons levels $\Theta_{AB}$ the shifted basis \eqref{eq:omega0tilde}.

In addition, these flux integrals $\Theta_{AB}$ in general induce in the 
3D M-theory effective action St\"uckelberg gaugings of the complexified 
K\"ahler moduli $T_A$ associated to the divisors $D_A$ as
\beq \label{eq:gaugingT_A}
	DT_A=dT_A+i\Theta_{AB}A^B\,.
\eeq
In this context, the couplings $\Theta_{AB}$ play the role of the
``embedding tensors''\footnote{We have chosen the conventions of
\cite{Grimm:2011sk} for the normalization of the $T_A$ so that no numerical factors of 2 appear in \eqref{eq:gaugingT_A}. 
Then $T_\alpha$
agrees with 4D Type IIB K\"ahler moduli and is related to 3D K\"ahler moduli $t_\alpha$ as $t_\alpha=2 T_\alpha$.}. 
We note that for the purpose of anomaly cancellation these gaugings are 
essential since the imaginary part of some $T_\alpha$ will play
the role of an axion with an anomalous gauge transformation under 
the $U(1)$ gauge symmetries in the theory. This
will lead, as we will discuss in section \ref{sec:GStermsInF}, to a 4D 
generalized Green-Schwarz mechanism. 

The Chern-Simons levels \eqref{eq:theta_AB} are key objects to study the physics of the vertical fluxes 
\eqref{eq:verticalFluxGeneral} in F-theory. 
In order to have a clear 4D F-theory interpretation we have to impose additional conditions 
on the $G_4$-flux in M-theory 
that we summarize in terms of the flux integrals $\Theta_{AB}$ in
\eqref{eq:theta_AB}. We require the following integrals to vanish 
\cite{Marsano:2011hv,Grimm:2011sk},
\bea \label{eq:vanishingThetas}
	\Theta_{0\alpha}=\Theta_{\alpha\beta}=\Theta_{i\alpha}=0\,,\\
	\Theta_{00}=\Theta_{0i}=0\,.\nn
\eea
We emphasize that these conditions have to be evaluated in the
basis \eqref{eq:omega0tilde} that relates the fields of the M-theory 
reduction in this basis correctly to the circle-reduced 4D fields.
The conditions \eqref{eq:vanishingThetas} on the $G_4$-flux are imposed in  
addition to the conventional M-theory conditions on allowed $G_4$-flux. We 
will show in section \ref{sec:strucOfFlux+anomalies}
by evaluating the $\Theta_{AB}$ for a general $G_4$-flux of the form 
\eqref{eq:verticalFluxGeneral}
that these can be always satisfied by restricting the flux numbers 
$m^{AB}$. We will exemplify this 
even further for concrete examples in section \ref{sec:examples}. 

The requirement $\Theta_{i\alpha}=0$ can be understood readily in the 
effective field theory. 
By imposing these conditions the gaugings of the $T_\alpha$   
by the gauge fields $A^i$ are absent according to \eqref{eq:gaugingT_A}. 
These gaugings would break the 
non-Abelian part in the gauge group $G$ in the corresponding 
F-theory compactification, 
that we want to retain e.g.~as a GUT group 
for phenomenological applications. 
As we will discuss below in section \ref{sec:ChiralityInF}, the 
non-vanishing $\Theta_{\Lambda\Sigma}$ encode the chirality 
of charged matter in 4D.
It is important to stress that we do not require a vanishing of the Chern-Simons levels 
\beq
  \Theta_{0m} = \frac12\int_{\hat X_4} \tilde \omega_0 \wedge \omega_m \wedge G_4\ , 
\eeq
where, as in \eqref{eq:vanishingThetas}, one uses the redefined 
$\tilde \omega_0$ given in \eqref{eq:omega0tilde}. 
As we will discuss in more detail in section \ref{sec:Theta0m}, 
the non-vanishing components $\Theta_{0m}$   
are crucial in the study of mixed Abelian-gravitational anomalies. Clearly, due to 
the fact that $B \cdot D_m = 0$, as stated already in \eqref{eq:vanishingInts_1}, the integral 
$\int_{\hat{X}_4}\omega_{\hat 0} \wedge \omega_m \wedge G_4$ vanishes trivially 
for all fluxes $G_4$. However, this is not generically true for 
$\Theta_{0m}$ as discussed below.

Let us note that one might think that the $\Theta_{0m}$ are on an equal footing with the $\Theta_{0i}$ on
the 3D Coulomb branch of the circle reduced theory. However, as we will show in section \ref{sec:Theta0m}
a non-vanishing $\Theta_{0m}$ is generated in the IR from integrating out
Kaluza-Klein states of 4D charged fermions while the $\Theta_{0i}$ are also zero at one loop. 
We will deduce that $\Theta_{0m}$ is 
precisely given by the 4D mixed Abelian-gravitational anomaly.

\subsection{Green-Schwarz terms in F-theory}
\label{sec:GStermsInF}

In the following we will determine all Green-Schwarz counter terms 
in \eqref{eq:GSterm}. More precisely, we will outline the steps 
to derive the coefficients $b^{\alpha}_I$, $b_{mn}^\alpha$ 
and $a^\alpha$ in 4D F-theory compactifications. We 
demonstrate that these coefficients are completely determined in terms 
of the intersections on the resolved fourfold $\hat{X}_4$ as well as 
by the canonical divisor, or equivalently the first Chern class, of the base $B_3$. A similar analysis for 4D compactifications without
$U(1)$ factors can be found in \cite{Grimm:2012yq}.

We start our discussion by noting that the prefactors of the first two 
terms in \eqref{eq:GSterm} are the imaginary parts of the 4D 
gauge coupling functions for the gauge fields $F^I$ of each non-Abelian 
factor $G_{(I)}$ respectively for the $F^m$ of each $U(1)_m$. These gauge 
couplings are given by the volume of the divisors $S_{(I)}^{\rm b}$ and 
$S_{(m)}^{\rm b}$ in the base $B_3$ 
introduced after \eqref{eq:def-Calpha} and \eqref{eq:def-Calpham}. This 
can be argued by performing the M-theory reduction on $\hat{X}_4$ in the 
non-Abelian case or by using intuition from weak coupling results in Type 
IIB, where the relevant divisors are those wrapped by D7-branes. In all 
these cases the divisors are determined uniquely on $\hat{X}_4$ from their 
defining intersection properties 
\eqref{eq:Cartan} and \eqref{eq:intsD_m}. The association 
of divisors wrapped by seven-branes to $U(1)$'s in F-theory might be 
unexpected and new since $U(1)$'s, as discussed in section 
\ref{sec:F+MresolvedFF} above, are not simply related to singularities in 
the elliptic fibration of $X_4$, but to rational sections of the elliptic 
fibration, i.e.~a non-trivial Mordell-Weil group. Mathematically, the 
relation of these rational sections to divisors 
in $B_3$ is more subtle but straightforwardly formulated in terms of the 
intersections \eqref{eq:intsD_m}
of the $D_m$ as elucidated further in the following.

We start with the determination of the coefficients $b^\alpha_I$. First we 
note that the axions $\rho_\alpha$ in \eqref{eq:GSterm} are in F-theory 
the imaginary parts of the K\"ahler moduli $T_\alpha$ associated to the 
divisors $D_\alpha$ in \eqref{eq:basisH^11}. 
Then, the coefficients $b^\alpha_I$ are just given as the coefficients
in the expansion \eqref{eq:def-Calpha}, and one finds
\beq \label{eq:matchC-b}
     b_I^\alpha = \delta^\alpha_{(I)}\ . 
\eeq 
Next we determine by the same logic the coefficients $b_{mn}^\alpha$ in \eqref{eq:GSterm}. 
As mentioned before this physically amounts to define seven-branes supporting 
$U(1)$-symmetries in F-theory.
As in \eqref{eq:def-Calpham} we denote by $S^{\rm b}_{(m)}$ the divisors that support 
$U(1)$-gauge factors. 
Thus, we see from the first equation
in \eqref{eq:intsD_m} that the coefficients $b_{mn}^\alpha$ must be diagonalizable, 
and we identify
\beq \label{eq:match_b_mn}
   b_{mn}^\alpha = \delta_{mn} \delta_{(m)}^\alpha \ ,
\eeq
where the coefficients $\delta_{(m)}^\alpha$ were introduced in 
\eqref{eq:def-Calpham}. Finally, we determine the coefficients $a^\alpha$. 
We just state the final result here and refer to 
\cite{Sadov:1996zm,Bonetti:2011mw,Grimm:2012yq} for the derivation in six- 
and four-dimensional F-theory. The coefficients read
\beq  \label{eq:defa^alpha}
  a^\alpha = K^\alpha\ , \qquad  \quad c_1(B_3) = - K^\alpha \omega_\alpha\ ,  
\eeq
where we expanded the first Chern class into a basis restricted to $B_3$.

To end this section let us introduce an alternative 
way to present the coefficients $b^\alpha_I, b^\alpha_{mn},a^\alpha$ in terms of intersection numbers of 
geometrical objects. In order to do so we have to use the basis of $h^{1,1}(B_3)$ curves $ \cC^\alpha_{\rm b}$
in $B_3$ and their intersection matrix $\eta_{\alpha}^{\ \, \beta}$ introduced in \eqref{eq:intmatrix} as well as 
the induced four-cycles $ \cC^\alpha$ in $\hat X_4$ 
with Poincar\'e dual four-forms $\tilde \omega^\alpha$.
Furthermore, we introduce the  push-forward $\pi_*$ from $H_{4}(\hat{X}_4)$ to $H_{2}(B_3)$ in homology 
induced by the projection $\pi$. Its action on surfaces $S$, and by Poincar\'e duality 
also on four-forms $\tilde{\omega}$ in $H^{4}(\hat{X}_4)$, is defined as
\beq	\label{eq:push-forward}
\pi_*(S)=D^{\rm b}_\beta\eta^{-1\beta}_\alpha\ (\cC^\alpha\cdot S)_{\hat X_4}\,,\qquad \pi_*(\tilde{\omega})=\omega^{\rm b}_{\beta}\eta^{-1\beta}_\alpha\int_{\hat{X}_4}\tilde{\omega}\wedge \tilde{\omega}^\alpha \,,
\eeq
where $\omega^{\rm b}_{\alpha}$ is a two-form in $H^{(1,1)}(B_3)$ dual to $D^{\rm b}_{\alpha}$ and where we introduced the 
inverse $\eta^{-1}$ of the intersection matrix \eqref{eq:intmatrix}.
Using these definitions and the equations of section \ref{sec:F+MresolvedFF} it is straightforward
to infer
\bea\label{eq:determingb_A}
	\delta^\alpha_I C^{(I)}_{i_Ij_I}  &=&- \eta^{-1\alpha}_\gamma \ D_{i_I}\cdot D_{j_I}\cdot \cC^\gamma  =-  \eta^{-1\alpha}_\gamma\ \pi_*(D_{i_I}\cdot D_{j_I})\cdot \cC^\gamma_{\rm b}  \,,\\
     \label{eq:determingb_m} 
	\delta^\alpha_m&=&- \eta^{-1\alpha}_\beta \ D_m\cdot D_m\cdot \cC^\beta=- \eta^{-1\alpha}_\beta\ \pi_*( D_m\cdot D_m)\cdot \cC_{\rm b}^\beta\ ,\\
	\label{eq:determininga}
        a^\alpha & = & \eta^{-1\alpha}_\beta\ B \cdot B \cdot \cC^\beta =  \eta^{-1\alpha}_\beta\ \pi_*(B \cdot B) \cdot \cC_{\rm b}^\beta\ ,
\eea
where no sum is performed over $I$ and we have used that $B$ satisfies $B\cdot B= -[c_1(B_3)]\cdot B$. This way of presenting the Green-Schwarz coefficients will be 
particularly useful when translating the anomaly conditions into purely geometric conditions involving 
$G_4$ in section \ref{sec:anomalycancellationinF}. It also facilitates the determination of the
coefficients $b^\alpha_I, b^\alpha_{mn},a^\alpha$ if only the Cartan divisors $D_i$ and the sections $D_m$ are known.

\section{One-loop Chern-Simons Terms and Their F-Theory Interpretation}
\label{sec:ChiralityInF}

In this section we study the 3D Chern-Simons terms in the duality \eqref{eq:3dDuality} 
between F-theory on $X_4\times S^1$ and M-theory on the smooth fourfold
$\hat{X}_4$. We describe the matching of the one-loop Chern-Simons term in the circle 
compactification of F-theory with the classical flux-induced Chern-Simons term in M-theory.
In section~\ref{sec:ChiralityFormula3D} we concentrate on the Chern-Simons terms 
for the 3D gauge fields inherited from 4D gauge fields. We recall that 
in the circle reduction such terms are generated at one-loop after 
integrating out charged matter that acquired a mass on the 3D Coulomb 
branch. A matching with the flux-induced Chern-Simons term in M-theory 
allows to infer the 4D chiral index counting the net number of chiral 
fermions. In section~\ref{sec:Theta0m} we focus on a new 
one-loop Chern-Simons term that has not been considered so far in three 
dimensions. It involves the Kaluza-Klein vector $A^0$, and we will show 
explicitly that it is induced at one loop by integrating out massive 
Kaluza-Klein modes of the charged fermions. It will be linked with the 4D 
gauge-gravitational anomalies in section~\ref{sec:anomalycancellationinF}.

\subsection{4D Chirality formulas from the 3D Coulomb branch}
\label{sec:ChiralityFormula3D}

We begin our discussion by first stating the expected form of the 4D 
chiral index formula for charged chiral matter in a representation 
$\mathbf{R}$. In F-theory on a singular elliptic Calabi-Yau fourfold $X_4$ 
charged chiral matter is induced by seven-brane flux which maps to 
vertical $G_4$-flux. The chiral index of charged matter in a 
representation $\mathbf{R}$ of the gauge group $G$ is given by the flux 
integral 
\cite{Donagi:2008ca,Hayashi:2008ba,Braun:2011zm,Marsano:2011hv,Krause:2011xj,Grimm:2011fx,Krause:2012yh,Kuntzler:2012bu} 
\beq \label{eq:chi(R)}
	\chi(\mathbf{R})= \int_{S_{\mathbf{R}}}G_4\,.
\eeq  
In this general form, without specifying a construction of the fluxes and 
the so-called matter surfaces $S_{\mathbf{R}}$, the expression \eqref{eq:chi(R)} is 
not surprising. If seven-brane fluxes have an $G_4$-flux image in M-theory then only 
a linear expression that vanishes for $G_4=0$ is conceivable. 
However, it is important to stress that for our construction 
both $G_4$ and $S_{\mathbf{R}}$ are both naturally defined on the smooth 
M-theory fourfold $\hat{X}_4$, not on the singular fourfold $X_4$. 
This might be counter-intuitive, since in an M-theory compactification 
on a smooth space no massless charged matter appears in the effective theory.  
Nevertheless, following the strategy of \cite{Grimm:2011fx}, we argue next that 
the formula \eqref{eq:chi(R)} can be derived by consideration of the three-dimensional $\mathcal{N}=2$ 
effective gauge theory and its dual formulation in terms of F-theory on $S^1$ on the one hand and M-theory on
$\hat{X}_4$ on the other hand.


We start on the F-theory side with a 4D $\mathcal{N}=1$ supergravity theory with
gauge group $G$ and chiral matter in a representation 
$\mathbf{R}$. Then we compactify this theory on $S^1$ and move onto the Coulomb
branch of the resulting 3D gauge theory. This breaks the non-Abelian 4D gauge
symmetry to its maximal torus, $G\rightarrow U(1)^{\text{rank}(G)}$, with Abelian
gauge fields $A^{\Lambda}$. The Coulomb branch parameters are given by the VEVs
of scalars $\zeta^\Lambda$ in the 3D $\mathcal{N}=2$ vector multiplets that are the
components $A^\Lambda_3$ of the 4D vectors 
along the $S^1$, i.e.~the holonomies 
\beq
    \zeta^\Lambda=\int_{S^1}A^\Lambda\,.
\eeq
Simultaneously the chiral matter receives mass-terms that are 
proportional to the Coulomb branch parameters $\zeta$. In the 3D effective action in
the IR at energy scales below $\zeta^\Lambda$ these massive fields have to be integrated
out and generate Chern-Simons terms for the Abelian vectors at one loop,
\beq \label{eq:CSterms}
	S^{(3)}_{CS}=\int \Theta^{\text{loop}}_{\Lambda\Gamma}A^\Lambda\wedge F^\Gamma\, ,
\eeq
where we use the conventions of \cite{Kao:1995gf}.
We note that these terms are classically absent, i.e.~not generated in the compactification
from 4D to 3D. In order to prepare for section \ref{sec:Theta0m} it is 
important to mention that in a Kaluza-Klein theory also excited modes 
of the 4D matter fields along $S^1$ are 
charged under $A^\Lambda$ and can, in principle, induce a one-loop 
contribution in \eqref{eq:CSterms}. In the following we will first discuss 
the one-loop term without these excited Kaluza-Klein modes and then 
comment on their inclusion.

Let us consider the one-loop Chern-Simons term induced by Kaluza-Klein 
zero modes of the charged matter fields that became massive in the Coulomb 
branch. We denote the real mass of the matter fermion $f$ by $m_f$. The 
Coulomb branch masses $m_f$ are given by 
\beq \label{eq:qdotxi}
	m_f = q_f \cdot 
	\zeta \equiv \sum_{\Gamma=1}^{\text{rank}(G)}(q_f)_\Gamma 
	\zeta^\Gamma\,,
\eeq
where $(q_f)_\Lambda$ denotes the charge of the fermion $f$ under the $U(1)$ vector field 
$A^\Lambda$.
The loop integral expression for the Chern-Simons level is \cite{Niemi:1983rq,Redlich:1983dv,Aharony:1997bx}
\beq \label{eq:theta_LambdaSigma}
	\Theta^{\text{loop}}_{\Lambda\Sigma}=\frac{1}
	{2}\sum_f(q_f)_\Lambda(q_f)_\Sigma \ \text{sign}(m_f)=\frac12\sum_{\mathbf{R}}n(\mathbf{R})\sum_{q\in \mathbf{R}}q_\Lambda q_\Sigma \ \text{sign}(q\cdot 
	\zeta)\,,
\eeq
with the sum taken over all fermions $f$ charged under $A^\Sigma$. In the 
second equality we split this sum 
over fermions into a sum over representations $\mathbf{R}$ and then over 
charges $q$ in that representation, where $n_\mathbf{R}$ denotes the 
multiplicity of $\mathbf{R}$. Thus, we see that the Chern-Simons terms are 
only proportional to the 4D chiralities 
$\chi(\mathbf{R})=n(\mathbf{R})-n(\mathbf{R}^*)$ since the weights of the 
complex conjugate representations $\mathbf{R}^*$ are 
$q(\mathbf{R}^*)=-q(\mathbf{R})$ and since
\eqref{eq:theta_LambdaSigma} is an odd function in the charges $q$.

In a next step we can consider the contributions of excited Kaluza-Klein 
modes. These can be equally charged under $A^\Lambda$ and hence contribute 
to the Chern-Simons term.The mass of the $n$th excited fermionic mode $f$ 
is given by 
\beq
   m_n^f = q_f \cdot \zeta + \frac{n}{r}\ .
\eeq
This expression has to be used in the one-loop Chern-Simons level 
\eqref{eq:theta_LambdaSigma}.
In the following we will consider the mass hierarchy 
\beq \label{mass_hierarchy}
  m_0^f = q_f \cdot \zeta \ <\ \frac{1}{r} = m_{\rm KK}\ , 
\eeq 
such that the Coulomb branch mass scale is below the Kaluza-Klein scale. 
In this case one can drop the contribution $q_f \cdot \zeta$ for all 
excited modes $n\neq0$, since the sign $\text{sign}(m^f_n)$ is 
determined by the contribution $n/r$ alone. For each Kaluza-Klein level 
$n>0$ one finds that there is a positive term that is canceled by a term 
arising from the level $-n$. This pairwise cancellation can be inferred 
physically from the fact that a Chern-Simons term arises from parity 
violation and the excited modes do not violate parity. Therefore,
we conclude that only the zero modes contribute non-trivially to this 
coupling. As we will see in section \ref{sec:Theta0m} the situation 
changes once we consider Chern-Simons terms involving the Kaluza-Klein 
vector under which the excited modes are charged. 

The above argument shows that a certain linear combination
of the CS-couplings in \eqref{eq:CSterms} yields the chiral index 
$\chi(\mathbf{R})$ of matter in the representation $\mathbf{R}$,
\beq \label{eq:chi(R)3D}
\chi(\mathbf{R})=t_{\mathbf{R}}^{\Lambda\Sigma}\Theta_{\Lambda\Sigma}\,,
\eeq
where the matrix $t_{\mathbf{R}}^{\Lambda\Sigma}$ roughly first projects 
the sum in \eqref{eq:theta_LambdaSigma} to a particular representation 
$\mathbf{R}$ and then cancels the sum over all charges $q$ in this 
representation $\mathbf{R}$.

Considering the same terms in the M-theory reduction, we have noted in section \ref{sec:F+MresolvedFF} 
that classical Chern-Simons terms for the vectors $A^A$ are given by the 
flux-integrals \eqref{eq:theta_AB}. By M-/F-theory duality \eqref{eq:3dDuality} these have to be 
identified precisely with the loop-generated CS-terms
\eqref{eq:theta_LambdaSigma} as \cite{Grimm:2011fx}
\beq	 \label{eq:matching1loop+classical}
	-\Theta^{\text{loop}}_{\Lambda\Sigma}\equiv\frac12\Theta_{\Lambda\Sigma}
	=\frac14\int_{\hat{X}_4}G_4\wedge\omega_\Lambda\wedge\omega_\Sigma\,.
\eeq
It is important to note that also the sign-function in 
\eqref{eq:theta_LambdaSigma} is then determined by the
geometry by a simple rule. A charge vector $\underline{q}$ is positive, i.e.~$\text{sign}(q\cdot \zeta)=1$, if the corresponding shrinkable curve 
in M-theory lies in the Mori cone of $\hat{X}_4$. Accordingly, we call
it negative, $\text{sign}(q\cdot \zeta)=-1$, if it does not lie in the 
Mori cone. Therefore, in combination with \eqref{eq:chi(R)3D} we obtain 
the F-theory chiral index \eqref{eq:chi(R)} from matching 3D CS-terms. 
Furthermore, the matter surfaces $S_{\mathbf{R}}$ 
are identified as the Poincar\'e duals of 
\beq
\big[S_{\mathbf{R}}\big]=t_{\mathbf{R}}^{\Lambda\Sigma}\omega_\Lambda\wedge\omega_\Sigma\,,
\eeq
where the brackets $[\cdot]$ indicate the action of Poincar\'e duality on $\hat{X}_4$..

\subsection{Graviphoton Chern-Simons terms with Kaluza-Klein modes}
\label{sec:Theta0m}

In this section we show that a mixed Chern-Simons term for the 3D graviphoton $A^0$ and the 4D $U(1)$-vectors
$A^m$ is generated at one-loop when integrating out all Kaluza-Klein states of 4D charged fermionic matter. 
The fact that Chern-Simons terms involving the Kaluza-Klein vector are induced 
by one-loop diagrams with excited Kaluza-Klein modes running in the loop was 
first noted in a five-dimensional context  \cite{Bonetti:2011mw,Bonetti:2012fn}. It was 
shown in these works that the five-dimensional Chern-Simons terms capture the 
information about six-dimensional gravitational anomalies. 

In the following we present the one loop computation in the 
three-dimensional setting. We find that the only non-vanishing 
Chern-Simons term involving the 3D graviphoton $A^0$ is  
\beq \label{eq:graviphotonCS}
	S^{(3)}_{CS}=\int \Theta^{\text{loop}}_{0m} A^0\wedge F^m\,,
	\qquad\Theta^{\text{loop}}_{0m}=-\frac1{12}\sum_fq_m^f=-\frac1{12}\sum_{\underline{q}} n(\underline{q})q_m\,.
\eeq
The coefficient $\Theta^{\text{loop}}_{0m}$ is precisely the mixed 
Abelian-gravitational anomaly encountered in 
\eqref{eq:Anomalies:Abgravitational}. In fact, we argue that 
the relevant 3D loop diagrams generating $\Theta^{\text{loop}}_{0m}$ can 
be understood as a dimensional reduction of the 4D anomalous triangle 
diagram for the mixed Abelian-gravitational anomalies. In addition, by the 
relation \eqref{eq:matching1loop+classical} between the Chern-Simons 
levels also to classical flux integrals \eqref{eq:theta_AB}, we can 
actually deduce in 3D the 4D cancellation condition 
\eqref{eq:Anomalies:Abgravitational} of mixed Abelian-gravitational 
anomalies. Indeed, anticipating the property 
$\Theta_{0m}=-\frac12 K^\alpha\Theta_{\alpha m}$ of 
the classical intersections on $\hat{X}_4$ that we will derive in 
\eqref{eq:ThetaABcalculated}, \eqref{eq:ThetaAB_abel_2} in section 
\ref{sec:strucOfFlux+anomalies}, we obtain the anomaly condition 
by the M-/F-theory duality \eqref{eq:3dDuality} as
\beq	 \label{eq:3DDerivationMixedAnomaly}
	\frac1{3}\sum_{\underline{q}} n(\underline{q})q_m=-4\Theta^{\text{loop}}_{0m}\equiv2\Theta_{0m}=
	 -K^\alpha\Theta_{\alpha m}\,.
\eeq
This agrees precisely with \eqref{eq:Anomalies:Abgravitational} using the identification $a^\alpha=K^\alpha$ made in 
\eqref{eq:defa^alpha}. In contrast to the non-Abelian anomalies, this is a direct derivation of 4D mixed anomaly cancellation
from 3D Chern-Simons terms.

\begin{figure}[ht!]
	\begin{center}
	\includegraphics[width=0.35\textwidth]{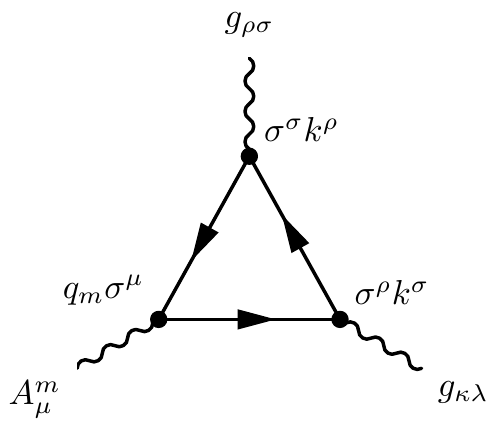}
	\end{center}
	\caption{One-loop triangle diagram responsible for the Abelian-gravitational anomaly. }
	\label{fig:triangleDiagram}
\end{figure}

We begin by understanding diagrammatically the connection of 4D anomalies to the Chern-Simons-levels 
$\Theta_{0m}$ and the relevance of KK-states running in the 3D loops.
The 4D mixed Abelian-gravitational anomaly is encoded in the one-loop triangle diagram in figure 
\ref{fig:triangleDiagram}, that leads to a violation of the corresponding current conservation law at the
quantum level. 
Here the external lines are given by two gravitons $g_{\mu\nu}$ and one 
$U(1)$-vector $A_\mu^m$. Massless charged Weyl fermions carrying 
$U(1)$-charges under the $A_\mu^m$ run in the loop. 
The vertex rules are attached to each vertex, where $k^\rho$ denotes the 
4-momentum of the fermions in the loop and we have used 
$\sigma^\mu=(-\mathbf{1},\sigma^i)$ with $\sigma^i$ denoting the 
Pauli-matrices.

\begin{figure}[ht!]
	\begin{center}
	\includegraphics[width=0.45\textwidth]{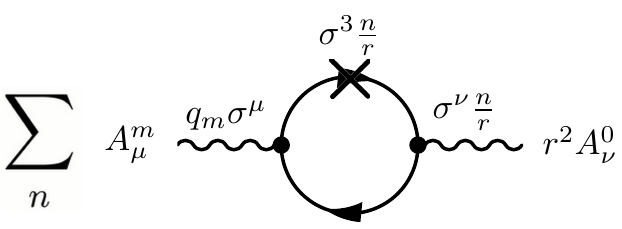}
	\end{center}
	\caption{One-loop diagram with KK-states in the loop generating the 3D Chern-Simons level $\Theta_{0m}$. }
	\label{fig:loopForTheta0m}
\end{figure}

For the dimensional reduction of the four-dimensional theory to three 
dimensions, one direction of 4D Minkowski space, say the third direction, 
is replaced by a circle of radius $r$. Thus one metric 
component reads $g_{33}=r^2$, that we interpret as a constant background 
field in 3D. From the isometries of $S^1$ we get in addition a new 3D 
vector $A_\mu^0\sim g_{\mu 3}$, the graviphoton. The heuristic dimensional 
reduction of the diagram \ref{fig:triangleDiagram} follows analogous to 
the situation in \cite{Bonetti:2012fn}. We replace in the diagram 
\ref{fig:triangleDiagram} one external graviton by the background field 
$r^2$ and the second graviton by the graviphoton as in figure 
\ref{fig:loopForTheta0m}.
We also introduced the momentum $k^3$ along the circle that is quantized as $k^3=\frac{n}{r}$. This gives a qualitative
idea of the 3D vertex rules, however, we refer to the remainder of this section for a more thorough derivation.

The background field $r$ acts like a mass term and together with the fact 
that only KK-states are charged under the graviphoton, this enforces that all KK-states of the 4D 
massless charged fermions run in the loop. Thus, effectively we obtain infinitely many one loop diagrams, 
one for each KK-state, with only two external lines, namely the Abelian gauge field $A^m_\mu$ and the 
graviphoton $A^0_\mu$. Each single of these infinitely many loop-diagrams is of the form of the 
loop-diagrams considered in \cite{Niemi:1983rq,Redlich:1983dv,Aharony:1997bx}. Thus, it is clear that a 
3D Chern-Simons term of the form \eqref{eq:graviphotonCS} is induced from the 4D mixed anomaly. The main 
subtlety that remains is to evaluate the infinite sum over KK-states appropriately. This sum has to be
regularized by zeta function regularization and will precisely account for the prefactor in \eqref{eq:graviphotonCS}.

We conclude this diagrammatic discussion by excluding further 3D 
Chern-Simons terms by a similar logic. 
As can be seen from the comparison to 4D anomaly graphs, neither the 
Chern-Simons level $\Theta_{00}$,
nor the levels $\Theta_{0i}$ can be induced at one loop. This is clear 
since the former would arise from the 
dimensional reduction of the gravitational anomaly, that is identically 
zero in 4D, and the latter from the 
mixed non-Abelian-gravitational anomaly, i.e.~either 
$\text{tr}R^2\text{tr}F$ or $\text{tr}R\text{tr}F^2$. 
However, these are identically zero by $\text{tr}F=0$ due to  the 
traceless condition of the generators of the 
non-Abelian gauge group in $G$ respectively the reality of the 
representations of the 4D Lorentz group 
implying $\text{tr}R=0$. The vanishing of $\Theta_{00}$ and $\Theta_{0i}$ can also be inferred directly in 
the 3D effective theory. For example, for $\Theta_{00}$ one can invoke a 
similar argument to the one of section \ref{sec:ChiralityFormula3D}, 
namely that all Kaluza-Klein modes pairwise cancel because of the sign-
function in the loop-correction \eqref{eq:theta_LambdaSigma}.   

Let us now come to the quantitative discussion and the derivation of the Chern-Simons
levels \eqref{eq:graviphotonCS}.
The Kaluza-Klein ansatz for the 4D metric on $\mathbb{R}^{(1,2)}\times S_r^1$ takes the form
\beq \label{eq:KKansatzg}
	g^{(4)}_{MN}(x^\mu,y)=\begin{pmatrix} g_{\mu\nu}+r^2A^0_\mu A^0_\nu & -r^2A^0_\mu\\
						-r^2A^0_\nu &  r^2
	\end{pmatrix}
\eeq
where $y$ denotes an angular coordinate of period $2\pi$, that is related to the coordinate on 
the circle of radius $r$  by $ry$ and $M,N$ denote 4D Minkowski indices for the following discussion
to avoid confusion. We denote the 3D Minkowski coordinates by $x^\mu$ and introduced the 3D metric 
$g_{\mu\nu}$ and the graviphoton $A_\mu^0$. For simplicity we assume no dynamics of the radial mode 
$r$, i.e.~$r$ is a constant. We note that the reduction ansatz 
\eqref{eq:KKansatzg} implies the vielbein
\beq
	e^a=\tilde{e}^a_\mu dx^\mu\,,\quad e^3=r(dy-A^0_\mu dx^\mu)\,,
\eeq
where we indicated a split $e_M^A=(e_M^a,e_M^3)$ of the 4D vierbein 
$e_M^A$, $A=0,\ldots, 3$ into the 3D dreibein 
$e_\mu^a$, $a=0,1,2$, and a 3D one-form $e^3$. For completeness we 
introduce the inverse metric $g_{(4)}^{MN}$ 
and the inverse vierbein $e_A^M=e_{A N}g_{(4)}^{NM}$ reading
\beq \label{eq:KKansatzg-1}
	g_{(4)}^{MN}=\begin{pmatrix} g^{\mu\nu} & (A^0)^{\mu}\\
						(A^0)^{\nu} &  \frac{1}{r^2}+A^0_\kappa (A^0)^{\kappa}
			     \end{pmatrix}\,,\qquad e_a=\tilde{e}^\nu_a\frac{\partial}{\partial x^\nu} 
                             +\tilde{e}_a^\nu A^0_\nu \frac{\partial}{\partial y}\,,\quad e_3=\frac1r\frac{\partial}{\partial y}
\eeq
where we raise and lower the indices $A$ and $a$ by the flat metric.

Next we specify the reduction ansatz for the 4D gauge theory. We start 
with vectors $A_\mu$ with KK-ansatz
\beq \label{eq:KKansatzA}
	A(x^\mu,y)=A_\mu dx^\mu+ \zeta r(dy- A^0_\mu dx^\mu )=A_\mu dx^\mu+ \zeta e^3\,,
\eeq
where $\zeta$ denotes the adjoint valued scalar in 3D.  Note that 
\eqref{eq:KKansatzg} and \eqref{eq:KKansatzA} imply 
that both $\zeta$ and $A^0$ have mass-dimension one by noting that $dy$ 
and the 3D metric have mass-dimension zero.
For 4D fermions $\Psi$ charged under the 4D $U(1)$-gauge symmetries, we specify their KK-ansatz as
\beq \label{eq:KKansatzPsi}
	\Psi(x,y)=\sum_n \Psi_n(x)e^{iy n}\,.
\eeq
The KK-tower of these states will be running in the loop and generate the 
Chern-Simons term $\Theta_{0m}$ as we will show next. It is interesting to 
note that while $T_\alpha$ transforms under the $U(1)$ as 
dictated by the gauge covariant derivative \eqref{eq:gaugingT_A} 
the fermionic partners will have an ordinary derivative 
\cite{Wess:1992cp}, $D_M=\partial_M$.

Considering massless 4D fermions the only terms in the 4D effective theory for the fermions $\Psi$ 
that are relevant for our discussion are the kinetic term. In the UV and considering a single fermion for 
simplicity this is the standard kinetic term
for a Weyl fermion reading
\beq \label{eq:kineticWeyl}
	\mathcal{L}_{\text{Weyl}}^{(4)}=-\frac{i}{\kappa_4^2}\bar{\Psi}\sigma^A e_A^M D_M \Psi\,,
\eeq
where $D_M=\partial_M+iA_M$ is the 4D covariant derivative, $\sigma^A=(-\mathbf{1},\sigma^i)$  and $\kappa_4^2$ the 4D Planck mass. We use \eqref{eq:KKansatzg-1}, \eqref{eq:KKansatzA} and \eqref{eq:KKansatzPsi} to reduce 
this to three dimensions. After integration over $y$  we obtain
\beq \label{eq:3DKKTheory}
	\mathcal{L}_{\text{KK}}^{(3)}=
	 \sum_{n=-\infty}^\infty \left[-i\bar{\Psi}_n\sigma^a \tilde{e}^\mu_a \mathcal{D}_\mu\Psi_n +\bar{\Psi}_n\sigma^3 
	 (\frac{n}{r}+q\cdot\zeta)\Psi_n\right]\,,
\eeq
where we used the shorthand notation \eqref{eq:qdotxi}.
Furthermore, we have assumed in addition that we are readily on the 3D Coulomb branch by switching on VEVs for 
the scalars $\zeta$ along the Cartan directions of $G$ breaking it into its maximal torus $U(1)^{\text{rank}(G)}$. 
The $U(1)$-charges under the remaining massless gauge fields on the Coulomb branch $A^\Lambda$ are 
denoted by  $q_\Lambda$. The covariant derivative for the KK-fermions is given by
\beq	 
	  \mathcal{D}_\mu\Psi_n=(\partial_\mu+iq_\Lambda A^\Lambda_\mu +inA_\mu^0)\Psi_n\,.
\eeq
From this we read off the 3D mass and the 3D charges\footnote{The 3D theory \eqref{eq:3DKKTheory} 
without a dynamical dilaton in \eqref{eq:KKansatzA} is automatically in the 3D Einstein frame since the factor $r$ is 
absorbed into the definition of the 3D Newtons constant multiplying the entire 3D Lagrangian. We have set it to one 
for convenience.} 
of the KK-fermions that are discretely 
labeled by the integral momentum number $n$ along the $S^1$ as 
\beq \label{eq:KKcharge+mass}
	q_0=n\,,\qquad m_n^f=\frac{n}{r}+q^f\cdot \zeta\,,
\eeq
where we label by $f$ all the $U(1)$-charges in the case that we have more than one fermion.
We see that the masses of the KK-states are offset from zero by 
the mass $m_0^f=\zeta\cdot q^f$ of the zero mode $\Psi_0$ on the Coulomb 
branch, see \eqref{eq:qdotxi}. 
The expressions \eqref{eq:KKcharge+mass} straightforwardly generalize for 
a more complicated 
constant background metric in the kinetic term \eqref{eq:kineticWeyl} for 
several fermions. However, the diagram we want to compute has two 
vertices and two fermion propagators and the metric and normalizations 
drop out. 

To determine the loop-induced Chern-Simons level $\Theta_{0m}$, we have to 
calculate the loop-diagram depicted in figure \ref{fig:loopForTheta0m}. As 
we have argued above heuristically, we have to integrate out all massive 
fermions coupling to $A^0$ and $A^m$. Indeed, we see from 
\eqref{eq:3DKKTheory} that the whole KK-tower of fermions with charge 
and mass \eqref{eq:KKcharge+mass} couples and runs in the loop. The vertex 
rules are as anticipated in figure \ref{fig:loopForTheta0m}. The idea is 
now to apply at each KK-mass-level $m^f_n$ the general formula 
\eqref{eq:theta_LambdaSigma} with the replacement of 
$\text{sign}(\zeta\cdot q^f)$ by $ \text{sign}(m^f_n)$.
Then we use $q_0=n$ to obtain the one-loop correction to the Chern-Simons 
level as
\beq \label{eq:Theta0m}
	\Theta^{\text{loop}}_{0m}=\frac12\sum_f\sum_{n=-\infty}^\infty  n q^f_m\, \text{sign}\Big(\zeta\cdot q^f+\frac{n}{r}\Big)\,.
\eeq

This sum is in general divergent and can be regulated using zeta function regularization. 
We recall the definition of the Riemann zeta function by the Dirichlet series
\beq \label{eq:Riemannzeta}
	\zeta(s)=\sum_{k=0}^\infty k^{-s}\,,\qquad \text{for }\ \ \text{Re}(s)> 1\,,
\eeq
In order to evaluate the infinite sum \eqref{eq:Theta0m} we need the following identity that holds for $\vert x\vert<1$,
\beq \label{eq:basicIdentity}
	\sum_n n \,\text{sign}\Big(x+n\Big)=
	\sum_n n \,\text{sign}(n)
	=2\sum_{n=1}^\infty n=2\zeta(-1)=-\frac{1}{6}\,. 
\eeq
We note that the first equality holds since $x$ lies in the interval 
$[0,1)$ and this allows us to replace $\text{sign}(n+x)=\text{sign}(n)$. 
Then in the second equality we split the sum into positive and negative 
$n$, that yields due to the sign$(n)$ the same sum. Finally we used the 
well-known result $\zeta(-1)=-\frac{1}{12}$ from analytic 
continuation of the Riemann zeta function \eqref{eq:Riemannzeta} for 
$\text{Re}(s)<1$.

Now we are prepared to perform the sum over $n$ in \eqref{eq:Theta0m}. We 
focus on each summand in the sum over fermions $f$ independently. 
Furthermore, we identify $x\equiv r m_0^f=r \zeta\cdot q^f$
in \eqref{eq:basicIdentity}.  We consider the effective theory in the IR 
at an energy scale, that is sufficiently smaller than
the mass-scale $m_{\text{KK}}$ of KK-states and the mass scale $m_0^f$ on 
the Coulomb branch. However, in order to 
apply the field theory analysis of 
\cite{Niemi:1983rq,Redlich:1983dv,Aharony:1997bx} and the result
\eqref{eq:theta_LambdaSigma} the masses $m^f_0$  of the massive fermions 
have to be smaller than the KK mass scale $m_{\text{KK}}$
as in \eqref{mass_hierarchy}. 
Since the mass scale $m_0^f$ depends on the position $\zeta^\Lambda$ on 
the 3D Coulomb branch, it can be made parametrically small. In this case 
we can trust the loop result \eqref{eq:theta_LambdaSigma}.\footnote{The 
effective field theory description used in \cite{Aharony:1997bx} 
breaks down for $x\geq 1$ since interactions with KK-states become as 
relevant as interactions with the $\Psi_0$. 
To find a relation similar to \eqref{eq:basicIdentity} when $x>1$
one uses the floor function $\lfloor x \rfloor$ that rounds down a given 
real number $x$. Then one obtains  $\sum_n n \,\text{sign}\Big(x+n\Big)= 
\zeta(-1,-\lfloor x\rfloor)+\zeta(-1,\lfloor x\rfloor)
-\lfloor x\rfloor=-\frac16-\lfloor x\rfloor(\lfloor x\rfloor+1)$, 
where the Hurwitz zeta function $\zeta(s,q)$ has been evaluated. One sees 
that the Chern-Simons terms incrementally jump by $2$ for each fermion 
with mass $m_0^f$ crossing the threshold $m_{\text{KK}}$.}
We then have $x\equiv r m_0^f<1$ and can apply \eqref{eq:basicIdentity} to 
readily obtain the Chern-Simons levels as
\bea \label{eq:Theta0mfinal}
	\Theta^{\text{loop}}_{0m}=-\frac1{12}\sum_fq_m^f=-\frac1{12}\sum_{\underline{q}} n(\underline{q})q_m\,.
\eea
Here we replaced the sum over individual fermions $f$ by a sum over 
charges $q$ where $n(\underline{q})$ denotes their multiplicities. We note that \eqref{eq:Theta0mfinal} is an odd function in the charges and 
as a result only sensitive to the chiral index 4D $\chi(\underline{q})=n(\underline{q})-n(-\underline{q})$ 
since 4D vector-like pairs of fermions cancel out.

We see that \eqref{eq:Theta0mfinal} is precisely the mixed Abelian-gravitational anomaly on the left hand side 
in \eqref{eq:Anomalies:Abgravitational}. This is what one expects since, by reversing the logic 
from the beginning of this section, in the limit $\zeta\rightarrow 0$ we have $x=0$ and sending $r\rightarrow\infty$
all KK-states become massless and we have to recover the 4D anomaly result.

\section{Anomaly Cancellation In F-Theory} \label{sec:anomalycancellationinF}

In this section we will discuss anomaly cancellation in F-theory. This 
involves relating, via the general formulas 
\eqref{eq:Anomalies:purenonAb}--\eqref{eq:Anomalies:Abgravitational} for 
anomaly cancellation in section \ref{sec:anomlies+cancellation}, the 
Green-Schwarz counter terms discussed in section \ref{sec:GStermsInF} to 
the 4D chiralities obtained from 3D Chern-Simons terms following section 
\ref{sec:ChiralityInF}. This will connect seemingly physically different 
object, the geometric intersections on $\hat{X}_4$ and flux integrals, 
with each other. The main challenge of this section will be to understand 
these relations directly from analyzing the geometrical structure of the 
resolution fourfold $\hat{X}_4$ on the one hand, and 3D Chern-Simons terms 
on the other hand. 
Anomaly cancellation in 6D F-theory compactifications has been under
intense investigation as reviewed in \cite{Taylor:2011wt}. 

\subsection{The geometric structure of F-theory anomaly cancellation}

We begin with an outline of the general geometric relations 
imposed on $\hat{X}_4$ by anomaly cancellation in F-theory. As we 
will argue in this section these relate intersections 
of resolution divisors and holomorphic curves over the matter curves 
$\Sigma_{\mathbf{R}}$ in the base $B_3$ on the one side to certain flux 
integrals on $\hat{X}_4$ on the other side. These geometric relations that 
we will discover by imposing 4D anomaly cancellation will be very similar 
to those found in 6D F-theory on Calabi-Yau threefolds \cite{Park:2011ji}. 
The crucial point in the analysis in 4D will 
be the necessity of the inclusion of $G_4$-flux on both sides of the 
discovered relations.

The 4D anomaly cancellation conditions 
\eqref{eq:Anomalies:purenonAb}--\eqref{eq:Anomalies:Abgravitational} can 
be directly translated into the geometry of the resolved fourfold 
$\hat{X}_4$. We denote the holomorphic curves in $\hat{X}_4$, that resolve 
singular elliptic fibers over codimension two\footnote{These 
curves are characterized by the fact that they are isolated over 
the codimension two curves $\Sigma_{\mathbf{R}}$ in $B_3$ with 
moduli space given by  $\Sigma_{\mathbf{R}}$. Curves 
corresponding to Yukawa couplings are isolated at points, i.e.~do 
exhibit the moduli space of a point.} in the base $B_3$ 
and that thus lie in the weight lattice of $G$, by $c$. Then,  
the 4D charge of a matter particle obtained from a wrapped 
M2-brane on $c$ under $A^\Lambda$ is given by
\beq \label{eq:M2charge}
	q_\Lambda=\int_{c}\omega_\Lambda\,.
\eeq 
This can be seen from reducing the electric coupling of the 
M2-brane to $C_3$ along $c$. Then the anomaly constraints relate 
sums over these holomorphic curves in the weight lattice of $G$ to 
certain flux integrals of $G_4$. It is important to emphasize that the 
following geometric relations only hold if the $G_4$-flux obeys all the 
conditions \eqref{eq:vanishingThetas} and not for generic $G_4$-flux in 
M-theory. The anomaly cancellation conditions 
\eqref{eq:Anomalies:purenonAb}-\eqref{eq:Anomalies:AbnonAb}, using 
\eqref{eq:M2charge} to express the charges, then translate into 
\bea \label{eq:anomalyCondCohomology_1}
    \frac13\sum_{S_{\mathbf{R}}}\sum_{c\,\subset S_\mathbf{R}}\int_{S_\mathbf{R}}G_4\int_{c}\omega_{(\Lambda}\int_{c}\omega_\Sigma
    \int_{c}\omega_{\Gamma)}&=&\frac{1}{2}\int_{\hat{X}_4}G_4\wedge 
    \omega_\alpha\wedge\omega_{(\Gamma} \eta^{-1\,\alpha}_\beta\int_{S^\beta} 
    \omega_\Lambda\wedge\omega_{\Sigma)}       \nn \\
    &=&\frac{1}{2}\int_{\hat{X}_4}G_4\wedge 
    \pi_*(\omega_{(\Lambda}\wedge\omega_\Sigma)\wedge\omega_{\Gamma)} 
    \,,
\eea
where we indicate a symmetrization of indices $\Lambda$, $\Sigma$ and 
$\Gamma$ by $(\cdot)$. The cancellation condition 
\eqref{eq:Anomalies:Abgravitational} of the Abelian-gravitational anomaly
yields the geometric relation
\bea \label{eq:anomalyCondCohomology_2}
    \frac1{3}\sum_{S_{\mathbf{R}}}\sum_{c\,\subset S_\mathbf{R}}\int_{S_\mathbf{R}}G_4\int_{c}\omega_\Lambda&=&\int_{\hat{X}_4}G_4\wedge 
    \omega_\alpha\wedge\omega_\Lambda \eta^{-1\,\alpha}_\beta\int_{S_{\rm b}^\beta} 
    c_1(B_3)\nn\\
    &=&\int_{\hat{X}_4}G_4\wedge 
    c_1(B_3)\wedge\omega_\Lambda\,.
\eea
In both equations we split the sum over curves $c$ into a sum over matter 
surfaces $S_\mathbf{R}$ and a sum over curves $c$ lying inside a 
particular $S_{\mathbf{R}}$. In addition we used the topological metric 
$\eta_{\alpha}^{\ \beta}$ respectively its inverse 
$\eta_{\alpha}^{-1\,\beta}$ on $B_3$ as introduced in \eqref{eq:intmatrix} 
and from the first to the second line in each equation the push-forward 
$\pi_*$ to $H^{(1,1)}(B_3)$ defined in \eqref{eq:push-forward}.

We can use Poincar\'e duality to 
rewrite the relation \eqref{eq:anomalyCondCohomology_1} and 
\eqref{eq:anomalyCondCohomology_2} more compactly in homology
as intersections of divisors and holomorphic curves $c$ in the weight 
lattice of $G$
\bea \label{eq:anomalyCondHomology}
	\frac{1}{3}\sum_{S_{\mathbf{R}}}\sum_{c\,\subset S_\mathbf{R}}(S_\mathbf{R}\cdot [G_4])(c\cdot D_{(\Lambda})(c\cdot D_\Sigma)
	(c\cdot D_{\Gamma)})&=&\frac{1}{2}[G_4]\cdot D_{(\Gamma}\cdot
    \pi_*(D_\Lambda\cdot D_{\Sigma)})\ ,\\
    \frac1{3}\sum_{S_{\mathbf{R}}}\sum_{c\,\subset S_\mathbf{R}}(S_\mathbf{R}\cdot [G_4])(c\cdot D_\Lambda)&=&[G_4]\cdot 
    [c_1(B_3)]\cdot D_\Lambda\, , \nn
\eea
where we again symmetrized the indices $\Lambda$, $\Sigma$ and $\Gamma$.

Both \eqref{eq:anomalyCondCohomology_1} and the first equation contains of 
\eqref{eq:anomalyCondHomology} capture all three different
types of gauge anomalies, i.e.~the purely Abelian as well as non-Abelian
anomalies and the mixed Abelian-non-Abelian anomalies. This can be 
seen by choosing $\Gamma, \Sigma, \Lambda$ as the indices of the 
Cartan divisors $D_{i_I}$ of $G_{(I)}$ and noting that the right 
hand side vanishes because of $\Theta_{i\alpha}=0$ in 
\eqref{eq:vanishingThetas} yielding the purely non-Abelian anomaly 
condition \eqref{eq:Anomalies:purenonAb}. The mixed 
Abelian-non-Abelian anomaly \eqref{eq:Anomalies:AbnonAb} is obtained 
by choosing  $\Gamma, \Sigma, \Lambda=i,j,m$. In order to see a matching
of the Lie algebra structure we expand following \cite{Park:2011ji} the 
Cartan generators in the GS-terms \eqref{eq:GSterm} involving 
$\text{tr}(F^I\wedge F^I)$ in a coroot basis $\mathcal{T}_{i_I}$. 
Since the anomalies have to hold for all Cartan generators and exploiting 
$\frac{1}{\lambda_I}\text{tr}(\mathcal{T}_{i_I}\mathcal{T}_{j_I})
=\mathcal{C}^{(I)}_{i_Ij_I}$ in the GS-terms we obtain a match with the 
right hand side of \eqref{eq:anomalyCondCohomology_1}, 
\eqref{eq:anomalyCondHomology}, where we employ \eqref{eq:Cartan} . 
The purely Abelian anomaly \eqref{eq:Anomalies:pureAb} 
is manifest for the choice  $\Gamma, \Sigma, \Lambda=m,n,k$. The second
equation summarizes all mixed Abelian-gravitational anomalies 
\eqref{eq:Anomalies:Abgravitational} as is evident by choosing 
$\Lambda=m$. 

As we have noted already in the introduction the 
conditions \eqref{eq:anomalyCondHomology} might generally hold also for 
$G_4$-flux that are not wedges of two two-forms as in 
\eqref{eq:verticalFluxGeneral}. These more general $G_4$-flux can 
correspond to seven-brane fluxes trivial in the ambient space. For these 
fluxes the right-hand side of  \eqref{eq:anomalyCondHomology} induced by 
the gaugings vanishes and the left side might pose a non-trivial 
constraint on the spectrum.

\subsection{The structure of F-theory fluxes and Green-Schwarz terms}
\label{sec:strucOfFlux+anomalies}

Next we discuss the general structure of the  $G_4$-flux
obeying the F-theory constraints \eqref{eq:vanishingThetas}. A very 
general, presumably the most general, vertical $G_4$-flux on $\hat{X}_4$ 
that can meet all conditions listed in section \ref{sec:flux_review} takes 
the form
\beq \label{eq:generalG4}
	G_4= \widehat G_4+ F^{(m)}\wedge \omega_m\,.
\eeq
Here the first term $\widehat G_4$ denotes a non-Abelian flux 
related to the non-Abelian gauge groups $G_{(I)}$. It only involves the 
two-forms $\omega_i, \omega_\alpha, \omega_0$ introduced in 
\eqref{eq:basisH^11}. 
The second part are Abelian fluxes that are of the form of wedge products 
of a two-form $F^{(m)}$ in $B_3$ with each $\omega_m$ dual
to the $U(1)$-Cartan divisors $D_m$ in \eqref{eq:basisH^11}. These fluxes
can be thought of as the lift of internal world-volume fluxes $F^{(m)}$ on 
each $U(1)$ seven-brane to $G_4$-flux.
The fluxes $F^{(m)}$ being $(1,1)$-forms on
$B_3$ can be expanded, after introducing in general arbitrary flux numbers 
$f^{(m)\alpha}$, as
\beq
	F^{(m)}=f^{(m)\alpha}\, \omega_\alpha\,.
\eeq	

The validity of the generic splitting \eqref{eq:generalG4} of $G_4$ into 
Abelian and non-Abelian fluxes can be motivated by moving in the 
complex structure moduli space of the Calabi-Yau fourfold. For this 
argument let us begin with a fourfold $\hat {Y}_4$ yielding a purely 
non-Abelian gauge group $G_{(I)}$ without any $U(1)$-factors. Then only 
the first flux in \eqref{eq:generalG4} is present, of which we assume that 
it obeys all consistency conditions \eqref{eq:vanishingThetas} on 
$\hat {Y}_4$. Upon specializing the complex structure 
of $\hat {Y}_4$ we can unhiggs a number of $U(1)$ symmetries by forming 
and resolving a singularities in $\hat {Y}_4$. This leads to a new 
fourfold $\hat{X}_4$ with extra divisors $D_m$. Then we can just pull-back 
the non-Abelian flux $\widehat G_4$ from $\hat {Y}_4$ to $\hat{X}_4$ and 
obtain, by virtue of the second and third equation in \eqref{eq:intsD_m},  
a valid flux that we again denoted by $\widehat G_4$. This pull-back flux 
is still the most general non-Abelian flux $\widehat G_4$ we can construct 
on $\hat{X}_4$, since $\widehat G_4$ by definition involves only the 
$\omega_i$, $\omega_\alpha$ and $\omega_0$ and since in the transition 
from $\hat{Y}_4 $ to $\hat{X}_4$ no any new Cartan divisors $D_i$ nor new 
base divisors $D_\alpha$ were induced.  
Thus, new fluxes on $\hat{X}_4$ that were not available  on $\hat{Y}_4$ 
are of the form $F^{(m)}\wedge \omega_m$\footnote{Although physically more 
elusive and not present in examples considered below, fluxes of the form 
$\omega_i\wedge\omega_m$ and $\omega_m\wedge\omega_n$ for more than one 
$U(1)$ are mathematically not excluded in general.}, where $F^{(m)}$ is 
a general two-form on $B_3$ as we will see below.

\subsubsection{Non-Abelian fluxes only}

We start by working out the conditions imposed by \eqref{eq:vanishingThetas} on the non-Abelian flux 
$\widehat G_4$ in \eqref{eq:generalG4} as well as the gaugings induced by it. We expand $\widehat G_4$ that in a basis of
cohomology of $H^{4}_V(\hat{X}_4)$ as
\beq \label{eq:G4expansion}
	\widehat G_4 =N^{ij}\, \omega_i\wedge\omega_j+f^{i\alpha}\, \omega_i\wedge\omega_\alpha
	+\tilde{N}_{\alpha}\, \tilde{\omega}^\alpha+N^{\alpha}\, \omega_0\wedge\omega_\alpha\,,
\eeq
where the $\tilde{\omega}^\alpha$ as before denote pull-backs from 
$H^{4}(B_3)$.
The coefficients $f^{i\alpha}$ have the interpretation 
of a seven-brane two-form flux $F_2^i$ in the direction of the i-th Cartan 
generator of $G$. The fluxes $N^{ij}$ are non-Abelian in nature as they 
are associated to products of two Cartan divisors.

With the expansion \eqref{eq:G4expansion} of $\widehat{G}_4$ we readily 
evaluate the constraints \eqref{eq:vanishingThetas}. 
Employing the conditions \eqref{eq:vanishingInts_0}, 
\eqref{eq:vanishingInts_1}, \eqref{eq:Cartan}, 
\eqref{eq:intsD_m}  and \eqref{eq:newIntsTildeB} we obtain after some 
algebra the following,
\bea \label{eq:ThetaABcalculated}
	\Theta_{00}&=&K^\alpha K^\beta\Theta_{\alpha\beta}\,,\qquad\qquad\qquad\qquad\quad\ \ \,\Theta_{0\alpha}
	=\frac12(\tilde{N}_\alpha+N^\gamma K^{\beta}\mathcal{K}_{\alpha\beta\gamma}- K^\beta 
	\Theta_{\alpha\beta})\,,\nn\\ 
	\Theta_{0\Lambda}&=&-\frac12 K^\alpha \Theta_{\alpha \Lambda}\,,\qquad\qquad\qquad\qquad\quad\ \
	\Theta_{\alpha\beta}=\frac12(N^\gamma-N^{ij}C_{ij}^\gamma)\mathcal{K}_{\alpha\beta\gamma} 	\,,\nn\\
	\Theta_{\alpha i}&=&\frac12(N^{j_1 j_2}\mathcal{K}_{ij_1j_2\alpha}-
	f^{j\beta}C_{ij}^\gamma \mathcal{K}_{\alpha\beta\gamma})\,,\qquad \quad
	\Theta_{\alpha m}=\frac12 N^{ij}\mathcal{K}_{ijm\alpha}\,,\nn\\
	\Theta_{ij}&=&\frac12(N^{i_1i_2}\mathcal{K}_{i_1i_2ij}+f^{i_1\alpha}\mathcal{K}_{i_1ij\alpha}
	-\tilde{N}_\alpha C_{ij}^\alpha)\,,\,\,\, 
	\Theta_{im}=\frac12(N^{i_1i_2}\mathcal{K}_{i_1i_2im}+f^{j\alpha}\mathcal{K}_{ijm\alpha})\,,\nn\\
	\Theta_{mn}&=&\frac12(N^{ij}\mathcal{K}_{ijmn}+f^{i\alpha}\mathcal{K}_{imn\alpha}-\tilde{N}_\alpha b_{mn}^\alpha) \,.
\eea
Here we used the definitions \eqref{eq:fourfoldintersect}, 
\eqref{eq:matchC-b}, \eqref{eq:match_b_mn} and 
$\mathcal{K}_{\alpha\beta\gamma}$ denotes the triple intersections on 
$B_3$ introduced in \eqref{eq:tripleintersect} and the $K^\alpha$ 
were defined by the expansion $c_1(B_3)=-K^\alpha\omega_\alpha$. We also 
introduced the shorthand notation $C^\gamma_{ij}=C_{ij}^{(I)}b^\gamma_I$.

Indeed, we can solve the constraints \eqref{eq:vanishingThetas} explicitly 
using the expressions \eqref{eq:ThetaABcalculated}. 
We obtain three sets of conditions from 
$\Theta_{\alpha\beta}=\Theta_{0\alpha}=\Theta_{\alpha i}=0$ 
that we solve to determine the flux numbers $N^\alpha$, $\tilde{N}_\alpha$ 
and $f^{i \alpha}$ in terms of the $N^{ij}$,
\bea \label{eq:vanishingThetasExplicit}
	&N^{\alpha}=N^{ij}C_{ij}^\alpha\,,\qquad\tilde{N}_{\gamma}
	= -\frac12 (N^\gamma+N^{ij}C_{ij}^\gamma)K^\beta\mathcal{K}_{\alpha\beta\gamma}
	=-N^{ij}C_{ij}^\gamma K^\beta\mathcal{K}_{\alpha\beta\gamma}\,,&\nn\\
	&f^{i\alpha}=N^{i_1i_2}\mathcal{K}_{i_1i_2j\alpha}(C^{-1}_{(I)})^{j i}\mathcal{K}_{S_{(I)}^{\rm b}}^{\beta\alpha}\,.&
\eea
Here we have introduced the inverse $\mathcal{K}_{S_{(I)}^{\rm b}}^{\beta\alpha}$ of the intersection form on the 
seven-brane divisor $S^{A}_{\rm b}$ as well as the inverse of the 
corresponding Cartan matrix $C^{-1}_{(I)}$.

It is satisfying to see that $\Theta_{0m}=-\frac12 K^\alpha \Theta_{\alpha m}$ is implied by \eqref{eq:ThetaABcalculated}. Thus, $\Theta_{0m}$ is
precisely the coefficient of the
Green-Schwarz counter term relevant for the cancellation of mixed Abelian-
gravitational anomalies in \eqref{eq:Anomalies:Abgravitational} 
respectively \eqref{eq:anomalyCondCohomology_2}. 
This is precisely what we expect from the discussion of section 
\ref{sec:Theta0m} where we related
the Chern-Simons level $\Theta_{0m}^{\text{loop}}$ induced by loops of KK-
fermions to precisely the mixed anomalies.
As we argued there further, the matching of the 3D Chern-Simons levels 
$\Theta_{0m}\equiv \Theta_{0m}^{\text{loop}}$ 
by M-/F-theory duality \eqref{eq:3dDuality} is a prove of the 4D anomaly 
cancellation condition. 

Before switching to the discussion of Abelian fluxes let us conclude by 
mentioning that by virtue of section \ref{sec:ChiralityInF} the flux 
integrals $\Theta_{\Sigma\Lambda}$ in the next to last and last line in 
\eqref{eq:ThetaABcalculated} encode the 4D chiralities. By means of 
\eqref{eq:vanishingThetasExplicit} the chiralities only depend on the
flux numbers $N^{ij}$. It would be interesting to explicitly express
these in terms of the $N^{ij}$.

\subsubsection{Inclusion of Abelian fluxes}

We begin by checking that Abelian fluxes of the form alluded to in \eqref{eq:generalG4} can be added
to a given non-Abelian flux $\widehat G_4$. For this we check that fluxes of the form 
$F^{(m)} \wedge \omega_m $ with $F^{(m)}$ obtained from elements in $H^{(1,1)}(B_3)$ 
obey all constraints in \eqref{eq:vanishingThetas}. 

As before we work out explicitly the flux integrals $\Theta_{AB}$. 
Using again the intersections \eqref{eq:intsD_m} and 
\eqref{eq:vanishingInts_0}, \eqref{eq:vanishingInts_1} as well as the 
shifted form $\tilde{\omega}_0$ defined in \eqref{eq:omega0tilde} we  
obtain \beq \label{eq:ThetaAB_abel_1}
	\Theta_{00}=\Theta_{0\alpha}=\Theta_{0i}=\Theta_{\alpha\beta}=\Theta_{\alpha i}=0\,,\quad 
\eeq
and 
\bea \label{eq:ThetaAB_abel_2}
        &\Theta_{0 m}=-\frac12 K^\alpha\Theta_{\alpha m}\ ,\qquad \quad \Theta_{\alpha m}=-\frac12 f^{(n)\beta} b_{nm}^\gamma\mathcal{K}_{\alpha\beta\gamma}\ ,&\nn\\
	& \Theta_{ij}=\frac12 f^{(m)\alpha}\, \mathcal{K}_{\alpha mij}\ ,\quad \Theta_{im}=\frac12 f^{(n) \alpha} \mathcal{K}_{\alpha mnj}\ ,\quad \Theta_{mn}=\frac12 f^{(p) \alpha} \mathcal{K}_{\alpha p mn}\ .&
\eea
We see that all conditions \eqref{eq:vanishingThetas} are automatically obeyed. We note that we again have 
$\Theta_{0 m}=-\frac12 K^\alpha\Theta_{\alpha m}$ in agreement with the discussion in section \ref{sec:Theta0m}, 
as in the non-Abelian case discussed above. As expected the new fluxes contribute to the $\Theta_{\Sigma\Lambda}$ 
and thus will generate additional 4D chiralities.

\section{Anomalies and GS-terms in Type IIB Orientifolds} \label{sec:orientifold_anomalies}

In this section we 
analyze the connection of F-theory anomaly cancellation to  
anomaly cancellation in Type IIB orientifold setups at weak string coupling. 
We first recall anomaly cancellation for compactifications with 
D7-branes and O7-planes in subsection \ref{sec:anom_review} by
largely following \cite{Plauschinn:2008yd}. The discussion of anomalies 
and their cancellation 
in Type II orientifold setups has been studied significantly in 
the past, as can be found in the reviews \cite{Cvetic:2001nr,MarchesanoBuznego:2003hp,Blumenhagen:2005mu,Blumenhagen:2006ci}. 
This comparison to F-theory is performed in subsection \ref{sec:F-match}, 
by relating the gaugings and Green-Schwarz counter terms of both setups.
We discuss the complications arising due to additional `geometrically 
massive' $U(1)$ gauge 
symmetries that are present in orientifold setups but integrated out in 
F-theory. A detailed comparison $G_4$-flux in M-theory dual to F-theory 
compactifications with $SU(N)$ and $SU(N)\times U(1)$ and D7-brane fluxes 
in the Type IIB orientifold limit has been addressed in 
\cite{Krause:2012yh}.

\subsection{Anomaly cancellation in Type IIB orientifolds} \label{sec:anom_review}

We begin with a review of anomaly cancellation in 
Type IIB orientifolds with intersecting D7-branes.
We consider Abelian or non-Abelian D7-branes on divisors $S^{\rm IIB}_{(\cA)}$ 
in the orientifold covering space $Y_3$ and denote the orientifold images of 
these D7-brane divisors by $S^{\rm IIB}_{(\cA')}$. The orientifold planes are 
located at the fix-point set of the orientifold involution denoted by 
$S^{\rm O7}$. The number of D7-branes on a stack are denoted by $N_{\cA}$. 
This includes cases where $N_\cA =1$ yielding a D7-brane with a $U(1)$ gauge 
group. In the following we will only focus on $U(N_{\cA})$ 
groups for simplicity, such that the 4D gauge group is given by
\beq \label{G_ori}
  G_{\rm ori} = U(N_1) \times\ldots  \times  U(N_{{\widehat n_G}}) \ , \qquad 
  {\cA} =1,\dots, \widehat n_G\ . 
\eeq
The cases $SO(N_{I})$ and $Sp(N_{I})$ should work out similarly and it would 
be interesting to perform such an analysis using F-theory. 

The Type IIB orientifold matter spectrum arises from strings stretching 
between D7-branes. Four representations are present: 
(1) the bi-fundamental representation 
$\mathbf{F}_{\bar \cA \cB} = (\bar N_{\cA},N_{\cB})$ corresponding to strings 
stretching from a D7-brane on $S_{(\cA)}^{\rm D7}$ to a D7-brane on 
$S_{(\cB)}^{\rm D7}$ with $\cA \neq \cB$; 
(2) the bi-fundamental representation $\mathbf{F}_{\cA \cB } = 
(N_{\cA},N_{\cB})$ for strings stretching between $S_{(\cA')}^{\rm D7}$ and 
$S_{(\cB)}^{\rm D7}$ with $\cA \neq \cB$; 
(3) the symmetric representation $\mathbf{S}_\cA$, (4) the anti-symmetric 
representation $\mathbf{A}_\cA$. 
The latter two arise from strings stretching between the brane and its 
orientifold image. For evaluating the anomaly cancellation conditions it will 
be crucial to give the $U(1)$-charges $q_{\cA}(\mathbf{R})$ 
under the $\cA$th $U(1)$ in \eqref{G_ori} of the various representations. 
Together with the dimension of the representations they are summarized in 
Table \ref{table:orientifold_sprectrum}. 

The chiral indices in the orientifold setup are denoted by $\chi(\mathbf{R})$. 
For the intersecting D7-brane setup one has the relations 
\beq \label{eq:index_relations}
 \chi(\mathbf{F}_{\cA \cB}) = \chi(\mathbf{F}_{\cB \cA}) \ ,     \quad 
 \chi(\mathbf{F}_{\bar \cA \cB})= - \chi(\mathbf{F}_{\bar \cB \cA})\ , \quad 
  \chi(\mathbf{F}_{\bar \cA \cA}) =0\,,\quad \chi(\mathbf{F}_{\cA\cA})=\chi(\mathbf{S}_\cA)+\chi(\mathbf{A}_\cA)\, .
\eeq
The first condition is trivially true since exchanging brane and image, which
is a symmetry of the theory, exchanges the fundamental representations. 
The second condition uses the 
fact that the bi-fundamental $(\bar{N}_{\cA},N_{\cB})$ is the 
complex conjugate of $(N_{\cA},\bar{N}_{\cB})$ and the chiral index changes
sign under the exchange of a representation and its complex conjugate. The 
third equality is true since since the representation 
$\mathbf{F}_{\bar{\cA}\cA}$ is real and the last equality makes use of the 
group theoretical decomposition $N_\cA\otimes N_\cA=\mathbf{S}_\cA\oplus \mathbf{A}_\cA$. To simplify our notation we will introduce the 
abbreviations
\beq
  \chi_{\cA \cB} = \chi(\mathbf{F}_{\cA \cB})\ ,\qquad 
  \chi_{\bar \cA \cB} = \chi(\mathbf{F}_{\bar \cA \cB})\ ,\qquad 
  \chi_{(\cA)} = \chi(\mathbf{S}_\cA)\ ,\qquad  
  \chi_{[ \cA]} = \chi(\mathbf{A}_\cA)\ .
\eeq
The chiralities  in Type IIB are calculated explicitly from 
a simple index depending on the D7-brane flux. 
In the following we will focus on D7-branes with $U(1)$-fluxes $\cF^{\cA}$
only, which are a non-trivial background on $S^{\rm IIB}_\cA$ in the $U(1)$-directions 
of \eqref{G_ori}. The indices then read 
\beq  \label{eq:I_abTypeIIB}
	I_{\cA\cB}=\int_{S^{\rm IIB}_{(\cA)}\cdot S^{\rm IIB}_{(\cB)}} (\cF^{\cA}-\cF^{\cB})\,
\eeq
for two D7-brane divisors $S^{\rm IIB}_{(\cA)},$ $S^{\rm IIB}_{(\cB)}$ with 
their intersection denoted by $S^{\rm IIB}_{(\cA)}\cdot S^{\rm IIB}_{(\cB)}$. 
For convenience we have set the string length $\ell_s=1$. 
In terms of these 
indices the chiralities are given by
\bea \label{eq:chiTypeIIB}
	\chi_{\bar\cA \cB} & =& I_{\cA\cB}\,,\qquad \chi_{\cA\cB}=I_{\cA'\cB}\,,\\
	\chi_{(\cA)} &=&\tfrac{1}{2}(I_{\cA'\cA}-2 I_{O7 \cA})\,,\quad \chi_{[\cA]}=\tfrac{1}{2}(I_{\cA'\cA}+2 I_{O7 \cA})\, . \nn
\eea
Here $\cA'$ denotes the orientifold image of a divisor $S^{\rm IIB}_{(\cA)}$
and $I_{O7 \cA}$ is obtained from \eqref{eq:I_abTypeIIB}  by 
using the orientifold divisor $S_{\rm O7}$ and noting that the are 
no fluxes on an O7-plane.
We note that from \eqref{eq:chiTypeIIB} and the symmetry properties of \eqref{eq:I_abTypeIIB}
we immediately obtain the relations \eqref{eq:index_relations}, where we here and
in the following have to make use of
\beq \label{eq:orientifoldAction}
	\cF^{\cA'} \wedge [S^{\rm IIB}_{\cA'}]=- \sigma^* (\cF^\cA\wedge [S^{\rm IIB}_{\cA}])\,, \qquad 
	A^\cA_{U(1)} = - \sigma^* A^{\cA'}_{U(1)}\,.
\eeq
The map $\sigma$ denotes the orientifold involution and $A^{\cA}_{U(1)}$ 
are the $U(1)$-gauge fields in the setup.
\begin{table}[h!]
\centering 
 \begin{tabular}{|c |c |c |c|c|c|c|}
 \hline
 \rule[-.2cm]{0cm}{0.7cm} $\mathbf{R}$ & rep.  & $U(1)_\cC$-charge  &   $ \text{dim}(\mathbf{R})$ &   $\quad V(\mathbf{R}) \quad $  &   $U(\mathbf{R})  $  &   $ \chi(\mathbf{R})$    \\ \hline \hline
 \rule[-.2cm]{0cm}{0.7cm}   $\mathbf{F}_{\bar \cA \cB} $  &  $(\bar N_\cA, N_\cB)_{\underline{q}}$ & $(\underline{q})_\cC=-\delta_{\cC \cA}  + \delta_{\cC \cB}$  & $N_\cA N_\cB$  & $(-1,1)$ & $(1,1)$ &  $\chi_{\bar \cA \cB}$  \\
 \rule[-.2cm]{0cm}{0.7cm}   $\mathbf{F}_{\cA \cB }$ &  $(N_\cA, N_\cB)_{\underline{q}}$ &  $(\underline{q})_\cC=\delta_{\cC \cA}  + \delta_{\cC \cB}$  & $N_\cA N_\cB$ & $(1,1)$ & $(1,1)$ &  $\chi_{\cA \cB}$  \\
 \rule[-.2cm]{0cm}{0.7cm}   $\mathbf{S}_\cA$ & $\text{Sym}(N_\cA)_{\underline{q}}$ & $(\underline{q})_\cC=2 \delta_{\cC \cA}$ & $\tfrac12 N_\cA (N_\cA + 1)$ & $N_A+4$ & $N_A+2$ &  $\chi_{( \cA)}$ \\
 \rule[-.2cm]{0cm}{0.7cm}   $\mathbf{A}_\cA$ & $\Lambda^2(N_\cA)_{\underline{q}}$ & $(\underline{q})_\cC=2 \delta_{\cC \cA}$ &  $\tfrac12 N_\cA (N_\cA - 1)$& $N_A-4$ & $N_A-2$ &  $\chi_{[ \cA]}$ \\
 \hline
\end{tabular}
\caption{A list of the matter representations arising from intersecting D7-
branes. The second column denotes the representations of $SU(N_\cA)\times 
SU(N_\cB)\times U(1)^{\hat{n}_G}$ respectively of $SU(N_\cA)\times 
U(1)^{\hat{n}_G}$ and the third column denotes the charges under the $\cC$th 
$U(1)$. Finally we summarize the dimension of $\mathbf{R}$ in the fourth, the 
values of the cubic Casimir and the index in \eqref{eq:traceRelations} in the fifth respectively
sixth columns and the chiral indices in the last.}
\label{table:orientifold_sprectrum}
\end{table}

Next we evaluate the anomalies on the left-hand side of 
\eqref{eq:Anomalies:purenonAb}-\eqref{eq:Anomalies:Abgravitational} 
for the 4D orientifold theory.
One evaluates for the weak coupling representations employing
the last relation in \eqref{eq:index_relations}
and Table \ref{table:orientifold_sprectrum} that 
\bea \label{eq:AnomaliesOrie}
   \text{tr}_\mathbf{f} (F^\cA)^3:& 
        & A^{F^3}_\cA = \sum_{\cC \neq \cA} N_\cC \big(\chi_{\bar \cC \cA}  
        +  \chi_{ \cC \cA}  \big)  + (N_\cA + 4) \chi_{(\cA)} + (N_\cA-4) 
        \chi_{[ \cA]} \,,   \nn \\[0cm] 
   F^\cA_{U(1)} (F^\cB_{U(1)})^2 :&  
        &\tfrac12 N_\cA \big[ N_\cB   \big(\chi_{\bar \cB \cA} +  
        \chi_{ \cB \cA}\big) (1-\delta_{\cA\cB})\\&&
         +  \tfrac{1}{3}\delta_{\cA\cB} \big(\sum_{\cC \neq \cA} N_\cC \big(\chi_{\bar \cC \cA}  
        +  \chi_{ \cC \cA}  \big)  + 4(N_\cA + 1) \chi_{(\cA)} +4 (N_\cA-1) 
        \chi_{[ \cA]}    \big) \big]\nn\\
          &&=\tfrac12 N_\cA \big[ N_\cB   \big(\chi_{\bar \cB \cA} +  
        \chi_{ \cB \cA}\big)    +  \tfrac13\delta_{\cA\cB} A^{F^3}_\cA \big]\,,\nn \\[.2cm]  
  F^\cA_{U(1)}\text{tr}_\mathbf{f} (F^\cB)^2: & 
        & \tfrac12\big[  N_\cA \big(\chi_{\bar \cB \cA} 
         +  \chi_{ \cB \cA} \big) (1-\delta_{\cA\cB})
         +  \delta_{\cA \cB} \big( \sum_{\cC \neq \cA} N_\cC \big(\chi_{\bar \cC \cA}  
        +  \chi_{ \cC \cA}  \big)  \\
        &&+ 2(N_\cA + 2) \chi_{(\cA)} + 2(N_\cA-2) 
        \chi_{[ \cA]} \big)\big]=\tfrac12 \big[N_\cA \big(\chi_{\bar \cB \cA} 
         +  \chi_{ \cB \cA} \big)+\delta_{\cA\cB} A^{F^3}_\cA  \big]\,, \nn   \\[.2cm] 
     F^\cA_{U(1)} \text{tr}R^2:&
        &  \tfrac{1}{48} N_\cA \Big[ A^{F^3}_\cA - 3 \chi_{(\cA)} + 3 \chi_{[\cA]}\Big]=\tfrac{1}{48} N_\cA \Big[ A^{F^3}_\cA +6I_{O7\cA}\Big]\ . 
\eea
Here $F^\cA$ denotes the field strength of $SU(N_{\cA})$
and $F^\cA_{U(1)}$ are the $U(1)$ field strengths of the 
Abelian factors appearing in \eqref{G_ori}.
Note that the contribution $\delta_{\cA \cB}$ in the second and third condition
arises from bi-fundamental matter from all intersections with D7-branes with the 
$\cA$-stack together with the symmetric and anti-symmetric representations 
arising from the $\cA$-stack itself. We have conveniently written the 
result using $A^{F^3}_\cA$.

These anomalies are canceled by a GS-mechanism in 4D when imposing 
D7-brane and D5-brane tadpole cancellation. For general Type
IIB Calabi-Yau orientifold setups with D7-branes this has been shown 
in \cite{Plauschinn:2008yd}. Since $ \text{tr}_\mathbf{f} (F^\cA)^3$
is non-factorizable one has 
\beq
  \text{D7-/D5-tadpole cancellation}\qquad \Rightarrow\qquad   A^{F^3}_\cA=0\ . 
\eeq
The remaining conditions 
then read
\bea \label{eq:AnomaliesOrie_2}
   F^\cA_{U(1)} (F^\cB_{U(1)})^2&:\quad  
        & \tfrac{1}{2} N_\cA  N_\cB \, \big[   I_{\cB \cA} +  
        I_{ \cB' \cA} \big]\,, \\[.2cm]  
%
%
%
%
  F^\cA_{U(1)} \text{tr}_\mathbf{f} (F^\cB)^2&:\quad 
        & \tfrac12 N_\cA \, \big[I_{\cB \cA} 
         +  I_{ \cB' \cA} \big]\,, \nn   \\[.2cm] 
       F^\cA_{U(1)} \text{tr}R^2&:\quad  
        &  \tfrac{1}{8}   N_\cA\, I_{O7 \cA} \ ,
      \nn
\eea
where we have used the relations \eqref{eq:chiTypeIIB} and 
\eqref{eq:orientifoldAction}. This concludes the evaluation of the 
one-loop anomaly.

Next we will recall the derivation of the gaugings and 
Green-Schwarz counter terms. There are two different types of axions 
relevant in the Green-Schwarz mechanism
in Type IIB orientifolds. Firstly, there are orientifold-even axions 
$\rho_\alpha$ arising 
from the expansion of the R--R-form $C_4$ into a basis of $H^4_+(Y_3)$. 
Secondly, there are orientifold-odd axions in the expansion of the 
R--R-form $C_2$ into a basis of $H^2_-(Y_3)$. They appear in the 
Green-Schwarz-terms of Type IIB as
\beq \label{eq:GStermIIB}
	S_{\rm GS}^{(4)}=-\frac{1}{4}\int (b^\alpha_\cA\rho_\alpha + b_{a\,\cA}c^a)\text{tr}(F^\cA\wedge F^\cA)
	                       + (b^{\alpha}_{\cA\cB}\rho_\alpha 
	                       + b_{a\,\cA\cB}c^a)F^\cA_{ U(1)}\wedge F^\cB_{U(1)}-\frac{1}{4}a^\alpha \rho_\alpha \text{tr}(R\wedge R)\,,
\eeq
where we identified 
$I \cong \cA$, $m,n \cong \cA,\cB$ and used $\lambda_\cA=1$ for 
$SU(N_\cA)$ in \eqref{eq:GSterm}.
The coupling coefficients $b_{\cA}$, $b_{\cA\cB}$ and $a$ determined in 
the following.

In order to obtain the gaugings and GS-terms in weakly coupled Type IIB 
setups one has to expand the D7-brane and O7-plane 
world-volume actions. 
The resulting gaugings of these axions can be extracted from St\"uckelberg couplings arising 
from  $C_4$ and $C_6$ in the D7-brane actions and are given by \cite{Jockers:2004yj} 
\beq \label{eq:TypeIIBgaugings}
 \nabla \rho_\alpha = d\rho_\alpha + \Theta_{\alpha \cA} \, A^{\cA}_{U(1)}\ , \qquad 
	\nabla c^a  =  dc^a + \Theta^{a}_{\ \cA}\, A^{\cA}_{U(1)}\ , \eeq
with constant gaugings specified by 
\beq \label{eq:ThetacAalphaTypeIIb}
  \Theta_{\alpha \cA}=2N_{\cA}
	\int_{S_{(\cA)}^{\rm IIB}}\mathcal{F}^{\cA}\wedge 
	\omega_\alpha 
	\ ,
	\qquad   
   \Theta^{a}_{\ \cA} = - 2N_{\cA} \delta_{(\cA)}^{a}\, .
\eeq
Here the factor of $2$ in the first equation appears since the same 
coupling is generated
twice, by the D7-brane on $S_{(\cA)}$ and by its image brane on 
$S_{(\cA')}$. The $A^\cA_{U(1)}$ are the $U(1)$'s in \eqref{G_ori}
that comprise combinations of the brane and image brane $U(1)'s$. 
The D7-brane field strength 
$F^{(\cA)}_{\rm D7}$ has a flux 
part $\mathcal{F}^{(\cA)}$ on $S^{\rm IIB}_{(\cA)}$, where we focus
on fluxes in the $U(1)_{\cA}$-direction only. We stress that the gaugings 
of $c^a$ are purely geometrical and independent of any brane flux. They 
contain the coefficients $ \delta_{(A)}^{a}$
arising in the expansion of the D7-brane locus $S_{(\cA)}^{\rm IIB}$ as
\beq \label{eq:Sexp_ori}
	\tfrac12(S_{(\cA)}^{\rm IIB} +S_{(\cA')}^{\rm IIB} )  = \delta_{(\cA)}^{\alpha}D_\alpha^{\rm IIB}\, ,\qquad  \tfrac12(S_{(\cA)}^{\rm IIB} - S_{(\cA')}^{\rm IIB})  =  \delta_{(\cA)}^{a}D_a^{\rm IIB}\,,\qquad  
	S_{\rm O7}=\delta_{\rm O7}^{\alpha}D_\alpha^{\rm IIB}\, ,
\eeq
where $D_\alpha^{\rm IIB},\ D_a^{\rm IIB}$ is a basis of $H_4^+(Y_3)$, 
$H_4^-(Y_3)$, respectively. The factor of $2$ in the second equation
in \eqref{eq:ThetacAalphaTypeIIb} is due to the factor $\frac12$ in 
\eqref{eq:Sexp_ori}.
For completeness we have displayed here the expansion of the O7-plane 
locus as well. 

The GS-counter terms can be extracted 
from the Chern-Simons world-volume actions of D7-branes and O7-planes.
One obtains the GS-terms \eqref{eq:GSterm} for both the $C_4$ axions 
$\rho_\alpha$ with the coefficients $b^\alpha_{\cA}$, $b_{\cA\cB}^\alpha$ 
and $a^\alpha$ and for the $C_2$ axions $c^a$ with  coefficients 
$b_{a\,\cA}$, $b_{a\,\cA\cB}$ and $a^\alpha$ as 
\bea \label{eq:bs_asTypeIIB}
	   b^\alpha_{\cA} \!\!\!&=&\!\!\! 2\delta^\alpha_{(\cA)}\,, \qquad b_{a\, \cA} =	2\int_{S_{(\cA)}^{\rm IIB}}\mathcal{F}^{\cA}\wedge 
	\omega_a\, , \qquad b_{\cA\cB}^\alpha =  N_\cA\delta_{\cA\cB}\, b^\alpha_{ \cA}\,, \qquad 
         b_{a\, \cA \cB}  = N_\cA \delta_{\cA \cB}\, b_{a\, \cA} \,\nn \\
          a^\alpha\!\! \! &=&\! \! \!
	-\tfrac{1}{6}\big(2\delta_{\rm O7}^\alpha+\sum_\cA 
	N_\cA\delta_{(\cA)}^\alpha\big)=-\delta_{O7}^\alpha\, .  
\eea
As before the factor of $2$ in $b_{a\,\cA}$ is due to the two identical 
contributions from $S_{(\cA)}$ and the 
image cycle $S_{(\cA')}$.
Here the first equality for $a^\alpha$ can be inferred by the reduction of 
the D7-brane world-volume theory on $S_{(\cA)}^{\rm IIB}$  and the image 
cycle $S_{(\cA')}^{\rm IIB}$ as well as the O7-plane action on 
$S_{\rm O7}$ using \eqref{eq:Sexp_ori} and 
$\mu_{\rm O7}=-8\mu_{\rm D7}$ on the orientifold double cover 
\cite{Grimm:2012yq}.
For the second equality we use the Type IIB  D7-brane tadpole, formulated 
again on the double cover, reading
\beq
	8S^{O7}=\sum_\cA N_{\cA}(S^{\rm IIB}_{(\cA)}+S^{\rm IIB}_{(\cA')})\,.
\eeq

Finally we can formulate the anomaly cancellation conditions of a Type IIB
orientifold. This means that the anomalies \eqref{eq:AnomaliesOrie} have 
to equal the GS-terms as required in 
\eqref{eq:Anomalies:purenonAb}-\eqref{eq:Anomalies:Abgravitational}.
With \eqref{eq:bs_asTypeIIB},  \eqref{eq:ThetacAalphaTypeIIb} and \eqref{eq:I_abTypeIIB}  these take 
the form 
\begin{align}
	& F^\cA_{U(1)}\, (F^\cB_{U(1)})^2: &\qquad  &
         \tfrac{1}{4}b^\alpha_{\cB\cB} \Theta^{\phantom{\alpha}}_{\cA\alpha} + \tfrac14 b^{\phantom{a}}_{a\, \cB \cB} \Theta^a_{\ \cA} =  \tfrac12 N_\cA N_\cB \, \big[I_{\cB \cA} 
         +  I_{ \cB' \cA} \big] \\[.2cm]  
  &  F^\cA_{U(1)}\text{tr}_\mathbf{f} (F^\cB)^2:&\qquad &
        \tfrac{1}{4}b^\alpha_{\cB} \Theta^{\phantom{\alpha}}_{\cA \alpha} + \tfrac14 b^{\phantom{a}}_{a\,  \cB} \Theta^a_{\ \cA}
        = \tfrac12 N_\cA \, \big[I_{\cB \cA} 
         +  I_{ \cB' \cA} \big] \,, \nn   \\[.2cm] 
&   F^\cA_{U(1)} \text{tr}R^2 :&\qquad  &
         -\tfrac{1}{16}a^\alpha\Theta_{\alpha\cA}=
        \tfrac{1}{8}   N_\cA\, I_{O7 \cA}\, .\nn
\end{align}
We note that there is no symmetrization over the indices $\cA$, $\cB$ in 
the first equation since the 
Green-Schwarz terms for the $U(1)$-field strengths 
$F_{ U(1)}^\cA$ in \eqref{eq:GStermIIB} are diagonal in $\cA$, $\cB$
due to \eqref{eq:bs_asTypeIIB}.
This shows agreement of the Green-Schwarz terms with the anomalies 
\eqref{eq:AnomaliesOrie_2}. Thus, anomaly cancellation in Type IIB 
with D7-branes and O7-planes with canceled tadpoles is guaranteed.

\subsection{Comparison with F-theory anomaly cancellation} \label{sec:F-match}

The complication in matching the F-theory setup with the corresponding weak 
coupling compactification arises from 
the fact that the $U(1)$'s in each of the $U(N_\cA)=U(1)_\cA\times 
SU(N_\cA)$ can be massive due to a geometric St\"uckelberg mechanism 
\cite{Jockers:2004yj} as investigated recently in \cite{Grimm:2011tb}. 
More precisely, the geometric gaugings \eqref{eq:TypeIIBgaugings}, 
\eqref{eq:ThetacAalphaTypeIIb} of the additional R--R two-form axions 
$c^a$ arising from cycles negative under the 
orientifold projection can render the combinations 
$\Theta^a_{\ \cA}A_{U(1)}^\cA$ of $U(1)$'s massive even in the absence 
of fluxes. This implies that in general 
\beq \label{G_ori_reduction}
   G \subset G_{\rm ori}\ ,\qquad n_{ U(1)} \leq \widehat n_{G}\ ,
\eeq
where $G$ is the F-theory gauge group 
$G= SU(N_1) \times \ldots  \times SU(N_{n_G}) \times U(1)_1 \times \ldots 
U(1)_{n_{U(1)}}$, and $n_G$ is the number of $U(N_\cA)$ factors in 
\eqref{G_ori} with $N_\cA>1$.

We denote the geometrically massive and massless $U(1)$'s by  
\bea
   \text{geom. massive:} \quad &A^M,& \quad M = 1,\ldots, \text{rank}(\Theta^a_{\ \ \cA})\, ,\\
   \text{geom. massless:}\quad &A^m,& \quad m = 1, \ldots,  n_{U(1)} \ , \nn 
\eea
where $n_{U(1)} \equiv \widehat n_G -  \text{rank} (\Theta^a_{\ \ \cA})$.
To perform the comparison with F-theory we thus 
have to perform a basis transformation of the
$U(1)$-gauge fields $A^\cA_{U(1)}$ to separate $A^M$ and $A^m$
in the Lagrangian.
Thus, we use the ansatz
\beq \label{eq:projections}
   A^{\cA}_{U(1)} = \pi_m^{\cA} A^m + \pi_M^\cA A^M \, , \qquad \quad  
   \pi^{\cA}_m \Theta^{a}_{\ \cA} = 0 \, ,
\eeq
where $\text{rank}(\pi_m^{\cA})=n_{U(1)}$ and $\text{rank}(\pi_m^{\cA}) + 
\text{rank}(\pi_M^{\cA}) = \widehat n_{G}$.
The vanishing condition tells us that the geometrically massless $U(1)$'s 
are in the kernel of the map $\Theta^{a}_{\ \cA}$. 

The map to F-theory can now be formulated using the matrices $\pi_m^{\cA} ,\pi_M^\cA$.
We identify the $A^m$ with the $U(1)$-vectors in 
\eqref{eq:C3expansion} in F-theory. 
The F-theory gaugings and Green-Schwarz terms are captured by 
\beq
   \Theta_{m \alpha} = \pi^\cA_m \Theta_{\cA \alpha}\ , \qquad  b^\alpha_m = \pi^\cA_m b^\alpha_\cA\ .
\eeq
Using these identifies we can project the orientifold Green-Schwarz terms to the 
geometrically massless modes as 
\begin{align} \label{proj_GSterms}
	& F^m_{U(1)}\, (F^n_{U(1)})^2: &\qquad  &
         \pi^{\cA}_m   \pi^{\cB}_n \big(    b^\alpha_{(\cA\cB} \Theta^{\phantom{\alpha}}_{\cB)\alpha} +  b^{\phantom{a}}_{a\, (\cA \cB} \Theta^a_{\ \cB)} \big) 
          =  b^\alpha_{(m n} \Theta^{\phantom{\alpha}}_{n)\alpha}   \\[.2cm]  
  &  F^m_{U(1)}\text{tr}_\mathbf{f} (F^I)^2:&\qquad &
      \pi^{\cA}_m   \big( b^\alpha_{I} \Theta^{\phantom{\alpha}}_{\cA \alpha} +  b^{\phantom{a}}_{a\,  I} \Theta^a_{\ \cA} \big)
        =  b^\alpha_{I} \Theta^{\phantom{\alpha}}_{m \alpha} \,, \nn   \\[.2cm] 
&   F^m_{U(1)} \text{tr}R^2 :&\qquad  &
         \pi^{\cA}_m a^\alpha\Theta_{\alpha\cA}= a^\alpha\Theta_{\alpha m }   \, ,\nn
\end{align}
where we have repeatedly used the condition $\pi^{\cA}_m \Theta^{a}_{\ \cA} = 0$ for the geometrically 
massless $U(1)$'s.
In order to complete the match we note that $a^\alpha$ agrees for the  
orientifold and F-theory setup. 
The Kodaira constraint, ensuring that $c_{1}(\hat X_4)=0$, that takes at weak coupling the form
\beq
	\label{eq:Kodaira}
	12 c_1(B_3)=[\Delta]=2[S^{\text{O7}}]+\sum_{I} 
	N_{\text{D7}_{I }}[S^{\text{IIB}}_{(I)}]+\sum_{m}
	[S^{\text{IIB}}_{(m)}] +  [\Delta'] \,.
\eeq
Here we split the discriminant $[\Delta]$ into a sum over D7-branes and 
O7-planes. It is important to note that the sum over D7-branes includes 
stacks with $N_{\text{D7}_I}$ D7-branes and single D7-branes corresponding 
to massless $U(1)$-symmetries. We claim that there is a residual 
discriminant $\Delta'$ corresponding to D7-branes with no associated 
geometrically massless $U(1)$'s. Thus, both geometrically massless and 
massive $U(1)$'s are captured by $\Delta$ in F-theory.

To conclude this section, we stress that the F-theory effective action 
obtained from a Calabi-Yau reduction of M-theory only includes the subset 
$A^m$ of $U(1)$'s that are geometrically massless but can become massive due to fluxes. 
Following the arguments of section \ref{sec:anomalycancellationinF}, together with 
the explicit checks for a number of examples in section~\ref{sec:examples},
we claim that this F-theory effective action is free of anomalies. 
The F-theory setup comprises a consistent reduction of the orientifold setup 
for scales where the geometrically massive $U(1)$'s are 
already integrated out. We have recalled that anomalies are canceled 
in orientifold compactifications with both types of $U(1)$'s due to 
tadpole cancellation. While we have not explicitly integrated out 
the geometrically massive $U(1)$'s one expects an anomaly free 
theory after this process. Indeed, the resulting theory should match 
the F-theory effective action determined via M-theory. 
We have then shown that the Green-Schwarz terms of the 
projected orientifold setup \eqref{proj_GSterms} indeed match the Green-Schwarz terms 
in F-theory determined in section \ref{sec:AnomaliesInFGUTs}.
To extend also the F-theory setup it would be very interesting to include 
the geometrically massive $U(1)$'s as suggested in \cite{Grimm:2011tb}. 
In this work it was argued that the geometrically massive $U(1)$'s can be included in 
the M-theory to F-theory limit when extending the M-theory reduction 
to include certain non-closed forms.


\section{4D Anomaly-Free F-Theory Compactifications}
\label{sec:examples}

In this section we study anomaly cancellation in concrete
examples. We consider compact elliptically fibered Calabi-Yau fourfolds
realized as toric hypersurfaces. We begin by showing anomaly cancellation
for a fourfold with base $B_3=\mathbb{P}^2\times \mathbb{P}^1$ and giving 
rise to an $SU(5)$ gauge symmetry in section \ref{sec:X4withSU5}.
Then we include $U(1)$ symmetries by the method of the $U(1)$-restricted 
Tate model in section \ref{sec:X4withSU5xU1}. We consider two fourfolds,
one with base $B_3=\mathbb{P}^2\times \mathbb{P}^1$ and another one with
$B_3=\text{Bl}_x(\mathbb{P}^3)$. In both cases we show that all anomalies
are cancelled by an implementation of the Green-Schwarz mechanism.
	
\subsection{F-theory with $SU(5)$ gauge symmetry}
\label{sec:X4withSU5}

Let us consider a concrete F-theory compactification to 4D with an $SU(5)$
gauge symmetry. The corresponding elliptically fibered Calabi-Yau fourfold 
$X_4$ has to be constructed so that an $SU(5)$ singularity in the elliptic 
fibration is formed over a divisor $S$ in the base $B_3$. In this simple 
case and assuming enhancement over codimension two in $B_3$ to $SU(6)$ and 
$SO(10)$ matter arises in the representations $\mathbf{5}$, $\mathbf{10}$ 
and its conjugate representations $\bar{\mathbf{5}}$,
$\overline{\mathbf{10}}$. Upon switching on suitable chirality inducing 
$G_4$-flux a chiral 4D matter spectrum is induced.

For this theory we evaluate the anomaly condition 
\eqref{eq:anomalyCondHomology}. Since the other anomalies are trivial, 
only the anomaly cancellation condition for the 
$SU(5)$ gauge anomaly poses a non-trivial constraint on the spectrum. 
As discussed above this anomaly is non-factorizable and 
can thus not be canceled by a GS-mechanism but has to cancel by 
itself. The anomaly condition then reads	
\beq \label{F^3-anomaly-SU5}
	\chi(\mathbf{5})+\chi(\mathbf{10})=0\,,
\eeq	
which is the well-known fact that in any consistent 
$SU(5)$ GUT theory one must have an equal number of 
$\mathbf{5}$'s and $\overline{\mathbf{10}}$'s. In evaluating 
\eqref{F^3-anomaly-SU5} we employed the general trace relations 
\eqref{eq:traceRelations} that we evaluated explicitly for $SU(5)$ using 
\eqref{F^3-trace-relations}.

We demonstrate that the anomaly condition \eqref{F^3-anomaly-SU5} is 
automatically obeyed in a concrete example of an F-theory
compactification with $SU(5)$ gauge symmetry. We consider the F-theory 
compactification on a Calabi-Yau fourfold $X_4$ elliptically fibered over 
the base $B_3=\mathbb{P}^2\times \mathbb{P}^1$. 
We realize it as a toric hypersurface in the five-dimensional toric 
ambient space $\mathcal{W}=\mathbb{P}(\mathcal{O}\oplus \mathcal{L}^2\oplus \mathcal{L}^2)$, where the 
line bundle $\mathcal{L}$ is given by the anti-canonical bundle 
$K^{-1}_{B_3}$ of $B_3$, $\mathcal{L}=K^{-1}_{B_3}$.
By specializing the complex structure of a generic elliptic fibration over $B_3$, we generate 
a single $SU(5)$ singularity over a divisor $\hat{S}$ in $B_3$,
that we choose to be the zero-section of the $\mathbb{P}^1$, i.e.~$\hat{S}=\mathbb{P}^2$. 
We readily blow-up this $SU(5)$ singularity torically by blowing up the toric ambient space
$\mathcal{W}$. This induces four new, exceptional divisors $D_i$. 
The Calabi-Yau hypersurface in this new toric space is the smooth Calabi-Yau fourfold $\hat{X}_4$.
The toric data specifying this smooth fourfold reads
\small
\begin{equation} \label{P1P2}
\begin{array}{|rrrrr|c|}
\hline
\multicolumn{5}{|c|}{\text{Points}}& D_A\\
\hline
\hline
   -1 &  0 & 0  & 0 & 0 &  \phantom{\int^X}\hspace{-1.5Em}\tilde{X}\\
     0 & -1 & 0 & 0 & 0 & Y\\
     3 &  2 &  0 & 0 & 0 & B\\
     3 &  2 & 1 & 0 & 0 &  \hat{S}\\
     3  & 2 & -1 & 0 & 0 & S\\
     3 &  2 & 0 & 1 & 1 & H\\ 
     3  & 2 & 0 &-1 & 0 & H\\
     3 &  2 & 0 & 0 &-1 & H\\
     2 &  1 & 1 & 0 & 0 & D_1\\
     1 & 1 & 1 & 0 &0 & D_4\\
     1  &  0 & 1 & 0 &0 & D_2\\
     0  & 0 & 1 & 0 & 0 & D_3\\
     \hline
\end{array}
\end{equation}
\normalsize
Each line denotes a point in $\mathbf{Z}^5$
representing the vertices defining the polyhedron of the toric ambient 
space $\mathcal{W}$. We introduced the generators of the divisor group, 
that are associated to these vertices. By linear equivalences, 
only $D_A=(B,S,H,D_1,D_2,D_3,D_4)$ are independent. Both $\tilde{X}$ and 
$Y$ as well as $S=\hat{S}+\sum_i D_i$, according to \eqref{eq:Shat} with 
$a_i=1$, are linear combinations of these generators. The basis of 
divisors on the base $B_3$ is given by $D_\alpha=(S,H)$.
It can easily be checked that the $D_i$ are the Cartan divisors of $SU(5)$ 
obeying the relation \eqref{eq:Cartan} for the Cartan matrix of $SU(5)$ 
and with the divisor $S$ as introduced in \eqref{P1P2}.

Omitting the details and referring to \cite{Grimm:2011fx} for the technicalities, the polyhedron 
defined by \eqref{P1P2} admits three different Calabi-Yau phases out of 54 different star-triangulations of the toric ambient space $\mathcal{W}$. Picking one particular triangulation,  
it can be seen that the relative 
Mori cone of $\hat{X}_4$\footnote{We note that we have to form the Mori cone of $\hat{X}_4$. It
is found by intersecting the various Mori cones of the different triangulations of the toric 
ambient space that lead to the same Calabi-Yau phase.} of effective curves contains some of the weights of 
the $SU(5)$ representations $\mathbf{R}=\mathbf{5},\overline{\mathbf{10}}$ and their complex 
conjugates. Thus, we see that indeed 6D matter arises precisely in these representations.
The 4D matter is induced by turning on a suitable $G_4$-flux meeting all the consistency conditions 
\eqref{eq:vanishingThetas}. Choosing the triangulation denoted as phase I 
in \cite{Grimm:2011fx} we start with the ansatz \eqref{eq:G4expansion} 
without Abelian fluxes, since there are no $U(1)$-symmetries by 
construction, and satisfy the equations 
\eqref{eq:vanishingThetasExplicit}. We obtain the 
one-parameter family of allowed flux that takes the form
\beq \label{eq:GfluxSU5}
	G_4(\lambda)=\lambda(8D_2D_4-4D_2^2-2D_2D_3+3D_3^2+9D_3 H)\,,
\eeq
with quantized flux parameter $\lambda$. We have suppressed the use of 
Poincar\'e duality in our notation of $G_4(\lambda)$
for the sake of the readability of this expression.

Next we introduce the four multiplicities $n(\mathbf{R})$ of the representations $\mathbf{R}=\mathbf{5},\,\mathbf{10},\,\overline{\mathbf{10}}\,,\overline{\mathbf{5}}$ as free variables and 
solves for them by  the matching \eqref{eq:matching1loop+classical} of 3D 
Chern-Simons terms. One finds that as expected only the chiralities 
$\chi(\bar{\mathbf{5}})=n(\bar{\mathbf{5}})-n(\mathbf{5})$
and $\chi(\mathbf{10})=n(\mathbf{10})-n(\overline{\mathbf{10}})$ can be determined as
\beq
	\chi(\bar{\mathbf{5}})=\chi(\mathbf{10})=162\lambda\,.
\eeq
This immediately confirms that the non-Abelian $SU(5)$ anomaly condition 
\eqref{F^3-anomaly-SU5}
is obeyed. Since this is the only possible anomaly the constructed $SU(5)$ 
GUT is anomaly-free.

\subsection{F-theory with $SU(5)\times U(1)$ gauge symmetry}
\label{sec:X4withSU5xU1}

As a more involved example we consider now an F-theory compactification with
gauge group $SU(5)\times U(1)$. Anomaly cancellation in this theory is 
richer due to the possibility of both mixed and purely Abelian anomalies that
might require a non-trivial GS-mechanism as discussed in section \ref{sec:anomlies+cancellation}.

We will construct the singular fourfold leading to $SU(5)\times U(1)$ 
gauge symmetry by starting with a singular fourfold with only an $SU(5)$ 
singularity. Then we apply the method of the $U(1)$-restricted Tate model 
\cite{Grimm:2010ez} to generate a new section of the elliptic fibration of 
the fourfold. This means that the complex structure of the 
given fourfold hast to be restricted even further so that not
only an $SU(5)$ is generated, but also the $I_1$ locus self-intersects in 
a curve. This additional singularity results in a new section and a 
corresponding $U(1)$ gauge symmetry after resolution. 
For this gauge group, four-dimensional matter can arise in representations 
$\mathbf{5}_q$, $\mathbf{10}_{q'}$ and their complex conjugates. In 
general there can be matter in the same $SU(5)$ representation but with 
different $U(1)$-charge.

In this section we will consider two concrete fourfolds, that do not 
differ by their fiber singularities but the choice of base $B_3$. The 
first example will be the fourfold of the last section \ref{sec:X4withSU5} 
with base $B_3=\mathbf{P}^2\times \mathbf{P}^1$ on which
we apply the method of the $U(1)$-restricted Tate model. In this case we 
will demonstrate the structure of the the $G_4$-flux \eqref{eq:generalG4}, 
i.e.~the split into a non-Abelian and Abelian flux, explicitly. As 
a second example we consider the elliptically fibered fourfold
with $SU(5)\times U(1)$ singularities over the base 
$B_3=\text{Bl}_x(\mathbf{P}^3)$, the blow-up of $\mathbf{P}^3$ at a 
generic point $x$. 

\subsubsection{An $SU(5)\times U(1)$ singularity over $B_3=\mathbf{P}^2\times \mathbf{P}^1$}
\label{sec:SU5U1overP2xP1}

The first concrete example we consider is the elliptically fibered 
Calabi-Yau fourfold $X_4$ with base $B_3=\mathbf{P}^2\times \mathbf{P}^1$ 
considered in the last section \ref{sec:X4withSU5}. 

In addition 
to the $SU(5)$ singularity we generate by a further specialization of the 
complex structure a self-intersection of the $I_1$-locus. 
As before we study the geometry $\hat{X}_4$ obtained after the toric 
blow-up of all singularities by blowing up the toric ambient space 
$\mathcal{W}$. Because of the extra singularity, we have to perform five 
blow-ups that yield five new, exceptional divisors, the four
Cartan divisors $D_i$ of the $SU(5)$  and a single new exceptional divisor 
$X$ from resolving the curve of self-intersection of the $I_1$ locus.
The Calabi-Yau hypersurface in this new toric space is again a smooth 
Calabi-Yau fourfold $\hat{X}_4$. The toric data of the blown-up ambient 
space is that of \eqref{P1P2} augmented by a single new point $[-1\ -1\ 0\ 0\ 0\vert X]$, where as before the first column denotes the 
vertex and the second the associated divisor $X$. The basis of divisors on $\hat{X}_4$ is that
of the last section with the one new divisor $X$,
\beq \label{eq:divsOnSU5xU1P2xP1}
	B\,,\quad D_\alpha=(H,S)\,,\quad D_i=(D_1,D_2,D_3,D_4)\,,\quad X\,.
\eeq
Here, the divisors are as before the zero section of the elliptic 
fibration $B$, the vertical divisors $D_\alpha$ and the four Cartan 
divisors $D_i$ of $SU(5)$. The new toric divisor $X$ introduces a new 
section of the elliptic fibration of $\hat{X}_4$. We confirm this by 
checking in a toric calculation the  necessary condition
\eqref{eq:defPropSections} for a section, namely 
\beq
	X^2\cdot \mathcal{C}^\alpha=-[c_1(B_3)]\cdot X\cdot 
	\mathcal{C}^\alpha\,.
\eeq
We have evaluated this in a basis of vertical four-cycles  
$\mathcal{C}^{\alpha}=(S\cdot H, H^2)$ on $\hat{X}_4$. From the section 
$X$ we then construct a divisor of type 
$D_m$ by applying the Shioda map \eqref{eq:ShiodaMap} to $X$. This divisor 
thus induces a $U(1)$-gauge symmetry in the effective theory by the 
reduction of $C_3$ in \eqref{eq:C3expansion} along it. We note that the 
Shioda map agrees with the construction of a divisor of type $D_m$ used in 
\cite{Grimm:2010ez}. 

It is instructive to perform the Shioda map for $X\equiv \hat{\sigma}_m$ 
in two steps. We first introduce a divisor \cite{Grimm:2010ez}
\beq \label{eq:defD_5}
	D_5=X-B-c_1(B_3)\,,
\eeq 
which takes into account only the first three terms in 
\eqref{eq:ShiodaMap}. 
Here we have used that the intersection $X\cdot B=0$ on $\hat{X}_4$, as 
confirmed in an explicit toric calculation, which 
immediately implies $X\cdot B\cdot \mathcal{C}^\alpha=0$. 
Then, we obtain a divisor of type $D_m$, denoted by
$B_X$ for reasons discussed in the next paragraph, as 
\beq \label{B_X_example2}
	B_X=-2 D_1-4 D_2-6 D_3-3 D_4-5 D_5=-4C^{5i}_{SO(10)} D_i\,.
\eeq  
This is the fourth term of the Shioda map, where we have changed the 
normalization so that $B_X\equiv -5D_m$ in \eqref{eq:ShiodaMap} for later 
convenience. Indeed we take again the above basis 
of vertical four-cycles $\mathcal{C}^{\alpha}$ and of vertical 
divisors $D_\alpha$ of \eqref{eq:divsOnSU5xU1P2xP1} to calculate the 
following intersection matrix \eqref{eq:intmatrix} and the intersections 
with $X\cdot D_i$, 
\beq \label{eq:intmatrix_example2}
	\eta_\alpha^{\,\ \beta}=\delta_{\alpha}^\beta\,,\qquad X\cdot D_i\cdot \mathcal{C}^{\alpha}=\delta_{3i}\delta^\alpha_2\,.
\eeq
Thus since only $X\cdot D_3\cdot H^2=1$ is non-vanishing we choose 
$\alpha=2$ when evaluating \eqref{eq:ShiodaMap} and note that since 
$S_{(1)}=S$ in \eqref{eq:def-Calpha} we obtain 
$(\delta_{(1)}^\beta\eta_{\beta}^{\,\ 2})=\delta_{(1)}^2=1$. 
Then we evaluate the inverse $C_{(1)}^{-1}$ of the Cartan matrix $C_{(1)}$ 
of $SU(5)$ of which we need according to the second equality in 
\eqref{eq:intmatrix_example2} the third line $(C_{(1)}^{-1})^{3i}$. We 
finally obtain
\beq
	D_m=D_5+\frac{1}{5}(2D_1+4D_2+6D_3+3D_4)\,,
\eeq	
which clearly agrees with \eqref{B_X_example2} up to a factor of $-5$.

We note that the coefficients of $B_X$ in \eqref{B_X_example2} also agree
with the entries $(C^{-1}_{SO(10)})^{5i}$ of the inverse of the Cartan matrix 
$C_{SO(10)}$ of $SO(10)$. This is pointing to an underlying $SO(10)$ 
structure hidden in the construction of the divisor $B_X$ via the Shioda map.
This has been noted earlier, but differently motivated and without reference 
to the Shioda map, in \cite{Grimm:2010ez} as we outline next. 
First, one notes that the divisor $D_5$ defined in \eqref{eq:defD_5} intersects 
with the Cartan divisors $D_i$, $i=1,2,3,4$, of the $SU(5)$ singularity
almost like the Cartan divisors of an underlying $SO(10)$ 
singularity. Indeed, one obtains in a concrete toric calculation the 
intersections of the $D_i$ of $SU(5)$ and $D_5$ as
\beq \label{eq:almostSO10ints}
	D_i\cdot D_j\cdot D_\alpha\cdot D_\beta= -(\tilde{C}_{SO(10)})_{ij}S\cdot B\cdot D_\alpha\cdot D_\beta\,,\quad 
	\tilde{C}_{SO(10)}=\begin{pmatrix}
		2 &-1&0&0&0\\
		-1 &2& -1& 0&0\\
		0&-1&2&-1&-1\\
		0&0&-1&2&0\\
		0&0&-1&0&*
	\end{pmatrix}\,.
\eeq
Here $i,j=1,\ldots,5$ and $\tilde{C}_{SO(10)}$ agrees with the Cartan matrix 
of $SO(10)$ up to the entry $(\tilde{C}_{SO(10)})_{55}$. This entry is the 
intersection of $B_5^2$ with the product of two vertical divisors 
$D_\alpha\cdot D_\beta$, which is not $-2$, as required for an $SO(10)$, but 
actually depends on the choice of $D_\alpha\cdot D_\beta$. Geometrically, 
this fact is, however, not surprising since there is only an $SU(5)$
singularity over $S$ and the enhancement to $SO(10)$ happens only over 
co-dimension two on the $\mathbf{10}$ matter curve. 
We see that the construction of the divisor $B_X$ in \eqref{B_X_example2} 
reflects this underlying $SO(10)$ structure up to the mentioned mismatch. 
In this interpretation $B_X$ agrees with the $U(1)$-factor $U(1)_X$ in 
the group theoretical breaking $SO(10)\rightarrow SU(5)\times U(1)_X$.
This analogy between geometry and group theory has been the motivation for 
the notation $B_X$ in the literature \cite{Grimm:2010ez,Krause:2011xj}.

Finally we double-check in an explicit toric calculation that $B_X$ is 
indeed of type $D_m$, i.e.~that it obeys the intersection properties 
\eqref{eq:intsD_m}. The first condition is easily proven in general by 
noting that the construction of $B_X$ agrees precisely with conventional 
group theory embedding of $U(1)_X$ into $SO(10)$ as noted before. 
Thus, it also follows directly from group theory arguments that 
$B_X$ commutes with the Cartan divisors $D_i$, $i=1,\ldots, 4$, of $SU(5)$ 
which in geometrical terms directly implies the first condition in 
\eqref{eq:intsD_m}. In addition, we confirm this expectation
from group theory by an explicit toric computation.
Next we have to find the divisor $S^{(1)}$ of \eqref{eq:intsD_m}, namely 
the vertical divisor supporting the seven brane inducing the 
$U(1)$-symmetry in F-theory. In the Calabi-Yau phase specified later in 
this section we explicitly find the solutions $b^\alpha_{1}$ to 
\eqref{eq:determingb_m} that determine $S^{(1)}$ completely to
\beq \label{D_U(1)expansion}
	S^{(1)}=150 H+70 S\,,
\eeq
where we expanded in the basis of divisors \eqref{eq:divsOnSU5xU1P2xP1} on 
$B_3$. Thus we have confirmed explicitly that the divisor $B_X$ is indeed 
of type $D_m$ fulfilling the intersection properties \eqref{eq:intsD_m}. 
In addition we checked that the intersections \eqref{eq:vanishingInts_0} 
and \eqref{eq:vanishingInts_1} vanish as expected.

In order to determine the occurring matter representations we have to choose a 
Calabi-Yau phase and to analyze its Mori cone.
The blown-up toric ambient has 108 different star-
triangulations that are grouped into six inequivalent Calabi-Yau phases. The 
Calabi-Yau phase we consider is directly related to the Calabi-Yau phase 
chosen in section \ref{sec:X4withSU5} by the geometric transition induced by 
blowing down $X$ and deforming away the $I_1$ self-intersection.
We specify it by its charge vectors, i.e.~the Mori cone, that we construct by 
intersecting the Mori cones of the ambient toric variety corresponding to the 
same Calabi-Yau phase. It reads
\begin{equation}
\label{eq:MoriConeSU5U1P2xP1}
\begin{array}{|r|rrr|rrrrr|}
\hline
&\multicolumn{8}{c|}{\text{Mori cone generators}}\\
\hline
\hline
	&\phantom{\int^X}\hspace{-1Em}\ell^1 &\ell^2 &\ell^3 &\ell^4 &\ell^5 &\ell^6 &\ell^7 &\ell^8\\ \hline
     \phantom{\int^X}\hspace{-1Em}\tilde{X} & 0 & 0 & 0 & 1 & 0 & 0 & 0 & 0\\
     Y & 0 & 0 & 0 & 1 &-1 & 1 & 0 & 0\\
     B &-3 & 1 &-2 & 0 & 0 & 0 & 0 & 0\\
\hat{S} & 0 &-2 & 1 & 0 & 0 & 0 & 1 & 0\\
     S & 0 & 0 & 1 & 0 & 0 & 0 & 0 & 0\\
     H & 1 & 0 & 0 & 0 & 0 & 0 & 0 & 0\\ 
     H & 1 & 0 & 0 & 0 & 0 & 0 & 0 & 0\\
     H & 1 & 0 & 0 & 0 & 0 & 0 & 0 & 0\\
     \hline
   D_1 & 0 & 1 & 0 & 0 & 0 & 0 &-2 & 1\\
   D_2 & 0 & 0 & 0 & 0 & 1 &-1 & 1 &-1\\
   D_3 & 0 & 0 & 0 & 0 &-1 & 0 & 0 & 1\\
   D_4 & 0 & 1 & 0 & 0 & 0 & 1 & 0 &-1\\
     X & 0 & 0 & 0 &-1 & 1 & 0 & 0 & 0\\
     \hline
\end{array}
\end{equation}

Here, we noted the corresponding divisor classes in the first row. The three vectors $\ell^1$, $\ell^2$ and $\ell^3$ correspond to the two curves in the
base $\mathbf{P}^2\times \mathbf{P}^1$ and the class of the elliptic fiber. The five vectors
$\ell^4$ to $\ell^8$ encode the rational curves originating from the resolution process. They
shrink to zero in the blow-down $\hat{X}_4\rightarrow X_4$ and constitute the generators 
of the relative Mori cone \cite{Intriligator:1997pq,Grimm:2011fx}, the 
complement of the Mori cone of $X_4$ in $\hat{X}_4$. Physically we can 
wrap M2-branes over these curves that correspond to the W-bosons, i.e.~the 
roots, of $SU(5)$. The roots further split into weights of matter 
representations corresponding to new $\mathbf{P}^1$'s 
isolated over higher co-dimension loci in the base $B_3$. Thus, these 
exceptional $\mathbf{P}^1$'s naturally take values in the weight lattice 
of $SU(5)$, with which we identify the relative Mori cone 
\cite{Grimm:2011fx}.

The matter representations can be read off from the Mori cone generators 
\eqref{eq:MoriConeSU5U1P2xP1}. Performing the basis transformation from $X$ to $B_X$
and focusing on the relative Mori cone and the last five entries of the charge vectors
$\ell^{i}$, $i=4,\ldots,8$, we obtain the vectors
\beq \label{eq:relMCSU5xU1P2xP1}
	\begin{array}{|r|rrrrr|}
\hline
	&\phantom{\int^X}\hspace{-1Em}\ell^4 &\ell^5 &\ell^6 &\ell^7 &\ell^8\\
	\hline
		D_1&-2 & 0 & 0 & 0 & 1\\
		D_2&1 & 0 & 1 &-1 &-1\\
		D_3&0 & 0 &-1 & 0 & 1\\
		D_4&0 & 0 & 0 & 1 &-1\\
		B_X&0 & 5 &-3 & 1 &-1\\
		\hline
	\end{array}
\eeq
The rows are now precisely the $U(1)$-charges of an M2-brane wrapping the respective curves
corresponding to $\ell^i$ with respect to the Cartan divisors $D_i$, that are identified with the 
negative of the simple roots $-\alpha_i$, and $B_X$. The columns are then 
just the Dynkin labels of the $SU(5)$ representation with $U(1)_X$-charge given by the last entry. 
We refer to appendix \ref{app:SU5reps} for the Dynkin labels the $\mathbf{5}$ and $\mathbf{10}$ 
representations.
We immediately identify the Dynkin labels of the following representations with the charge vectors,
\beq \label{eq:repsSU5xU1P2xP1}
	\sum_i\ell^i\subset \overline{\mathbf{5}}_{2}\,,\quad\ell^5\subset \mathbf{1}_{5}\,,\quad \ell^6\subset \overline{\mathbf{5}}_{-3}\,,\quad \ell^7\subset \mathbf{10}_{1}\quad \ell^8\subset \overline{\mathbf{10}}_{-1}\,.
\eeq
The other weights of the respective representations, with some 
of them lying in the Mori cone, can be constructed by adding the simple roots $\alpha_i$.

Having determined the spectrum \eqref{eq:repsSU5xU1P2xP1} of 
six-dimensional matter we have to construct appropriate $G_4$-flux that 
potentially induces 4D chiral matter. We start with a general vertical
flux of the general form \eqref{eq:verticalFluxGeneral} and choose the 
flux quanta $m^{AB}$ such that \eqref{eq:vanishingThetas} are obeyed. In 
fact, for the Calabi-Yau fourfold at hand, there are only eleven 
independent surfaces $S$, i.e.~only eleven independent 
combinations $\omega_A\wedge\omega_B$ and as many flux parameters 
$m^{AB}$. We choose the basis of surfaces given by
\beq
	D_3^2,\ D_2\cdot D_3,\ D_2^2,\ D_2\cdot D_4,\ D_4^2,\ D_3\cdot H,\ H^2,\ \hat{S} \cdot H ,\ B\cdot H,
	\ B_X\cdot H,\ B_X\cdot S\,.
\eeq
In addition to the nine 
independent surfaces on the fourfold of the last section 
\ref{sec:X4withSU5} with only an $SU(5)$-singularity 
we have found two new surfaces $B_X\cdot H,\, B_X\cdot S$. 
The allowed $G_4$-flux then reads
\beq \label{eq:GfluxSU5xU1P2xP1}
	G_4=G^{\text{nA}}_4(\lambda)+(\alpha H+\beta S)\cdot B_X\,,
\eeq 
where as before we suppressed the use of Poincar\'e duality. The flux 
$G^{\text{nA}}_4(\lambda)$ denotes the one-
parameter non-Abelian flux \eqref{eq:GfluxSU5} already present on the Calabi-Yau fourfold before the formation 
of an $U(1)$-singularity and $\alpha$, $\beta$ denote two quantized flux numbers. 
We note that \eqref{eq:GfluxSU5xU1P2xP1} is precisely of the form 
\eqref{eq:generalG4} with one non-Abelian flux part $G_4(\lambda)$ induced 
from the fourfold without the $U(1)$-symmetry and 
thus independent of the divisor $B_X$ and a second, Abelian flux of the 
form $F\wedge[B_X]$ with $F$ denoting a general two-form flux on the base 
$B_3$. Fluxes of this type have been considered previously
in \cite{Braun:2011zm,Marsano:2011hv,Krause:2011xj,Krause:2012yh}.

Next we calculate the 4D chiralities using M-/F-theory duality for the CS-terms 
\eqref{eq:matching1loop+classical}. We calculate both the loop-induced CS-terms 
$\Theta_{\Lambda\Gamma}^{\text{loop}}$ and the classical 
flux integrals $\Theta_{\Lambda\Gamma}$. For the former we only have to determine the sign-function in \eqref{eq:theta_LambdaSigma}, since we have already identified the charges in
\eqref{eq:repsSU5xU1P2xP1}.
As explained in section \ref{sec:ChiralityInF} this is achieved by just testing whether a curve associated
to a given charge $\underline{q}$ lies in the Mori cone of $\hat{X}_4$. In this case we call
$\underline{q}$ positive, otherwise negative. This allows us to readily evaluate the loop-induced
CS-levels, suppressing the superscript for brevity, as
\beq
\small
\begin{array}{lll}
	\Theta_{11}=
	-2\Theta_{12}=-2\Theta_{23}=-\Theta_{44}=\chi(\bar{\mathbf{5}}_{-3}) +\chi(\bar{\mathbf{5}}_{2}) 
	+\chi(\overline{\mathbf{10}}_{-1})\,,&&\\[0.3Em] \Theta_{24}= \Theta_{25}=\Theta_{33}=
	\frac{1}{2}\Theta_{45}=\chi(\overline{\mathbf{10}}_{-1})\,,&\hspace{-3.7cm}\Theta_{22}=\chi(\bar{\mathbf{5}}_{-3}) 
	+ \chi(\bar{\mathbf{5}}_{2}) \,,&\hspace{-1.8cm}\Theta_{35}=3\chi(\bar{\mathbf{5}}_{-3}) 
	- 2\chi(\bar{\mathbf{5}}_{2})\,,\\[0.3Em]
	\Theta_{34}=\frac12(\chi(\bar{\mathbf{5}}_{-3}) + \chi(\bar{\mathbf{5}}_{2}) 
	- \chi(\overline{\mathbf{10}}_{-1}))\,,&\hspace{-3.7cm}\Theta_{55}=\frac92\chi(\bar{\mathbf{5}}_{-3}) + 2\chi(\bar{\mathbf{5}}_{2}) + \frac{25}{2}\chi(\mathbf{1}_{5})\, &
\end{array}
\eeq
with all other CS-levels vanishing and invoking symmetry of the matrix $\Theta_{\Lambda\Gamma}$. As expected the CS-levels are functions only of the 4D chiralities $\chi(\mathbf{R})$. 
Similarly, we evaluate the classical CS-levels \eqref{eq:theta_AB} to obtain
 \beq
 \small
	\Theta_{\Lambda\Sigma}=\begin{pmatrix}
	 0 &  0 & 0 & 0 & 0\\
     0 &-162\lambda + 3\alpha & 0 & 162\lambda - 3\alpha & 162\lambda - 3\alpha\\
     0 & 0 &162\lambda - 3\alpha & -162\lambda + 3\alpha&-486\lambda - 141\alpha\\
     0 & 162\lambda - 3\alpha & -162\lambda + 3\alpha & 0 & 324\lambda - 6\alpha\\
     0 & 162\lambda - 3\alpha & -486\lambda - 141\alpha & 324\lambda - 6\alpha& -729\lambda + 5751\alpha +6750\beta
	\end{pmatrix}\,
\eeq
depending on the three flux quanta. Equating the two expressions according 
to \eqref{eq:matching1loop+classical} we solve for the 4D chiralities as
\bea \label{eq:chiralitiesSU5xU1P2xP1}
	\chi(\bar{\mathbf{5}}_{-3})&=&162\lambda + 27\alpha\,,\quad \chi(\bar{\mathbf{5}}_{2})=-30\alpha\,, \nn\\ \chi(\overline{\mathbf{10}}_{-1})&=&-162\lambda + 3\alpha,\, \quad\chi(\mathbf{1}_{5})=-465\alpha - 540\beta
\eea

We are now prepared to determine the anomalies for the spectrum 
\eqref{eq:chiralitiesSU5xU1P2xP1}. Evaluating the left hand side of \eqref{eq:Anomalies:purenonAb}-\eqref{eq:Anomalies:Abgravitational} we
obtain
\bea \label{eq:anomaliesSU5xU1P2xP1}
	\text{tr}_\mathbf{f} F^3\!&\!\!:\!\!&\!\chi(\overline{\mathbf{10}}_{-1}) + \chi(\bar{\mathbf{5}}_{2}) 
	+\chi(\bar{\mathbf{5}}_{-3})=0\,,\\
	F^m F^n F^k\!&\!\!:\!\!&\!\frac{1}
{6}(-10\chi(\overline{\mathbf{10}}_{-1})+40\chi(\bar{\mathbf{5}}_{2}) 
-135 \chi(\bar{\mathbf{5}}_{-3})+125\chi(\mathbf{1}_5))\nn\\&&\nn=-3375\lambda-10500\alpha-11250\beta\nn\\
F^m\text{tr}_{\mathbf{f}}(F^I)^2\!&\!\!:\!\!&\!
\frac12(-3\chi(\overline{\mathbf{10}}_{-1})+2\chi(\bar{\mathbf{5}}_{2}) -3 \chi(\bar{\mathbf{5}}_{-3}))
=-75\alpha\nn\\
F^m\text{tr}R^2\!&\!\!:\!\!&\!\frac{1}{48}(-10\chi(\overline{\mathbf{10}}_{-1})+10\chi(\bar{\mathbf{5}}_{2})-15\chi(\bar{\mathbf{5}}_{-3})+5\chi(\mathbf{1}_{5}) )\nn\\&&=-\frac{1}{8}(135\lambda+510\alpha+450\beta)\,,\nn
\eea
where we made use of the group theory relations \eqref{F^3-trace-relations} and \eqref{F^2trace-relations} in
appendix \ref{app:ReviewOfAnomalies} for $SU(5)$. We note that since the non-Abelian anomaly is 
automatically canceled, we can potentially obtain a non-anomalous theory by a GS-mechanism.

We will show next that the chirality inducing $G_4$-flux \eqref{eq:GfluxSU5xU1P2xP1} indeed induces the appropriate
gaugings for a working GS-mechanism. We evaluate the $G_4$-flux induced 
gaugings of the K\"ahler moduli associated to the divisors
$D_\alpha=(H,S)$ in \eqref{eq:divsOnSU5xU1P2xP1} with respect to the $U(1)$ corresponding to $B_X$ as
\beq\label{eq:gaugingsSU5xU1P2xP1}
	\Theta_{m\alpha}=(\Theta_{m 1},\Theta_{m2})=( -90\lambda - 140\alpha - 300\beta,-300\alpha)\,.
\eeq
Next we note that the canonical bundle of the base $B_3$ is given by
$K_{B_3}=\mathcal{O}(-3H-2S)$.
Recalling that the $SU(5)$-singularity is located over the 
divisor $S$ by construction and the $U(1)$-divisor $S^{(1)}$ has been determined in \eqref{D_U(1)expansion},
we obtain the contribution of the GS-terms to the anomalies on the right hand side of 
\eqref{eq:Anomalies:purenonAb}-\eqref{eq:Anomalies:Abgravitational} as
\bea
	F^m F^n F^k&:&\frac1{4}(150\,\Theta^{\phantom{\cA}}_{m 1}+70\, \Theta_{m2})=-3375\lambda-10500\alpha-11250\beta\\
F^m\text{tr}_{\mathbf{f}}(F^I)^2&:&\frac{1}{4\lambda_I} \Theta_{m2}
=-75\alpha\nn\\
F^m\text{tr}R^2&:&-\frac1{16}(-3\Theta_{m 1}-2\Theta_{m 2})=-\frac{1}{8}(135\lambda+510\alpha+450\beta)\,,\nn
\eea
where we used $\lambda_I=1$ for $SU(N)$. By comparing this with the anomalies 
\eqref{eq:anomaliesSU5xU1P2xP1}
we see that the anomaly cancellation conditions 
\eqref{eq:Anomalies:purenonAb}-\eqref{eq:Anomalies:Abgravitational} are obeyed 
and the theory is anomaly-free.

\subsubsection{An $SU(5)\times U(1)$ singularity over $B_3=\text{Bl}_x(\mathbb{P}^3)$}

In this section we will consider another F-theory compactification on an 
elliptically fibered fourfold with an $SU(5)\times U(1)$ 
gauge symmetry and with base $B_3=\text{Bl}_x(\mathbb{P}^3)$,
i.e.~the blow-up of $\mathbb{P}^3$ at a point $x$. Since this discussion 
will be very similar to the one in the last section, section 
\ref{sec:SU5U1overP2xP1}, we will be as brief as possible here. We
refer the reader to \cite{Grimm:2011fx} for a more detailed discussion of 
the geometry.

We directly start with the smooth fourfold $\hat{X}_4$ that is obtained by resolving all singularities
in the singular fourfold $X_4$ with an $SU(5)$-singularity over a divisor $S$ and a codimension two singularity from the self-intersection of the $I_1$-locus. The fourfold $\hat{X}_4$ is realized as a toric Calabi-Yau 
hypersurface in a five-dimensional toric ambient space defined by a polyhedron with the following vertices
\small
\begin{equation} \label{P3BU}
\begin{array}{|rrrrr|c|}
\hline
\multicolumn{5}{|c|}{\text{Points}}& D_A\\
\hline
\hline
   -1 &  0 & 0  & 0 & 0 &  \phantom{\int^X}\hspace{-1.5Em}\tilde{X}\\
     0 & -1 & 0 & 0 & 0 & Y\\
     3 &  2 &  0 & 0 & 0 & B\\
     3 &  2 & 1 & 1 & 1 &  H\\
     3 &  2 & -1 & 0 & 0 & H+S\\
     3 &  2 & 0 & -1 & 0 & H+S\\ 
     3  & 2 & 0 &0 & -1 & H\\
     3 &  2 & 1 & 1 &0 & \hat{S}\\
     2 &  1 & 1 & 1 & 0 & D_1\\
     1 & 1 & 1 & 1 &0 & D_4\\
     1  &  0 & 1 & 1 &0 & D_2\\
     0  & 0 & 1 & 1 & 0 & D_3\\
     -1 & -1 & 0 & 0 &0 & X\\
     \hline
\end{array}
\end{equation}
\normalsize
Here we present as before the coordinates of the vertices as the 
five-dimensional row vectors and the corresponding toric divisors. 
The polyhedron admits 348 star-triangulations that correspond to twelve 
independent Calabi-Yau phases when restricted to the anti-canonical 
divisor in the toric ambient space.

A basis of divisors with all other divisors in \eqref{P3BU} by linear equivalences reads
\beq \label{eq:divsOnSU5xU1P3BU}
	B\,,\quad D_\alpha=(H,S)\,,\quad D_i=(D_1,D_2,D_3,D_4)\,,\quad X\,
\eeq
with $B$ being the base, $D_\alpha$ the vertical divisors, the $D_i$ denoting the Cartan divisors of $SU(5)$
and $X$ the exceptional divisor generated by the blow-up in the construction of the $U(1)$-restricted Tate model. We readily check the necessary condition 
\eqref{eq:defPropSections} for $X$ being a section.
The divisor $S$ is related to the divisor $\hat{S}$ as $S=\hat{S}+\sum_i D_i$. A divisor of type $D_m$
is constructed from $X$ as before in section \ref{sec:SU5U1overP2xP1} by applying the Shioda map \eqref{eq:ShiodaMap}. Again, we first construct $D_5=X-B-\left[c_1(B_3)\right]$, that intersects
the Cartan divisors $D_i$ as in \eqref{eq:almostSO10ints} resembling an $SO(10)$ singularity, and then form
$B_X=-2D_1-4D_2-6D_3-3D_4-5B_5$. Here we choose a basis of vertical 
four-cycles given by $\mathcal{C}^\alpha=(H\cdot S,S^2)$ and evaluate in the 
basis of vertical divisors $D_\alpha$ and Cartan divisors $D_i$ in 
\eqref{eq:divsOnSU5xU1P3BU}
\beq
	\eta_{\alpha}^{\,\ \beta}=\begin{pmatrix}
		0 & 1\\
		1& -2
	\end{pmatrix}\,,\quad X\cdot D_3\cdot \mathcal{C}^1=1\,,\quad X\cdot D_3\cdot \mathcal{C}^2=-2\,,
\eeq
with all other intersections $X\cdot D_i\cdot \mathcal{C}^\alpha=0$. Thus, we can choose any $\alpha$ to evaluate \eqref{eq:ShiodaMap} and obtain  $D_m=-\frac{1}{5}B_X$.
It can then be shown as in the last section that $B_X$ obeys the intersections
properties \eqref{eq:intsD_m} and thus generates a $U(1)$-symmetry in the low-energy effective action.

In particular we use \eqref{eq:intsD_m} and \eqref{eq:determingb_m} to determine in a concrete toric calculation in
a Calabi-Yau phase specified later in this section  the 
coefficients $b^\alpha_{1}$ in \eqref{eq:determingb_m}. The corresponding divisor  $S^{(1)}$
reads
\beq \label{D_U(1)expansion_2}
	S^{(1)}=200 H+120 S\,.
 \eeq
In addition, we check explicitly that the intersection properties \eqref{eq:intsD_m} are obeyed. 

Again we determine the 4D chiral matter by first looking at the Mori cone \eqref{eq:MoriConeSU5U1P3BU}
to determine the 6D matter representations. For the Calabi-Yau phase we consider it reads
\begin{equation}
\label{eq:MoriConeSU5U1P3BU}
\begin{array}{|r|rrr|rrrrr|}
\hline
&\multicolumn{8}{c|}{\text{Mori cone generators}}\\
\hline
\hline
	&\phantom{\int^X}\hspace{-1Em}\ell^1 &\ell^2 &\ell^3 &\ell^4 &\ell^5 &\ell^6 &\ell^7 &\ell^8\\ \hline
     \phantom{\int^X}\hspace{-1Em}\tilde{X} & 0 & 0 & 0 & 0 & 0 & 0 & 0 & 1\\
     Y & 0 & 0 & 0 & 1 & 0 & 0 & -1 & 1\\
     B &-3 &-1 & 1 & 0 & 0 & 0 &0 & 0\\
     H & 0 & 1 & 0 & 0 & 0 & 0 & 0 & 0\\
    H+S& 1 & 0 & 0 & 0 & 0 & 0 & 0 & 0\\
    H+S& 1 & 0 & 0 & 0 & 0 & 0 & 0 & 0\\ 
     H & 0 & 1 & 0 & 0 & 0 & 0 & 0 & 0\\
\hat{S}& 1 &-1 &-2 & 0 & 0 & 1 & 0 & 0\\
     \hline
   D_1 & 0 & 0 & 1 & 0 & 1 &-2 & 0 & 0\\
   D_2 & 0 & 0 & 0 &-1 &-1 & 1 & 1 &0\\
   D_3 & 0 & 0 & 0 & 0 & 1 & 0 &-1 & 0\\
   D_4 & 0 & 0 & 1 & 1 &-1 & 0 & 0 &0\\
     X & 0 & 0 & 0 & 0 & 0 & 0 & 1 & -1\\
     \hline
\end{array}
\end{equation}
As one can see by analyzing this Mori cone as in 
\eqref{eq:relMCSU5xU1P2xP1} and \eqref{eq:repsSU5xU1P2xP1} 
in the last section, cf.~\cite{Grimm:2011fx} for details, the occurring 
matter representations are as before given by
\beq \label{eq:repsSU5xU1P3BU}
	\overline{\mathbf{10}}_{-1}\,,\quad \overline{\mathbf{5}}_{-3}\,,\quad\overline{\mathbf{5}}_{2}\,,\quad\mathbf{1}_{5}\, .
\eeq
Next we construct the chirality inducing vertical $G_4$-flux starting with 
the ansatz \eqref{eq:verticalFluxGeneral}. There are 15 independent 
four-forms in $H^{(2,2)}_V(\hat{X}_4)$ that are Poincar\'e dual to the 
surfaces
\beq
	H\cdot D_i\,,\ S\cdot D_i\,,\ D_4^2\,,\ D_3\cdot D_4\,,\ B\cdot \hat{S}\,,B^2\,,\ H\cdot S\,,\ B_X\cdot 
	H\,,\  B_X\cdot S\,,
\eeq
where $i$ runs from $i=1,\ldots, 4$. Accordingly, we have only 15 
independent flux numbers $m^{AB}$ that are constrained further
by the conditions \eqref{eq:vanishingThetas}. We obtain the $G_4$-flux 
meeting all these constraints as
\bea \label{eq:GfluxSU5xU1P3BU}
	G_4&=&\lambda[ c_1(B_3)\cdot(D_1+2D_2-2D_3-6D_4)+5 D_4\cdot (D_3+2D_4)]+(\alpha H+\beta S)\cdot B_X\nn\\
	&\equiv & G_4^{\text{nA}}(\lambda)+(\alpha H+\beta S)\cdot B_X
\eea
where we implicitly apply Poincar\'e duality. Again we identified
a non-Abelian flux $G_4^{\text{nA}}(\lambda)$ depending on a single flux 
parameter $\lambda$ and two Abelian fluxes with flux numbers $\alpha$, 
$\beta$ in accordance with the general structure \eqref{eq:generalG4}.

As a next step, we determine the 4D chiralities by the matching 
\eqref{eq:matching1loop+classical} of the CS-levels in 3D. We evaluate 
first the one-loop CS-terms \eqref{eq:theta_LambdaSigma} yielding
\beq 
\small
\begin{array}{lll}
\Theta_{11}=-2\Theta_{12}=-2\Theta_{23}=-\Theta_{44}=\chi(\bar{\mathbf{5}}_{-3}) + \chi(\bar{\mathbf{5}}_{2})  +\chi(\overline{\mathbf{10}}_{-1})\,,&
\\[0.3Em] \Theta_{24}=\Theta_{25}=\Theta_{33}=\frac12\Theta_{45}=\chi(\overline{\mathbf{10}}_{-1}) \,, & \hspace{-3.7cm}\Theta_{22}=\chi(\bar{\mathbf{5}}_{-3}) + \chi(\bar{\mathbf{5}}_{2})\,,& \hspace{-2cm}\Theta_{35}=3\chi(\bar{\mathbf{5}}_{-3}) - 2\chi(\bar{\mathbf{5}}_{2})\,, \\[0.3Em]
\Theta_{34}=\frac12\chi(\bar{\mathbf{5}}_{-3}) + \frac12\chi(\bar{\mathbf{5}}_{2})  - \frac12\chi(\overline{\mathbf{10}}_{-1}) \,,& \hspace{-3.7cm}\Theta_{55}=\frac{9}{2}\chi(\bar{\mathbf{5}}_{-3}) + 2\chi(\bar{\mathbf{5}}_{2})  + \frac{25}{2}\chi(\mathbf{1}_5)  \,, &
\end{array}
\eeq
where we suppressed the superscript for brevity.
All other CS-levels vanish and the remaining ones are determined by symmetry of the matrix 
$\Theta^{\text{loop}}_{\Lambda\Gamma}$. As expected the CS-levels are functions only of the 4D chiralities $\chi(\mathbf{R})$. 
Then, we evaluate the classical CS-levels \eqref{eq:theta_AB} to obtain
 \beq
 \small
	\Theta_{\Lambda\Sigma}=\begin{pmatrix}
	 0 &  0& 0 &0 &  0\\
	 0&92\lambda+ 3\alpha - 2\beta &0&-92\lambda - 3\alpha + 2\beta&-92\lambda - 3\alpha + 2\beta\\
    0 &0 &-92\lambda - 3\alpha + 2\beta2 &92\lambda + 3\alpha - 2\beta&256\lambda- 111\alpha + 34\beta\\
    0 &-92\lambda - 3\alpha + 2\beta&92\lambda + 3\alpha - 2\beta &0&-184\lambda - 6\alpha + 4\beta\\
0 & -92\lambda - 3\alpha + 2\beta&256\lambda - 111\alpha + 34\beta&-184\lambda - 6\alpha + 4\beta& 404\lambda + 3891\alpha + 5886\beta
	\end{pmatrix}\,
\eeq
that are functions of the three flux quanta. Identifying the two 
expressions as in \eqref{eq:matching1loop+classical} we determine the 4D 
chiralities as
\bea \label{eq:chiralitiesSU5xU1P3BU}
	\chi(\bar{\mathbf{5}}_{-3})&=&-88\lambda + 21\alpha - 6\beta\,,\quad \chi(\bar{\mathbf{5}}_{2})=-4\lambda - 24\alpha + 8\beta\,, \nn\\ \chi(\overline{\mathbf{10}}_{-1})&=&92\lambda + 3\alpha- 2\beta ,\, \quad\chi(\mathbf{1}_{5})=-315\alpha - 470\beta
\eea

We can now evaluate the anomalies of the F-theory compactification
under consideration. Inserting the chiralities 
\eqref{eq:chiralitiesSU5xU1P3BU} into the left hand side of the anomaly 
constraints 
\eqref{eq:Anomalies:purenonAb}-\eqref{eq:Anomalies:Abgravitational} we 
obtain, using again the group theory relation of appendix 
\ref{app:ReviewOfAnomalies},
\bea	\label{eq:anomaliesSU5xU1P3BU}
	\text{tr}_\mathbf{f} F^3\!&\!\!:\!\!&\!\chi(\overline{\mathbf{10}}_{-1}) + \chi(\bar{\mathbf{5}}_{2}) 
	+ \chi(\bar{\mathbf{5}}_{-3})=0\,,\\
	F^m F^n F^k\!&\!\!:\!\!&\!\frac{1}
{6}(-10\chi(\overline{\mathbf{10}}_{-1})+40\chi(\bar{\mathbf{5}}_{2}) -135 \chi(\bar{\mathbf{5}}_{-3})+125\chi(\mathbf{1}_5))=1800\lambda-7200\alpha-9600\beta\nn\\
F^m\text{tr}_{\mathbf{f}}(F^I)^2\!&\!\!:\!\!&\!\frac12(-3\chi(\overline{\mathbf{10}}_{-1})+2\chi(\bar{\mathbf{5}}_{2}) -3 \chi(\bar{\mathbf{5}}_{-3}))
=-10\lambda-60\alpha+20\beta\nn\\
F^m\text{tr}R^2\!&\!\!:\!\!&\!\frac{1}
{48}(-10\chi(\overline{\mathbf{10}}_{-1})+10\chi(\bar{\mathbf{5}}_{2})-15\chi(\bar{\mathbf{5}}_{-3})+5\chi(\mathbf{1}_5) )=\frac{1}{2}(15\lambda-90\alpha-90\beta)\,,\nn
\eea
The anomaly is obviously completely factorized and thus 
of the appropriate form to be canceled by the GS-counter terms 
\eqref{eq:GSterm}.

Next we will show that the chirality-inducing $G_4$-flux induces the appropriate gaugings of 4D axion so that
the GS-mechanism works to cancel the anomalies.
We obtain the gaugings induced by the $G_4$-flux \eqref{eq:GfluxSU5xU1P3BU} explicitly as
\beq \label{eq:gaugingsSU5xU1P3BU}
	\Theta_{m\alpha }=(\Theta_{m1},\Theta_{m2})=(60\lambda - 240\beta,-40\lambda - 240\alpha + 80\beta)\,
\eeq
in the basis $D_\alpha=(H,S)$. The determination of the GS-terms on the right hand side of the anomaly condition is then completed
by noting that the $SU(5)$-singularity is supported over $S$, the $U(1)$-divisor has the expansion \eqref{D_U(1)expansion_2} 
and the canonical bundle of the base $B_3$  reads $K_{B_3}=\mathcal{O}(-4H-3S)$. Thus, we obtain using $\lambda_A=1$ for $SU(N)$
\bea
	F^m F^n F^k&:&\frac1{4}(200\,\Theta_{m 1}+120\, \Theta_{m2})\\
	&& =
	1800\lambda-7200\alpha-9600\beta\\
F^m\text{tr}_{\mathbf{f}}(F^I)^2&:&\frac{1}{4\lambda_I} \Theta_{m2}
=-10\lambda - 60\alpha + 20\beta\nn\\
F^m\text{tr}R^2&:&-\frac1{16}(-4\Theta_{m 1}-3\Theta_{m 2})=\frac{1}{2}(15\lambda-90\alpha-90\beta)\,.\nn
\eea
Comparing this to \eqref{eq:anomaliesSU5xU1P3BU} we immediately see that by the GS-mechanism all
anomalies are canceled and we obtain an anomaly-free effective theory.

\section{Conclusions}

In this work we have addressed anomaly cancellation in four-dimensional 
$\cN=1$ F-theory compactifications via their three-dimensional dual 
$\cN=2$ M-theory reductions. The M-theory compactification was performed 
on a resolved Calabi-Yau fourfold and included a non-trivial $G_4$-flux 
background. To match the M-theory configuration the four-dimensional 
F-theory setup was dimensionally reduced on an extra circle. The 
three-dimensional effective theories have been compared in the Coulomb 
branch and we argued that it is crucial to properly integrate out massive 
charged matter and excited Kaluza-Klein modes in the circle 
compactification. We focused on globally consistent chiral 
compactifications with both non-Abelian and Abelian $U(1)$ gauge 
symmetries due to space-filling seven-branes.

The investigation of some key geometric properties of divisors in 
resolved Calabi-Yau fourfolds was our first focus. We gave a detailed 
account of the intersection structure and the appropriate identifications 
to lift the 3D M-theory effective action to a 4D F-theory setup. This 
included an extended introduction to the geometry of seven-branes with 
$U(1)$ gauge symmetries, for example, introducing the Shioda map. 
Also the geometric definition of the $U(1)$ vectors $A^0,A^\alpha$, 
corresponding to the Kaluza-Klein vector and the duals of the K\"ahler 
moduli axions in the 4D to 3D reduction, turned out to be crucial
in the discussion of 4D anomalies from a 3D point of view.

The intriguing interplay of geometry and couplings in the effective theory 
becomes particularly apparent when aiming to demonstrate that 
F-theory compactifications automatically cancel all 4D gauge and 
mixed gauge-gravitational anomalies. Two ingredients are necessary: 
(1) The F-theory compactification has to be defined via a singular 
Calabi-Yau fourfold that admits a Calabi-Yau resolution of, at least, 
all gauge group and matter singularities. (2) A $G_4$-flux has to be 
defined on the resolved Calabi-Yau fourfold in such a way that it can be 
lifted to a 4D F-theory compactification, i.e.~it has to obey the 
vanishing conditions \eqref{eq:vanishingThetas}. 

To explicitly study anomaly cancellation in F-theory we had to 
explore the map between M-theory and F-theory and the corresponding 
3D and circle compactified 4D effective actions. 
We argued that it suffices to concentrate on deriving  
the coefficients of the 3D Chern-Simons terms and the coefficients of 
the four-dimensional topological terms $F\wedge F$ and $R\wedge R$.
The 3D Chern-Simons terms in the M-theory reduction are induced by 
$G_4$-flux. On the circle compactified F-theory side, they admit two
very different origins. They arise either from classical 4D gaugings, 
or are induced at one loop by integrating out matter fields 
that are massive in the 3D Coulomb branch or that arise as Kaluza-Klein 
modes.

We particularly focused on Chern-Simons terms involving the 
Kaluza-Klein vector $A^0$  of the 4D metric. While this term has been 
previously neglected we showed that it can be generated at one loop by 
integrating out Kaluza-Klein modes of 4D chiral matter. If this matter is 
charged under an $U(1)$ vector $A^m$ descending from four dimensions a 
one-loop Chern-Simons term $\Theta_{0m}\, A^0 \wedge F^m$ is induced with 
$\Theta_{0m}$ precisely measuring the 4D mixed Abelian-gravitational 
anomaly. On the M-theory side, geometry predicts that this coefficient is 
equal to the $U(1)$ gaugings contracted with the coefficient of the 
topological term $R\wedge R$ given by the first Chern class of the base 
$B_3$. This is precisely the matching enforced by the generalized 
Green-Schwarz mechanism. Hence, this term can be either used to justify 
anomaly cancellation for an F-theory setup with 
a given spectrum or used to infer information about an unknown spectrum 
after specifying the $G_4$-flux. 

The mixed Abelian-gravitational anomaly allows for the most direct match 
of the 3D and 4D data. The study of gauge anomalies is complicated by 
the fact that the chiral index, counting the net number of matter 
fields in a given representation, has to be extracted from the one-loop 
result of the 3D Chern-Simons terms. We demonstrated that this can be done 
for a number of explicit examples. Indeed, in these geometrically realized 
setups one can check complete anomaly cancellation for each $G_4$-flux 
lifting to F-theory. Our examples included gauge groups $SU(5)$ and 
$SU(5)\times U(1)$, all relevant for GUT model building. Needless to say 
our strategy is not fixed to specific examples and anomaly 
cancellation will be guaranteed generally. An inversion of the 
one-loop result for the Coulomb branch Chern-Simons couplings in terms of 
the chiral indices would shed additional light on the mechanisms ensuring
anomaly-freedom in F-theory.

Turning the story around, one can also use anomaly cancellation to derive 
a set of very non-trivial conditions on the geometry of resolved 
Calabi-Yau fourfolds. These have been given in 
\eqref{eq:anomalyCondHomology} and display conditions on the 
intersections of the resolution spheres supporting M2-branes in the 
M-theory Coulomb branch. We have shown that these equations are satisfied 
for our examples, but a direct general geometrical proof would be 
desirable. Further support from a physical point of view for automatic 
anomaly cancellation in F-theory can also be given in the weak string 
coupling limit. We have recalled that in this Type IIB limit the 
cancellation of D7-brane and induced D5-brane tadpoles suffices to 
guarantee the absence of 4D anomalies. Absence of anomalies in F-theory 
is then simply a consequence of the fact that the construction of the 
elliptically fibered Calabi-Yau fourfold ensures the cancellation of 
seven-brane tadpoles, while the choice of a harmonic $G_4$ flux guarantees 
the absence of 5-brane tadpoles.

In our analysis of the weakly coupled setup we also noted that general 
D7-brane stacks can support geometrically massive $U(1)$ gauge symmetries. 
In the F-theory effective action derived via M-theory these $U(1)$ gauge 
bosons and corresponding St\"uckelberg axions are already integrated
out and do not appear in the low energy spectrum. We have shown that the 
projection to the reduced low energy spectrum allows to consistently match 
Type IIB anomaly cancellation and F-theory anomaly cancellation. 
It would be interesting to include these additional
massive $U(1)$ also into the F-theory setup and extend the analysis 
accordingly.

Let us end by some additional observations and remarks. 
It is interesting to recall that 4D anomalies are only induced in the 
presence of $G_4$-flux and their cancellation employs the Green-Schwarz 
mechanism only in the presence of gauged $U(1)$ symmetries. In contrast, 
the coefficient $b^\alpha_I, b^\alpha_m$ and 
$a^\alpha$ in front of the 4D topological terms $F\wedge F$ and 
$R\wedge R$ can be extracted generally for any F-theory geometry without 
making reference to the flux $G_4$. 
Also anomalies are automatically satisfied for any $G_4$-flux that lifts 
to F-theory.  Following \cite{Grimm:2012yq} it is thus natural to 
conjecture that the geometric constraints are actually stronger than the 
4D low-energy constraints. The $G_4$ fluxes can be used to 
probe these geometric constraints and translate them into actual 4D 
anomaly cancellation conditions. It would be 
interesting to further investigate the theoretical underpinnings of these 
observations.

Let us close by emphasizing that in this work only the match of discrete 
data of topological couplings in 4D and 3D have been of importance. Indeed 
one expects that this information is most uncontrollably extracted for a 
dimensional reduction and can be traced through dualities. 
For a complete study of the effective action, however, one will also need 
to face the computation of less protected $\cN=1$ data such as the K\"ahler 
potential. While the classical matching can be found in 
\cite{Grimm:2010ks} it remains to show how corrections can be reliably 
computed in the M-theory to F-theory duality.

\subsubsection*{Acknowledgments}

We gratefully acknowledge discussions with Massimo Bianchi, Federico Bonetti, Yi-Zen Chu, Babak Haghighat, Jim Halverson, 
Stefan Hohenegger, Andreas Kapfer, Jan Keitel, Albrecht Klemm, Eran Palti, Daniel Park, Erich Poppitz, Raffaele Savelli, 
Sakura Sch\"afer-Nameki, Wati Taylor, 
and Timo Weigand. We like to especially thank Eran Palti and Sakura Sch\"afer-Nameki  for 
enlightening discussions about anomaly cancellation in F-theory including hypercharge flux.
The authors like to thanks the CERN Theory group and the Simons Center at Stony Brook
for hospitality and support. T.G.~also acknowledges hospitality and support
by the UPenn Theory group at Philadelphia. This work is supported in part 
by DOE grant DE-SC0007901, the Fay R. and Eugene L. Langberg Endowed Chair 
and the Slovenian Research Agency (ARRS).

\appendix

\section{Review of anomalies and the generalized Green-Schwarz mechanism}
\label{app:ReviewOfAnomalies}

In this appendix we discuss aspects of anomalies of local
symmetries and their cancellation via the generalized 
Green-Schwarz mechanism in four-dimensional 
effective theories coupled to gravity. The following discussion applies for 
a gauge theory with a generic semi-simple non-Abelian gauge 
group, i.e.~a product of simple gauge groups $G_{(I)}$, with any number $n_{U(1)}$ of $U(1)$-factors,
\beq
	G=G_1\times\cdots \times G_{(n_G)}\times U(1)_1\times\cdots\times U(1)_{n_{U(1)}}\,.
\eeq

In any quantum field theory in $D$ dimensions the local anomalies 
are characterized in a gauge invariant way by the 
anomaly polynomial $I_{D+2}$ of degree $D/2+1$ in the field 
strength $F_G$  of $G$ and the curvature two-form $R$ \cite{AlvarezGaume:1983ig}. 
The anomaly polynomial encodes all the information of the anomalous one-loop
diagram signalling the presence of the anomaly.
Writing the spectrum  of the four-dimensional effective theory only in terms 
left-handed Weyl fermions in a representation $\repr$ of the gauge group $G$, 
the induced four-dimensional anomaly takes the form
\beq \label{I6-anomaly}
  I_6 = \sum_{\repr} n(\repr)\, I_{1/2}(\repr)\,.
\eeq
Here $n(\repr)$ is the number of left-handed Weyl fermions in the 
representation $\repr$ and we introduce the anomaly polynomial 
$I_{1/2}(\repr)$ of a single left-handed Weyl fermion 
in a representation $\repr$  as\footnote{Here we use the sign conventions of 
the fourth reference in \cite{Anomaly-reviews} for the anomaly polynomial, however
with $\mathrm{A}=iA$ and $\mathrm{F}=iF$ in eq.~(8.25) of this reference due to a 
different convention for the covariant derivative compared to our choice in 
\eqref{eq:kineticWeyl}.}
\beq \label{I1/2}
	I_{1/2}(\repr)=\text{tr}_{\repr}(e^{F_G})\hat{A}(R)\,.
\eeq
The right hand side of this equation has to be read as a formal
polynomial in $F$ and $R$ and we denote by $\text{tr}_{\repr}$ the 
trace in the representation $\repr$.
Note that in four dimensions anomalies are induced only by 
massless Weyl-fermions. The anomalies of self-dual tensors 
and left-handed gravitinos are trivial since 
they only contribute purely gravitational anomalies that are absent in four 
dimensions by symmetry. 
We emphasize that only the chiral index 
$\chi(\repr) = n(\repr)-n(\repr^*)$ appears in 
\eqref{I6-anomaly} because a left-handed Weyl fermion in the 
complex conjugate representation $\repr^*$ is equivalent to a 
right-handed Weyl fermion in the representation $\repr$ that 
contributes with $-I_{1/2}(\repr)$ to the anomaly polynomial. This implies 
that the anomaly of real representations vanishes identically.
Expanding the general polynomial \eqref{I1/2} to third order in $F_G$ 
and $R$ we obtain
\beq \label{I12-4d}
   I_{1/2}(\repr) = \frac{1}{6} \text{tr}_{\repr} (F_G^3) + 
   \frac{1}{48} \text{tr}_{\repr} (F_G) \text{tr} R^2\,,
\eeq
where $\text{tr}$ denotes the trace in the fundamental 
representation of the four-dimensional Lorentz group SO$(3,1)$.

There can be a tree-level contribution to the anomaly polynomial due
to gauged scalars, the axions $\rho_\cA$, with an anomalous transformation
and anomalous couplings in the effective Lagrangian. This additional 
contribution can cancel the total anomaly polynomial \eqref{I6-anomaly} of Weyl-fermions  
rendering the theory anomaly-free. However, this cancellation mechanism, known as the 
generalized Green-Schwarz mechanism, is only possible if the anomaly polynomial 
$I_6$ factorizes appropriately. The requirement for this factorization is that on the one hand
the cubic non-Abelian anomaly in $tr_\repr F_G^3$ that
it is obviously non-factorizable has to vanish, as discussed below. This
puts a stringent condition on the spectrum of charged chiral matter. 
On the other hand, for the
cancellation of the factorizable terms, relations between
the anomalous couplings of the axions, their gaugings and 
the chiral indices $\chi(\repr)$ have to hold as we will derive next. 

The Green-Schwarz mechanism is a combination of two ingredients, firstly the presence 
of appropriate couplings of the axions $\rho_\cA$ and secondly gaugings of their shift symmetries.
The general form of the couplings of the axions
 that are topological and denoted as the Green-Schwarz counter terms reads in the conventions of 
\cite{Grimm:2012yq} as
\beq \label{GS-term}
     S^{(4)}_{\rm GS} = -\frac{1}{8}\int   \frac{2}{\lambda_I} b_I^\mathcal{\cA} \rho_{\mathcal{A}}  
     \text{tr}_{\textbf{f}} F^I\wedge F^I+2 b_{mn}^{\mathcal{A}} \rho_{\mathcal{A}}   F^m\wedge F^n  -  \frac{1}{2} a^\mathcal{A} \rho_\mathcal{A} \text{tr} R\wedge R \ ,
\eeq
where $\text{tr}_{\textbf{f}}$ denotes the trace in the fundamental 
representation $\textbf{f}$ of the gauge
group factor $G_{(I)}$, $F^I$ is the corresponding gauge field strength of 
the non-Abelian gauge group $G_{(I)}$, $F^m$ the 
$U(1)$-field strength and $\lambda_I=2c_{G_{(I)}}/V(\mathbf{adj})$ 
for $c_{G_{(I)}}$ the dual Coxeter number of the group factor $G_{(I)}$ 
and $V(\mathbf{adj})$ defined in 
\eqref{trF^3}.  For example we have $\lambda=1$ for $SU(N)$ and $U(1)$.
The constants $b^\mathcal{A}_I$ and $a^\mathcal{A}$ as well as the number of axions $\rho_{\mathcal{A}}$ 
depend on details of the considered effective theory. In 
an effective action from F-theory $b_I^\mathcal{A}$, $b_{ij}^A$ and $a^\mathcal{A}$ have been shown in 
\cite{Grimm:2012yq} to be determined 
by the F-theory geometry as discussed in section \ref{sec:AnomaliesInFGUTs} and the number of axions is
$h^{1,1}(B_3)+1$, where $B_3$ is the base of the F-theory elliptic fibration.

The Green-Schwarz terms can have non-trivial gauge transformations due to possible gaugings of the shift
symmetries of the axion $\rho_A$ by the $U(1)$-vectors in the theory. These gaugings are specified by the
covariant derivatives 
\beq \label{rhoGauging}
   \cD \rho_\mathcal{A} = d \rho_\mathcal{A} + \Theta_{\mathcal{A} m} 
   A^m\,,
\eeq
where the constant $\Theta_{\mathcal{A} m}$ encode the combination of 
$U(1)$-vectors $A^m$, $m=1,\ldots,n_{U(1)}$ gauging a given shift symmetry. 
We note that 
the gaugings \eqref{rhoGauging} in string theory are typically seen in the dual action where
the axions $\rho_\cA$ have been dualized into two-forms $C^\mathcal{A}$. In terms of these two-forms 
a gauging is signalled by a coupling of the form $\Theta_{\mathcal{A} m}C^\mathcal{A}\wedge F^m_{U(1)}$. 
Upon eliminating the $C^\cA$ from the action we it can be shown that the covariant derivatives 
\eqref{rhoGauging} are generated.

Thus we see that the Green-Schwarz counter terms \eqref{GS-term} in combination with the gaugings
\eqref{rhoGauging} can lead to an anomalous variation of the classical effective action.
The contribution of this classical non-gauge invariance of the Green-Schwarz terms to the 
anomaly polynomial is determined by the descend equation. First, we perform 
a gauge transformation of \eqref{GS-term} with \eqref{rhoGauging} to obtain  
the anomalous variation as
\beq \label{varGSaction}
	\delta S^{(4)}_{\rm GS}=-\frac18\int \Big(\frac{2}{\lambda_I}b_I^\mathcal{A}  
	\text{tr}_{\textbf{f}}(F^I)^2+2 b_{mn}^{\mathcal{A}}    F^m\wedge F^n  -  \frac12 a^\mathcal{A} \text{tr} R^2\Big) 
	\Theta_{\mathcal{A} k}\lambda^k\,
\eeq
assuming that the other terms in the action are gauge invariant.
According to the rules of  the descend equation we then have to replace  the gauge parameter 
$\lambda^m$ by the vector potential $A^m$ since $\delta 
A_{U(1)}^m=d\lambda^m$ and since the other terms in 
\eqref{varGSaction} are gauge invariant. Finally, we take the 
exterior derivative to obtain the 
contribution to the anomaly polynomial as
\beq \label{I6-GS}
	I_6^{\rm GS}=-\frac{1}{8}\Big(\frac{2}{\lambda_I}b_I^\mathcal{A}  
	\text{tr}_{\textbf{f}}(F^I)^2  +2 b_{mn}^{\mathcal{A}}    F^m\wedge F^n-   \frac{1}{2}a^\mathcal{A} \text{tr} R^2\Big) 
	\Theta_{\mathcal{A} k}F^k\,.
\eeq
Thus, taken the contribution $I_6^{\rm GS}$ to the anomaly polynomial into 
account all anomalies are canceled, i.e.~$I_6+I_{6}^{\text GS}=0$, 
if the anomaly polynomial of chiral matter in 
\eqref{I6-anomaly} factorizes 
like \eqref{I6-GS} with coefficients equal to 
$b^\mathcal{A}\Theta_{\mathcal{A}m}$ respectively 
$a^\mathcal{A}\Theta_{\mathcal{A}m}$.
This implies that relations between the gaugings, encoded in the 
$\Theta_{\mathcal{A} m}$, the couplings $a^\mathcal{A}$, 
$b^\mathcal{A}$ and the chiral indices
$\chi(\repr)$ have to hold.

In the following we work out in detail these 
relations for the four possible types of anomalies in four dimensions.

\subsection{Purely non-Abelian anomaly}

In order to evaluate the anomaly polynomial for purely non-Abelian 
factors, e.g.~for $SU(N)$, one rewrites all traces $\text{tr}_{\repr}$ in 
\eqref{I6-anomaly}, \eqref{I12-4d} into traces over the fundamental 
representation $\mathbf{f}$ using
\beq \label{trF^3}
     \text{tr}_{\repr} F^3 = V(\repr)  \text{tr}_{\mathbf{f}} 
     F^3\,,
\eeq
where we assume that an independent third order Casimir operator 
exists, what is the case e.g.~for $SU(N)$ with $N\geq 3$.
Since this anomaly is non-factorizable, it has to cancel by itself 
as
\beq \label{vanishingCubic}
  \sum_{\repr} n(\repr) V(\repr)= 0 \ ,
\eeq
For example, for $SU(N)$ the trace relations take the form
\bea \label{F^3-trace-relations}
   \text{tr}_{\mathbf{as}_2} F^3 &=& (N-4) \text{tr}_{\mathbf{f}} 
   F^3 \ , 
   \qquad \qquad N >2\ ,\\
   \text{tr}_{\mathbf{as}_3} F^3 &=& \frac12 (N^2 - 9N +18) 
   \text{tr}_{\mathbf{f}} F^3 \ , \qquad N >5\ . \nn
\eea
where $\mathbf{as}_2$, $\mathbf{as}_3$ is the second and third 
rank anti-symmetric tensor representation. 
In theories without $U(1)$'s the condition \eqref{vanishingCubic} is the only 
anomaly in four dimensions since by $tr_\repr F=0$ there can be 
no mixed non-Abelian-gravitational anomaly in \eqref{I12-4d}.

\subsection{Abelian and mixed anomalies}

In the presence of $U(1)$ factors in $G$ the four-dimensional 
anomaly cancellation gets slightly richer. To see this it is 
customary to split the field strength of $G$ accordingly in 
$U(1)$-factors with field strengths $F^m$ and a purely non-Abelian gauge 
field $F$ as
\beq
   F_G = F +  \sum_{m=1}^{n_{U(1)}} F^m\,.
\eeq
Upon inserting this split into the \eqref{I12-4d} we obtain
\bea
	I_{1/2}(\repr)&=&\frac{1}{6}\Big(\text{tr}_{\repA} (F^3)+
	\text{dim}(\repA)\Big(\sum_m q_m^{a} F^m\Big)^3\Big)\\
	&+&\frac{1}{2}\sum_m q_m^{a} F^m\Big(\text{tr}_{\repA} 
	(F^2)+\frac{1}{24} \text{dim}(\repA)\text{tr}R^2\Big)\,, 
	\nn
\eea
where we denoted the representation $\repr$ by 
$\repr=\repA_{\underline{q}}$, with irreducible 
representations $\repA$, of the purely non-Abelian gauge factor and with 
$U(1)$ charges $\underline{q}=(q_1,\ldots,q_k)$ under the $n_{U(1)}$ 
$U(1)$ factors. Inserting this into \eqref{I6-anomaly} we obtain 
the total anomaly $I_6$ of chiral matter as
\beq	
I_6=\sum_{\repA}\sum_{\underline{q}}n(\repA_{\underline{q}})I_{1/2}
(\repA_{\underline{q}})\,,
\eeq
where we split the sum over the non-Abelian representations 
$\repA$ and their possibly different $U(1)$ charges 
$\underline{q}$.

Now, taking the Green-Schwarz terms \eqref{GS-term} into account
we can write the anomaly cancellation conditions as
\bea	\label{mixed-anomalies}
F^mF^nF^k&:&\frac{1}
{6}\sum_{\underline{q}}n(\underline{q})q_{(m}q_nq_{k)}
=\frac1{4}b_{(mn}^\mathcal{A}\Theta_{k)\mathcal{A} }\\
F^m\text{tr}_{\mathbf{f}}(F^I)^2&:&\frac12\sum_{\repA^I}\sum_{\underline{q}}
n(\repA^I_{\underline{q}})U(\repA)q_m
=\frac{1}{4\lambda_I} 
b_{I}^\mathcal{A}\Theta_{\mathcal{A}m}\nn\\
F^m\text{tr}R^2&:&\frac{1}
{48}\sum_{\underline{q}}n(\underline{q})q_m=
-\frac{1}{16}a^\mathcal{A}\Theta_{\mathcal{A}m}\nn
\eea
where we now wrote the representation $\repr$ in terms of 
representations of $G_{(I)}$ and the $U(1)$-factors as
$\repr=(\repA^1,\ldots,\repA^{n_G})_{\underline{q}}$. In addition 
we split the non-Abelian field strength $F$ into its field 
strengths $F^I$ for each group factor $G_{(I)}$,
$F=\sum_I F^I$. Furthermore, we introduced the number of chiral 
multiplets $n(\underline{q})$ respectively 
$n(\repA^I_{\underline{q}})$ with charges $\underline{q}$ 
respectively in the representation $\mathbf{r}^I_{\underline{q}}$. 
Furthermore we used $\text{tr}F^I=0$ when evaluating 
\bea
\text{tr}_\repr F\!\!&\!\!=\!\!&\!\!\sum_m\text{tr}_\repr F^m=\text{dim}(\repr)\sum_m q_m F^m\,,\\
\text{tr}_\repr F^3\!\!&\!\!=\!\!&\!\!\text{dim}(\repr)\Big(\sum_I \frac{1}{\text{dim}(\repA^I)}\text{tr}_{\repA^I}(F^I)^3+3\sum_{I,m} \frac{q_m}{\text{dim}(\repA^I)}\text{tr}_{\repA^I}(F^I)^2F^m+\sum_{m,k,n}q_mq_nq_k F^mF^nF^k\Big) \,,\nn
\eea
Finally, we expressed all traces as traces over 
the fundamental representation employing \eqref{trF^3} and 
\beq
	\text{tr}_{\repr}
	(F^2)=U(\repr)\text{tr}_{\mathbf{f}}(F^2)\,.
\eeq

For later reference  we summarize the following trace relations between the fundamental representation 
and the rank two and
rank three anti-symmetric tensor representation of $SU(N)$,
\bea \label{F^2trace-relations}
	\tr F^2&=&2N \text{tr}_{\mathbf{f}}F^2\,,\\
   \text{tr}_{\mathbf{as}_2} F^2 &=& (N-2)  \text{tr}_{\mathbf{f}} F^2 \ ,\nn\\
   \text{tr}_{\mathbf{as}_3} F^2 &=&\frac{1}{2}(N^2-5N+6) \text{tr}_{\mathbf{f}} F^2\, . \nn
\eea
Here we denote by $\tr$ the trace in the adjoint representation
of $SU(N)$.

\section{Dynkin labels of $SU(5)$ representations}
\label{app:SU5reps}

The Cartan matrix of $SU(5)$ reads
\beq
	C_{ij}^{SU(5)}=\begin{pmatrix}
	2&-1&0&0\\
	-1&2&-1&0\\
	0&-1&2&-1\\
	0&0&-1&2
	\end{pmatrix}\,.
\eeq
The Dynkin labels of the $\mathbf{5}$ of $SU(5)$ read
\beq
	\mathbf{5}=\{(1, 0, 0, 0),\, (-1, 1, 0, 0),\, (0, -1, 1, 0),\, (0, 0, -1, 1),\, (0, 0, 0, -1)\}\,.
\eeq
The weights of the representation are obtained by subtracting successively the rows of the 
Cartan matrix of $SU(5)$ from the highest weight $\Lambda=(1,0,0,0)$.

The Dynkin labels of the $\mathbf{10}$ of $SU(5)$ are constructed from the highest weight 
$\Lambda=(0,1,0,0)$ and read
\bea
	\mathbf{10}&=&\{(0, 1, 0, 0),\, (1, -1, 1, 0),\, (-1, 0, 1, 0)\,, (1, 0, -1, 1),\, 
	(-1, 1, -1, 1),\,\nn\\&\phantom{=}& (1, 0, 0, -1),\, (0, -1, 0, 1),\,(-1, 1, 0, -1),\, (0, -1, 1, -1),\, (0, 0, -1, 0)\}\,.
\eea
The complex conjugates are obtained from the highest weights $\Lambda=(0,0,1,0)$ and $\Lambda=(0,0,0,1)$
for $\overline{\mathbf{10}}$ respectively $\bar{\mathbf{5}}$. Their Dynkin labels are the negative of the Dynkin
labels of the $\mathbf{10}$ and $\mathbf{5}$ by complex conjugation.

\end{document}